\documentclass[]{aa}
\usepackage{graphicx}
\usepackage[figuresright]{rotating}
\usepackage[authoryear]{natbib}
\usepackage{rotating}
\usepackage{subfigure}
\usepackage{tabularx}
\usepackage{txfonts}
\usepackage{longtable,lscape}

\newcommand{\vect}[1]{{\vec{#1}}}
\newcommand{\mydotfill}{\leaders\hbox to 2pt{\hss.\hss}\hfill\phantom{.}}

\usepackage{dblfloatfix}
\begin{document}
\title{Comparing the statistics of interstellar turbulence in simulations and observations:}
\subtitle{Solenoidal versus compressive turbulence forcing}
\titlerunning{Turbulence forcing in simulations and observations}
\author{C.~Federrath\inst{1,2,3} \and J.~Roman-Duval\inst{4,5} \and R.~S.~Klessen\inst{1} \and W.~Schmidt\inst{6,7} \and M.-M.~Mac Low\inst{2,3}}
\authorrunning{Federrath et al.}
%\offprints{C.~Federrath}
\institute{Zentrum f\"ur Astronomie der Universit\"at Heidelberg, Institut f\"ur Theoretische Astrophysik, Albert-Ueberle-Str.~2, D-69120 Heidelberg, Germany\\
\email{chfeder@ita.uni-heidelberg.de}
\and
Max-Planck-Institut f\"ur Astronomie, K\"onigstuhl 17, D-69117 Heidelberg, Germany
\and
Department of Astrophysics, American Museum of Natural History, Central Park West at 79th Street, New York, NY 10024-5192, USA
\and
Space Telescope Science Institute, 3700 San Martin Drive, Baltimore, MD 21218, USA
\and
Astronomy Department, Boston University, 725 Commonwealth Avenue, Boston, MA 02215, USA
\and
Institut f\"ur Astrophysik, Universit\"at G\"ottingen, Friedrich-Hund-Platz 1, D-37077 G\"ottingen, Germany
\and
Lehrstuhl f\"ur Astronomie, Institut f\"ur Theoretische Physik und Astrophysik, Universit\"at W\"urzburg, Am Hubland, D-97074 W\"urzburg, Germany
}
\date{Received month day, year; accepted month day, year}

\abstract{Density and velocity fluctuations on virtually all scales observed with modern telescopes show that molecular clouds (MCs) are turbulent. The forcing and structural characteristics of this turbulence are, however, still poorly understood.}{To shed light on this subject, we study two limiting cases of turbulence forcing in numerical experiments: solenoidal (divergence-free) forcing and compressive (curl-free) forcing, and compare our results to observations.}{We solve the equations of hydrodynamics on grids with up to $1024^3$ cells for purely solenoidal and purely compressive forcing. Eleven lower-resolution models with different forcing mixtures are also analysed.}{Using Fourier spectra and $\Delta$-variance, we find velocity dispersion--size relations consistent with observations and independent numerical simulations, irrespective of the type of forcing. However, compressive forcing yields stronger compression at the same RMS Mach number than solenoidal forcing, resulting in a three times larger standard deviation of volumetric and column density probability distributions (PDFs). We compare our results to different characterisations of several observed regions, and find evidence of different forcing functions. Column density PDFs in the \object{Perseus MC} suggest the presence of a mainly compressive forcing agent within a shell, driven by a massive star. Although the PDFs are close to log-normal, they have non-Gaussian skewness and kurtosis caused by intermittency. Centroid velocity increments measured in the \object{Polaris Flare} on intermediate scales agree with solenoidal forcing on that scale. However, $\Delta$-variance analysis of the column density in the \object{Polaris Flare} suggests that turbulence is driven on large scales, with a significant compressive component on the forcing scale. This indicates that, although likely driven with mostly compressive modes on large scales, turbulence can behave like solenoidal turbulence on smaller scales. Principal component analysis of \object{G216-2.5} and most of the \object{Rosette MC} agree with solenoidal forcing, but the interior of an ionised shell within the \object{Rosette MC} displays clear signatures of compressive forcing.}
{The strong dependence of the density PDF on the type of forcing must be taken into account in any theory using the PDF to predict properties of star formation. We supply a quantitative description of this dependence. We find that different observed regions show evidence of different mixtures of compressive and solenoidal forcing, with more compressive forcing occurring primarily in swept-up shells. Finally, we emphasise the role of the sonic scale for protostellar core formation, because core formation close to the sonic scale would naturally explain the observed subsonic velocity dispersions of protostellar cores.}

\keywords{hydrodynamics --- ISM: clouds --- ISM: kinematics and dynamics --- methods: numerical --- methods: statistical --- turbulence}

\maketitle

\section{Introduction}
Studying the density and velocity distributions of interstellar gas provides essential information about virtually all physical processes relevant to the dynamical evolution of the interstellar medium (ISM). Along with gravity, magnetic fields and the thermodynamics of the gas, supersonic turbulence plays a fundamental role in determining the density and velocity statistics of the ISM \citep[e.g.,][]{ScaloEtAl1998}. Thus, supersonic turbulence is considered a key process for star formation \citep{MacLowKlessen2004,ElmegreenScalo2004,ScaloElmegreen2004,McKeeOstriker2007}.

In this paper, we continue our analysis of the density probability distribution function (PDF) obtained in numerical experiments of driven supersonic isothermal turbulence.  Understanding the density PDF and its turbulent origin is essential, because it is a key ingredient for analytical models of star formation: The turbulent density PDF is used to explain the stellar initial mass function \citep{PadoanNordlund2002,HennebelleChabrier2008,HennebelleChabrier2009}, the star formation rate \citep{KrumholzMcKee2005,KrumholzMcKeeTumlinson2009,PadoanNordlund2009}, the star formation efficiency \citep{Elmegreen2008}, and the Kennicutt-Schmidt relation on galactic scales \citep{Elmegreen2002,Kravtsov2003,Tassis2007}. In \citet{FederrathKlessenSchmidt2008}, we found that supersonic turbulence driven by a purely compressive (curl-free) force field yields a density PDF with roughly three times larger standard deviation compared to solenoidal (divergence-free) turbulence forcing, which strongly affects the results obtained in these analytical models. Here, we want to compare our results for the density PDF to observations of column density PDFs \citep[e.g.,][]{GoodmanPinedaSchnee2009}.

Moreover, in \citet{FederrathKlessenSchmidt2009} we investigated the fractal density distribution of our two models with solenoidal and compressive turbulence forcing, which showed that compressive forcing yields a significantly lower fractal dimension ($D_\mathrm{f}\approx2.3$) compared to solenoidal forcing ($D_\mathrm{f}\approx2.6$). In the present contribution, we consider the scaling of centroid velocity increments computed for these models, and we compare them to observations of the \object{Polaris Flare} by \citet[][]{HilyBlantFalgaronePety2008}. We additionally used principal component analysis and compared our results to observations of the \object{G216-2.5} (\object{Maddalena's Cloud}) and the \object{Rosette MC} by \citet{HeyerWilliamsBrunt2006}.

Our results indicate that interstellar turbulence is driven by mixtures of solenoidal and compressive forcing. The ratio between solenoidal and compressive modes of the turbulence forcing may vary strongly across different regions of the ISM. This provides an explanation for the apparent lack of correlation between turbulent density and velocity dispersions found in observations \citep[e.g.,][]{GoodmanPinedaSchnee2009,PinedaCaselliGoodman2008}. We conclude that solenoidal forcing is more likely to be realised in quiescent regions with low star formation activity as in the Polaris Flare and in Maddalena's Cloud. On the other hand, in regions of enhanced stellar feedback, compressive forcing leads to larger standard deviations of the density PDFs, as seen in one of the subregions of the \object{Perseus MC} surrounding a central B star. Moreover, compressive forcing exhibits a higher scaling exponent of principal component analysis than solenoidal forcing. This higher scaling exponent is consistent with the measured scaling exponent for the interior of an ionising shell in the \object{Rosette MC}.

In~\S~\ref{sec:methods}, we explain the numerical setup and turbulence forcing used for the present study. We discuss our results obtained using PDFs, centroid velocity increments, principal component analysis, Fourier spectrum functions, and $\Delta$-variance analyses in~\S~\ref{sec:pdfs}, \ref{sec:cvis}, \ref{sec:pca}, \ref{sec:spectra}, and~\ref{sec:deltavar}, respectively. In each of these sections, we compare the turbulence statistics obtained for solenoidal and compressive forcing with observational data available in the literature. In~\S~\ref{sec:sonicscale}, we discuss the possibility that transonic pre-stellar cores typically form close to the sonic scale in a globally supersonic, turbulent medium. Section~\ref{sec:limitations} provides a list of the limitations in our comparison of numerical simulations with observations. A summary of our results and conclusions is given in~\S~\ref{sec:conclusions}.

\section{Simulations and methods} \label{sec:methods}
The piecewise parabolic method \citep{ColellaWoodward1984}, implemented in the astrophysical code FLASH3 \citep{FryxellEtAl2000,DubeyEtAl2008} was used to integrate the equations of hydrodynamics on three-dimensional (3D) periodic uniform grids with $256^3$, $512^3$, and $1024^3$ grid points. Since isothermal gas is assumed throughout this study, it is convenient to define
\begin{equation}
s\equiv\ln{\frac{\rho}{\left<\rho\right>}} \label{eq:sdefinition}
\end{equation}
as the natural logarithm of the density divided by the mean density $\left<\rho\right>$ in the system. For isothermal gas, the pressure, $P=\rho c_\mathrm{s}^2$, is proportional to the density $\rho$ with the constant sound speed $c_\mathrm{s}$. The equations of hydrodynamics solved here are consequently given by
\begin{eqnarray}
\frac{\partial s}{\partial t} + (\vect{v}\cdot\nabla)s & = & -\nabla\cdot\vect{v} \label{eq:hydro1} \\
\frac{\partial\vect{v}}{\partial t} + (\vect{v}\cdot\nabla)\vect{v} & = & -c_\mathrm{s}^2\,\nabla{s} + \vect{f}\;, \label{eq:hydro2}
\end{eqnarray}
where $\vect{v}$ denotes the velocity of the gas. An energy equation is not needed, because the gas is isothermal. The assumption of isothermal gas is very crude, but may still provide an adequate physical approximation to the real thermodynamics in dense molecular gas \citep{WolfireEtAl1995,PavlovskiSmithMacLow2006}. We discuss further limitations of our simulations in~\S~\ref{sec:limitations}. The stochastic forcing term $\vect{f}$ is used to drive turbulent motions.

\subsection{Forcing module} \label{sec:forcing}
Equations~(\ref{eq:hydro1}) and~(\ref{eq:hydro2}) have been solved before in the context of molecular cloud dynamics, studying compressible turbulence with either solenoidal (divergence-free) forcing or with a 2:1 mixture of solenoidal to compressive modes in the turbulence forcing \citep[e.g.,][]{PadoanNordlundJones1997,StoneOstrikerGammie1998,MacLowEtAl1998,MacLow1999,KlessenHeitschMacLow2000,HeitschMacLowKlessen2001,Klessen2001,BoldyrevNordlundPadoan2002,LiKlessenMacLow2003,PadoanJimenezNordlundBoldyrev2004,JappsenEtAl2005,BallesterosEtAl2006,KritsukEtAl2007,DibEtAl2008,KissmannEtAl2008,OffnerKleinMcKee2008,SchmidtEtAl2009}. The case of a 2:1 mixture of solenoidal to compressive modes is the natural result obtained for 3D forcing, if no Helmholtz decomposition (see below) is performed. Then, the solenoidal modes occupy two of the three available spatial dimensions on average, while the compressive modes only occupy one \citep{ElmegreenScalo2004,FederrathKlessenSchmidt2008}. In the present study, the solenoidal forcing case is thus also used as a control run for comparison with previous studies using solenoidal forcing. However, we additionally applied purely compressive (curl-free) forcing and analysed the resulting turbulence statistics in detail. Each simulation at a resolution of $1024^3$ grid cells consumed roughly $100\,000\:\mathrm{CPU\:hours}$. Therefore, we concentrated on two extreme cases of turbulence forcing with high resolution: (1) the widely adopted purely solenoidal forcing ($\nabla\cdot\vect{f}=0$), and (2) purely compressive forcing ($\nabla\times\vect{f}=0$). However, we also studied eleven simulations at numerical resolution of $256^3$ in which we smoothly varied the forcing from purely solenoidal to purely compressive by producing eleven different forcing mixtures.

The forcing term $\vect{f}$ is often modelled with a spatially static pattern, for which the amplitude is adjusted in time following the methods introduced by \citet{MacLowEtAl1998} and \citet{StoneOstrikerGammie1998}. This results in a roughly constant energy input on large scales. Other studies model the random forcing term $\vect{f}$ such that it can vary in time \emph{and} space \citep[e.g.,][]{PadoanJimenezNordlundBoldyrev2004,KritsukEtAl2007,FederrathKlessenSchmidt2008,SchmidtEtAl2009}. Here, we used the Ornstein-Uhlenbeck (OU) process to model $\vect{f}$, which belongs to the latter type. The OU process is a well-defined stochastic process with a finite autocorrelation timescale. It can be used to excite turbulent motions in 3D, 2D, and 1D simulations as explained in \citet{EswaranPope1988} and \citet{SchmidtHillebrandtNiemeyer2006}. Using an OU process enables us to control the autocorrelation timescale $T$ of the forcing. The concept of using the OU process to excite turbulence and the projections in Fourier space necessary to get solenoidal and compressive force fields are described below.

The OU process is a stochastic differential equation describing the evolution of the forcing term $\widehat{\vect{f}}$ in Fourier space ($k$-space):
\begin{equation} \label{eq:ou}
\mathrm{d}\widehat{\vect{f}}\,(\vect{k},t) = f_{0\,}(\vect{k})\;\mathcal{\underline{P}}^{\,\zeta}(\vect{k})\;\mathrm{d}\mathcal{\vect{W}}(t)\,-\,\widehat{\vect{f}}\,(\vect{k},t)\,\frac{\mathrm{d}t}{T}\;.
\end{equation}
The first term on the right hand side is a diffusion term. This term is modelled using a Wiener process $\mathcal{\vect{W}}(t)$, which adds a Gaussian random increment to the vector field given in the previous time step $\mathrm{d}t$. Wiener processes are random processes, such that
\begin{equation}
\mathcal{\vect{W}}(t)-\mathcal{\vect{W}}(t-\mathrm{d}t)=\mathcal{\vect{N}}(0,\mathrm{d}t)\;,
\end{equation}
where $\mathcal{\vect{N}}(0,\mathrm{d}t)$ denotes the 3D, 2D, or 1D version of a Gaussian distribution with zero mean and standard deviation $\mathrm{d}t$. This is followed by a projection with the projection tensor $\mathcal{\underline{P}}^{\,\zeta}(\vect{k})$ in Fourier space. In index notation, the projection operator reads
\begin{equation} \label{eq:projectionoperator}
\mathcal{P}_{ij}^\zeta\,(\vect{k}) = \zeta\,\mathcal{P}_{ij}^\perp\,(\vect{k})+(1-\zeta)\,\mathcal{P}_{ij}^\parallel\,(\vect{k}) = \zeta\,\delta_{ij}+(1-2\zeta)\,\frac{k_i k_j}{|k|^2}\;,
\end{equation}
where $\delta_{ij}$ is the Kronecker symbol, and $\mathcal{P}_{ij}^\perp = \delta_{ij} - k_i k_j / k^2$ and $\mathcal{P}_{ij}^\parallel = k_i k_j / k^2$ are the fully solenoidal and the fully compressive projection operators, respectively. The projection operator serves to construct a purely solenoidal force field by setting $\zeta=1$. For $\zeta=0$, a purely compressive force field is obtained. Any combination of solenoidal and compressive modes can be constructed by choosing $\zeta\in[0,1]$. By changing the parameter $\zeta$, we can thus set the power of compressive modes with respect to the total power of the forcing. The analytical ratio of compressive power to total power can be derived from equation~(\ref{eq:projectionoperator}) by evaluating the norm of the compressive component of the projection tensor,
\begin{equation} \label{eq:Flong}
\left|(1-\zeta)\,\mathcal{P}_{ij}^\parallel\right|^2 = (1-\zeta)^2\;,
\end{equation}
and by evaluating the norm of the full projection tensor
\begin{equation} \label{eq:Ftot}
\left|\mathcal{P}_{ij}^\zeta\right|^2 = 1-2\zeta+D\zeta^2\;.
\end{equation}
The result of the last equation depends on the dimensionality $D=1,2,3$ of the forcing, because the norm of the Kronecker symbol $|\delta_{ij}|=1,\,2$ and 3 in one, two and three dimensions, respectively. The ratio of equations~(\ref{eq:Flong}) and~(\ref{eq:Ftot}) gives the ratio of compressive forcing power $F_\mathrm{long}$ to the total forcing power $F_\mathrm{tot}$ as a function of the parameter $\zeta$:
\begin{equation} \label{eq:forcing_ratio}
\frac{F_\mathrm{long}}{F_\mathrm{tot}} = \frac{(1-\zeta)^2}{1-2\zeta+D\zeta^2}\;.
\end{equation}

\begin{figure}[t]
\centerline{\includegraphics[width=1.0\linewidth]{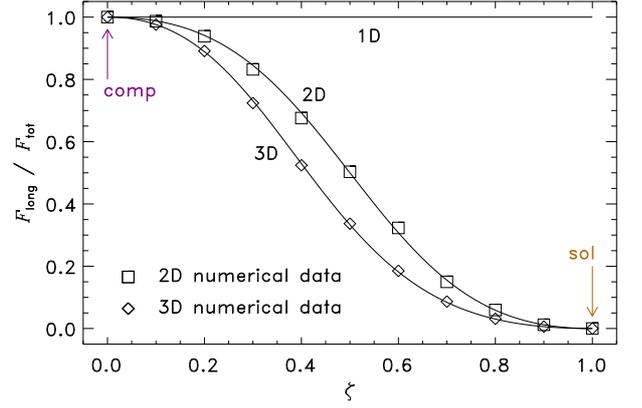}}
\caption{Ratio of compressive power to the total power in the turbulence force field. The solid lines labelled with 1D, 2D, and 3D show the analytical expectation for this ratio, equation~(\ref{eq:forcing_ratio}), as a function of the forcing parameter $\zeta$ for one-, two- and three-dimensional forcing, respectively. The diamonds and squares show results of numerical simulations in 3D and 2D with $\zeta=[0,1]$, separated by $\Delta\zeta=0.1$. Those models were run at a numerical resolution of $256^3$ and $1024^2$ grid points in 3D and 2D, respectively. The two extreme forcing cases of purely solenoidal forcing ($\zeta=1$) and purely compressive forcing ($\zeta=0$) are indicated as ''sol'' and ''comp'', respectively. Note that in any 1D model, all power is in the compressive component, and thus $F_\mathrm{long}/F_\mathrm{tot}=1$, independent of $\zeta$.}
\label{fig:forcing_ratio}
\end{figure}

Figure~\ref{fig:forcing_ratio} provides a graphical representation of this ratio for the 1D, 2D, and 3D case. For comparison, we plot numerical values of the forcing ratio obtained in eleven 3D and 2D hydrodynamical runs with resolutions of $256^3$ and $1024^2$ grid points, in which we have varied the forcing parameter $\zeta$ from purely compressive forcing ($\zeta=0$) to purely solenoidal forcing ($\zeta=1$) in the range $\zeta=[0,1]$, separated by $\Delta\zeta=0.1$. Note that a natural mixture of forcing modes is obtained for $\zeta=0.5$, which leads to $F_\mathrm{long}/F_\mathrm{tot}=1/3$ for 3D turbulence, and $F_\mathrm{long}/F_\mathrm{tot}=1/2$ for 2D turbulence. A simple way to understand this natural ratio is to consider longitudinal and transverse waves. In 3D, the longitudinal waves occupy one of the three spatial dimensions, while the transverse waves occupy two of the three on average. Thus, the longitudinal (compressive) part has a power of 1/3, while the transverse (solenoidal) part has a power of 2/3 in 3D. In 2D, the natural ratio is 1/2, because longitudinal and transverse waves are evenly distributed in two dimensions.

The second term on the right hand side of equation~(\ref{eq:ou}) is a drift term, which models the exponentially decaying correlation of the force field with itself. Thus, the autocorrelation timescale of the forcing is denoted by $T$. We set the autocorrelation timescale equal to the dynamical timescale $T=L/(2V)$ on the scale of energy injection, where $L$ is the size of the computational domain, $V=c_\mathrm{s}\mathcal{M}$ and $\mathcal{M}\approx5.5$ is the RMS Mach number in all runs. The autocorrelation timescale is therefore equal to the decay time constant in supersonic hydrodynamic and magnetohydrodynamic turbulence driven on large scales \citep{StoneOstrikerGammie1998,MacLow1999}. The forcing amplitude $f_0(\vect{k})$ is a paraboloid in 3D Fourier space, only containing power on the largest scales in a small interval of wavenumbers $1<|\vect{k}|<3$ peaking at $k=2$, which corresponds to half of the box size $L/2$. The effects of varying the scale of energy input were investigated by \citet{MacLow1999}, \citet{KlessenHeitschMacLow2000}, \citet{HeitschMacLowKlessen2001} and \citet{VazquezBallesterosKlessen2003}. Here, we consider large-scale stochastic forcing, which is closer to the observational data \citep[e.g.,][]{OssenkopfMacLow2002,BruntHeyerMacLow2009}. This type of forcing models the kinetic energy input from large-scale turbulent fluctuations, breaking up into smaller structures. Kinetic energy cascades down to smaller and smaller scales, and thereby effectively drives turbulent fluctuations on scales smaller than the turbulence injection scale.

% This approach together with the assumption of periodic boundary conditions implies that the most reasonable physical correspondence of our scale free simulations to real molecular clouds is that we are concentrating on a small subpart (e.g., $L\!\sim\!1\,\mathrm{pc}$) of a molecular cloud studying turbulence statistics with high resolution.

We have verified that our results are not sensitive to the general approach of using an Ornstein-Uhlenbeck process for the turbulence forcing. For instance, we have used an almost static forcing pattern, which is obtained in the limit $T\to\infty$ in test simulations. We have furthermore checked that the particular choice of Fourier amplitudes did not affect our results by using a band spectrum instead of a parabolic forcing spectrum. Varying these parameters did not strongly affect our results. In contrast, changing $\zeta$ from $\zeta=1$ (solenoidal forcing) to $\zeta=0$ (compressive forcing) always led to significant changes in the turbulence statistics.

\subsection{Initial conditions and post-processing} \label{sec:initandpostprocess}
Starting from a uniform density distribution and zero velocities, the forcing excites turbulent motions. Equations~(\ref{eq:hydro1}) and~(\ref{eq:hydro2}) have been evolved for ten dynamical times $T$, which allows us to study a large sample of realisations of the turbulent flow. Compressible turbulence reached a statistically invariant state within $2\,T$ \citep{FederrathKlessenSchmidt2009}. This allows us to average all statistical measures over $8\,T$ separated by $0.1\,T$ in the fully developed regime. We are thus able to average over 81 different realisations of the turbulence to improve statistical significance. The 1-$\sigma$ temporal fluctuations obtained from this averaging procedure are indicated as error bars for the PDFs, centroid velocity increments, principal component analysis, Fourier spectra and $\Delta$-variance analyses in the following sections and in all figures showing error bars throughout this study. The forcing amplitude was adjusted to excite a turbulent flow with an RMS Mach number $\mathcal{M}\approx5.5$ in all cases. We use the RMS Mach number as the control parameter, because this dimensionless number determines most of the physical properties of scale-invariant turbulent flows and is often used to derive important flow statistics such as the standard deviation of the density distribution. However, in the next section we show that the latter depends sensitively on the turbulence forcing parameter $\zeta$ as well.

\begin{figure*}[tp]
\begin{center}
\begin{tabular}{rr}
\includegraphics[width=0.44\linewidth]{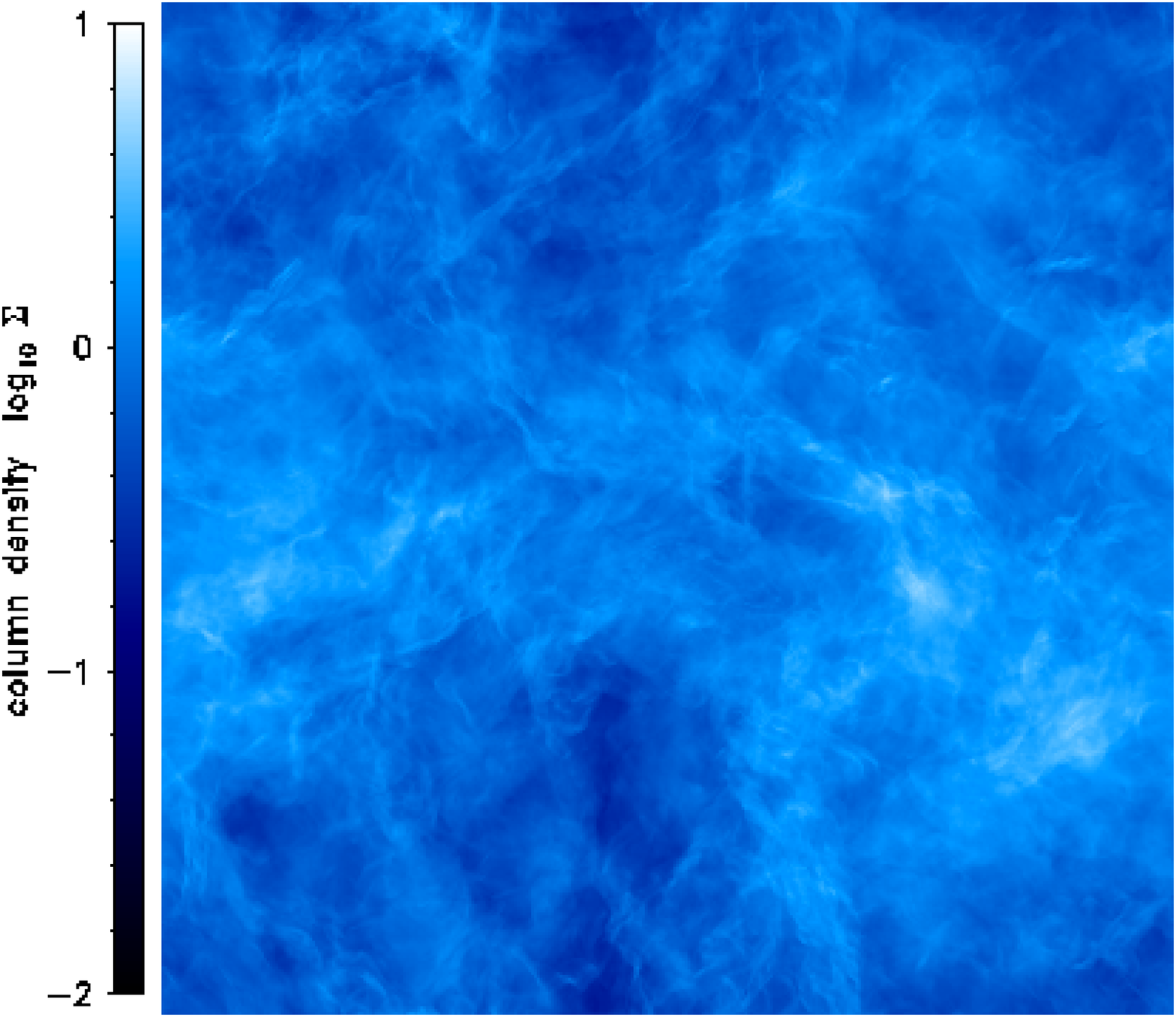} &
\includegraphics[width=0.44\linewidth]{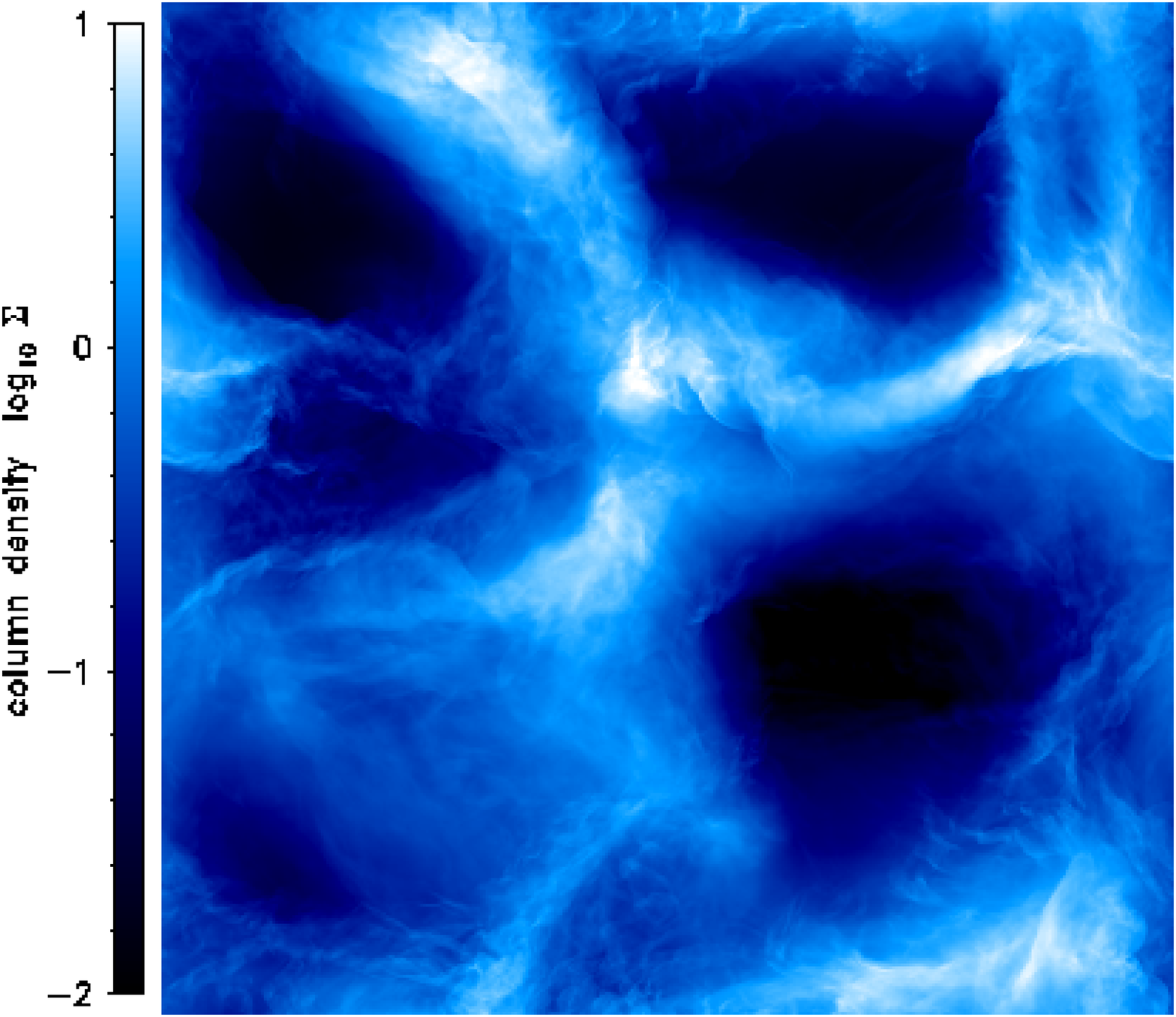} \\
\includegraphics[width=0.44\linewidth]{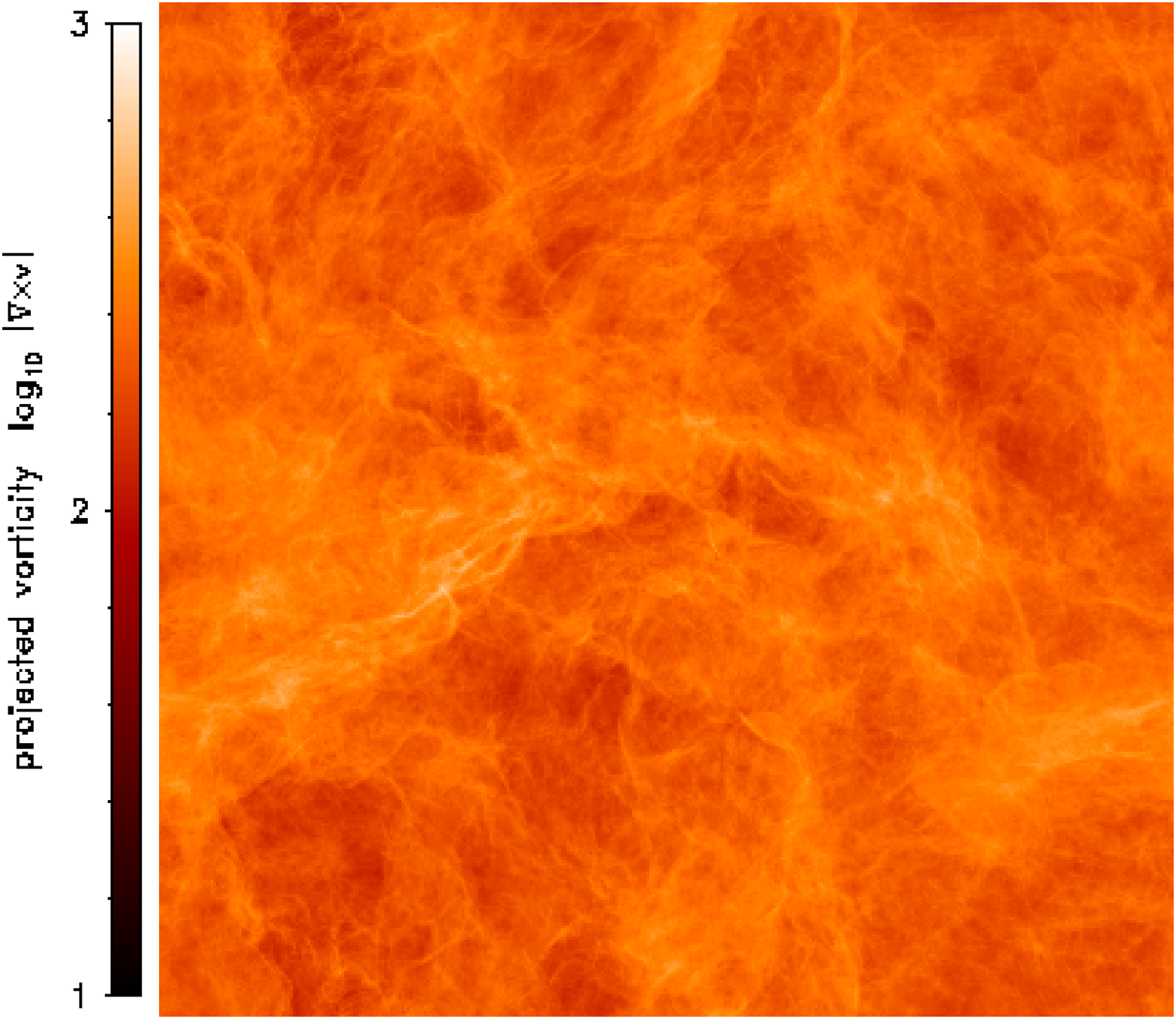} &
\includegraphics[width=0.44\linewidth]{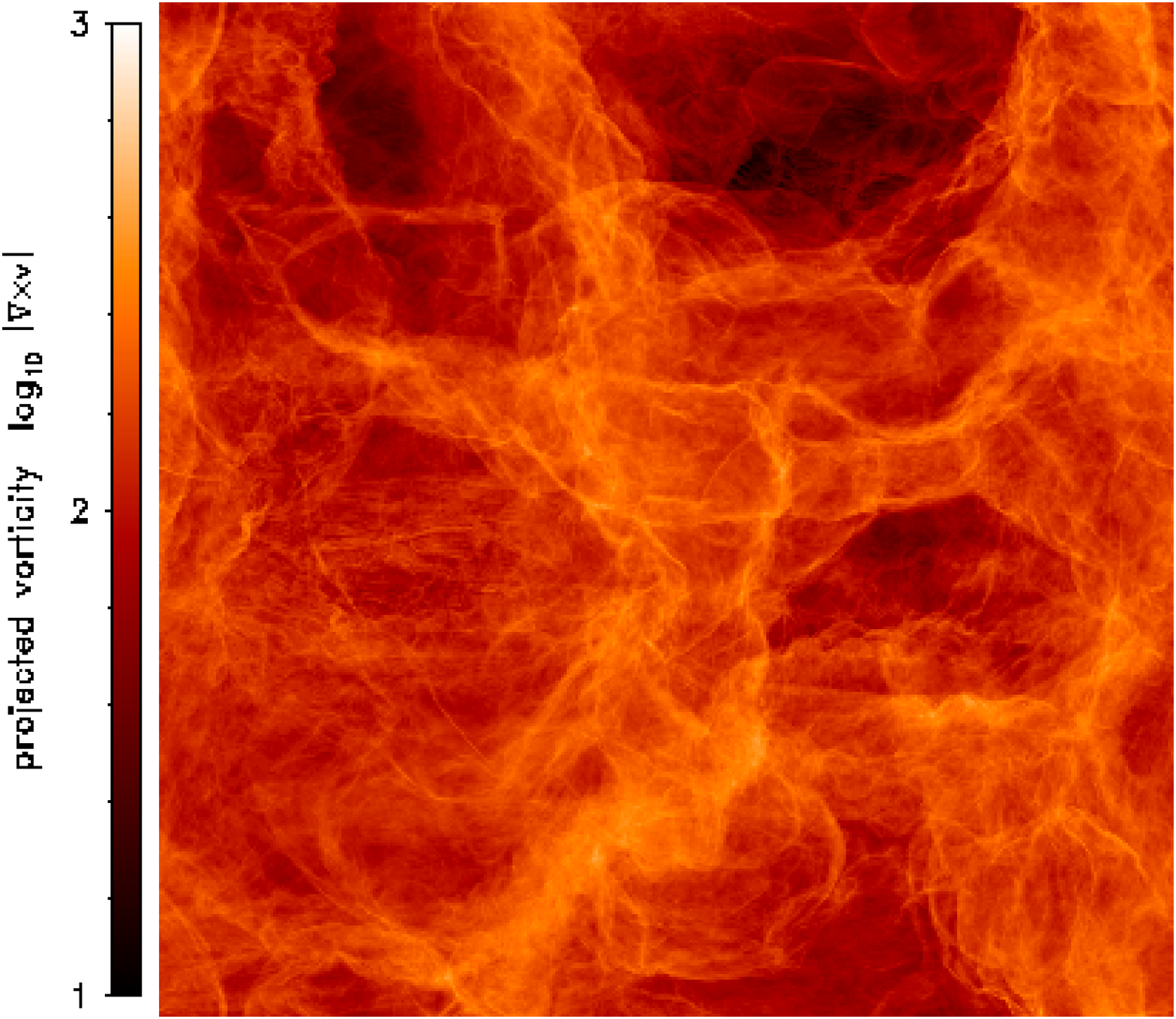} \\
\includegraphics[width=0.44\linewidth]{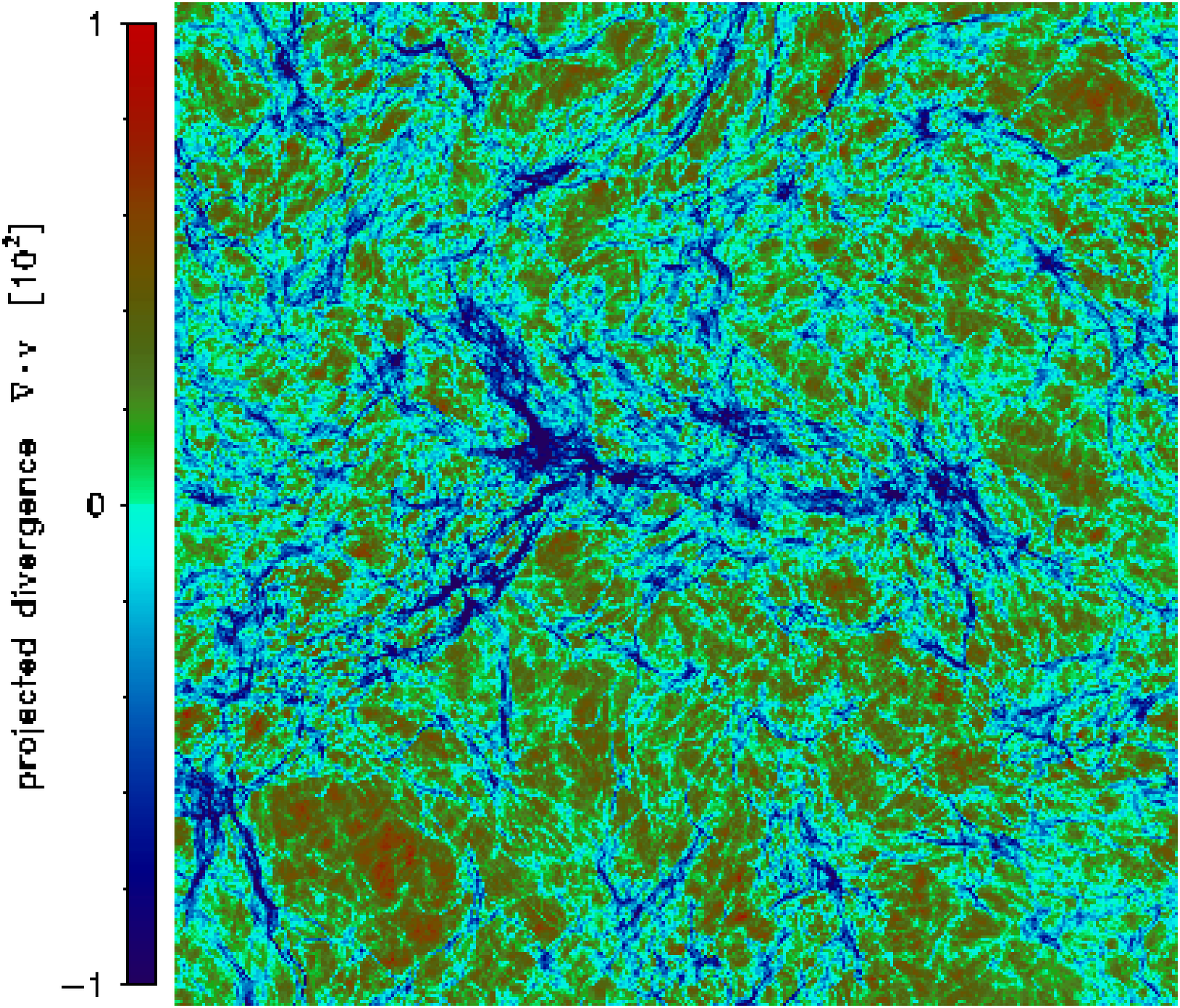} &
\includegraphics[width=0.44\linewidth]{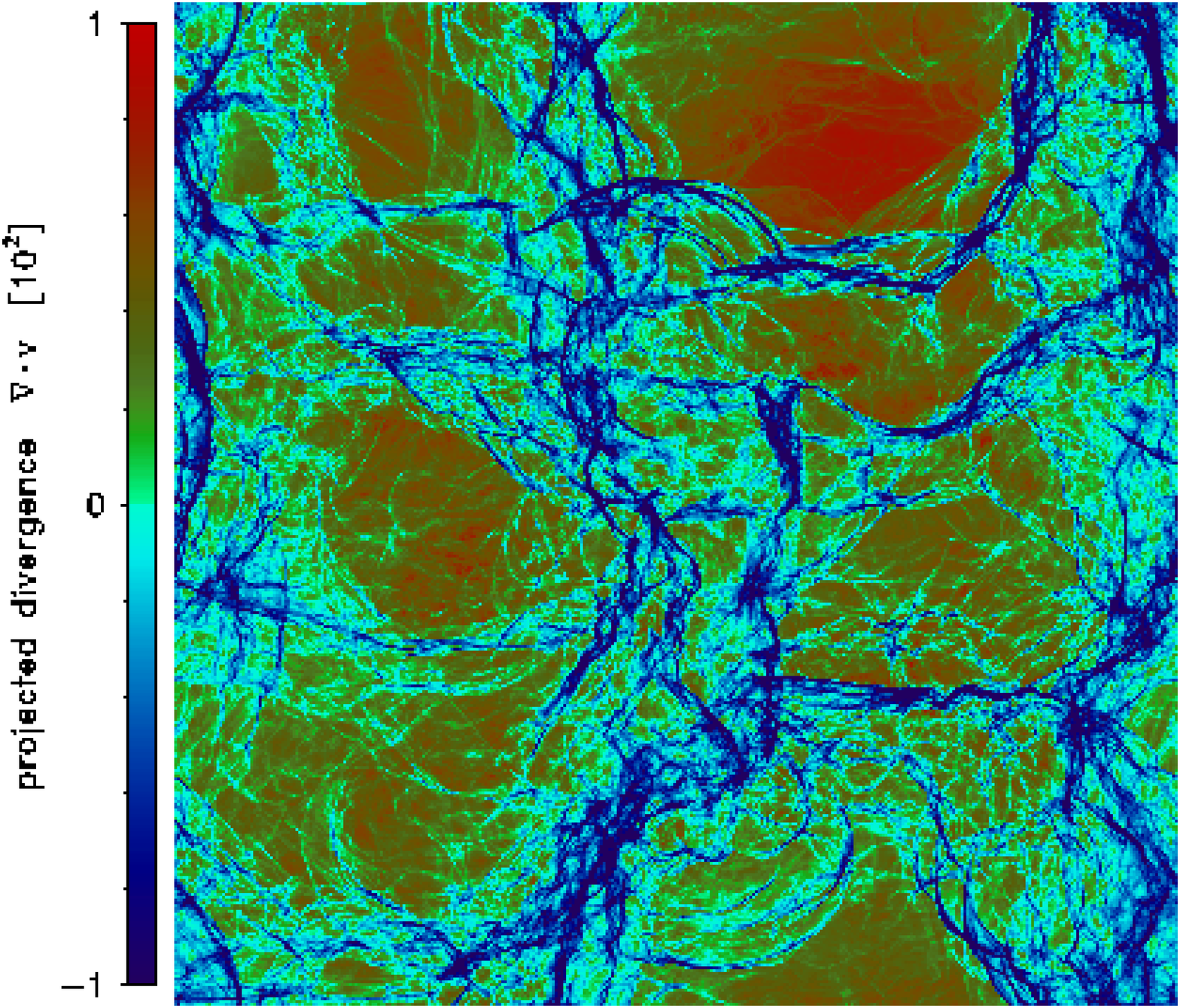}
\end{tabular}
\end{center}
\caption{Maps showing density (\emph{top}), vorticity (\emph{middle}) and divergence (\emph{bottom}) in projection along the $z$-axis at time $t=2\,T$ as an example for the regime of statistically fully developed, compressible turbulence for solenoidal forcing (\emph{left}) and compressive forcing (\emph{right}). \emph{Top panels:} Column density fields in units of the mean column density. Both maps show three orders of magnitude in column density with the same scaling and magnitudes for direct comparison. \emph{Middle panels:} Projections of the modulus of the vorticity $\left|\nabla\times\vect{v}\right|$. Regions of intense vorticity appear to be elongated filamentary structures often coinciding with positions of intersecting shocks. \emph{Bottom panels:} Projections of the divergence of the velocity field $\nabla\cdot\vect{v}$ showing the positions of shocks. Negative divergence corresponds to compression, while positive divergence corresponds to rarefaction.}
\label{fig:snapshots}
\end{figure*}

\begin{figure}[t]
\centerline{\includegraphics[width=1.0\linewidth]{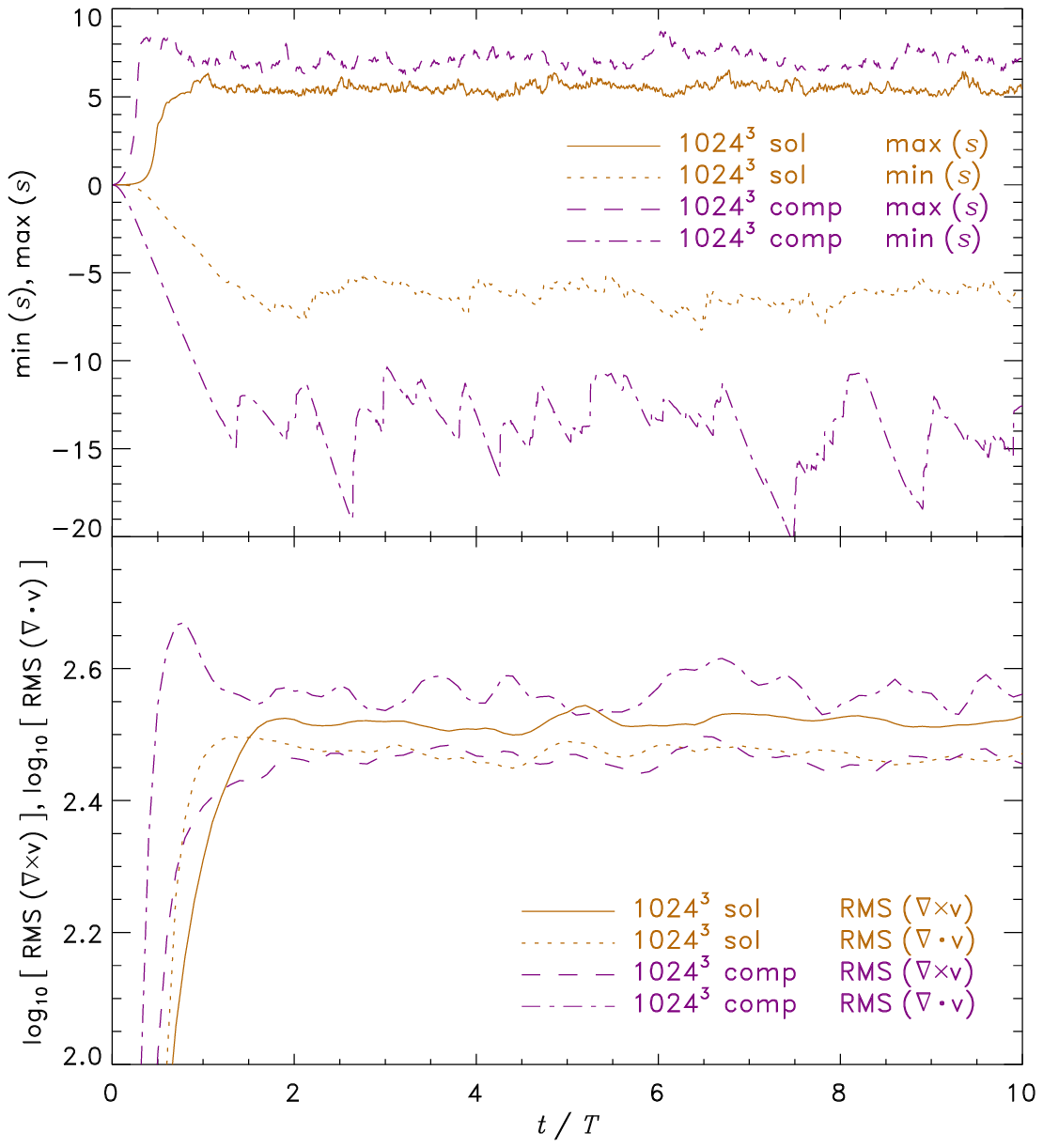}}
\caption{\emph{Top panel}: Minimum and maximum logarithmic density $s=\ln\left(\rho/\left<\rho\right>\right)$ as a function of the dynamical time $T$. Note that compressive forcing yields much stronger compression and rarefaction compared to solenoidal forcing, although the RMS Mach number is roughly the same in both cases \citep[see][Fig.~2]{FederrathKlessenSchmidt2009}. \emph{Bottom panel}: RMS vorticity $\langle(\nabla\times\vect{v})^2\rangle^{1/2}$ and RMS divergence $\langle(\nabla\cdot\vect{v})^2\rangle^{1/2}$ as a function of the dynamical time. Within the first $2\,T$, a statistically steady state was reached for both solenoidal (sol) and compressive (comp) forcing. This allows us to average statistical measures (probability density functions, centroid velocity increments, principal component analysis, Fourier spectra and $\Delta$-variances) in the range $2 \leq t/T \leq 10$ to improve statistical significance of our results and to estimate the amplitude of temporal fluctuations (snapshot-to-snapshot variations) between different realisations of the turbulence.}
\label{fig:evol}
\end{figure}

Figure~\ref{fig:snapshots} (top panels) shows column density fields projected along the $z$-axis from a randomly selected snapshot at time $t=2\,T$ in the regime of fully developed, statistically stationary turbulence for solenoidal (left) versus compressive forcing (right). This regime was reached after 2 dynamical times $T$, which is shown in Figure~\ref{fig:evol} for the minimum and maximum logarithmic densities $s$ (top panel) and RMS curl and divergence of the velocity field (bottom panel) as a function of the dynamical time. It is evident that compressive forcing produces higher density contrasts, resulting in higher density peaks and bigger voids compared to solenoidal forcing.

\section{The probability density function of the gas density} \label{sec:pdfs}
It is interesting to study the probability distribution of turbulent density fluctuations, because it is a key ingredient for many analytical models of star formation: it is used to explain the stellar initial mass function \citep{PadoanNordlund2002,HennebelleChabrier2008,HennebelleChabrier2009}, the star formation rate \citep{KrumholzMcKee2005,KrumholzMcKeeTumlinson2009,PadoanNordlund2009}, the star formation efficiency \citep{Elmegreen2008}, and the Kennicutt-Schmidt relation on galactic scales \citep{Elmegreen2002,Kravtsov2003,Tassis2007}.

The probability to find a volume with gas density in the range $[\rho,\rho\!+\!\mathrm{d}\rho]$ is given by the integral over the volume-weighted probability density function (PDF) of the gas density: $\int_{\rho}^{\rho+\mathrm{d}\rho}\,p_\rho(\rho')\,\mathrm{d}\rho'$. Thus, the PDF $p$ describes a \emph{probability density}, which has dimensions of probability divided by gas density in the case of $p_\rho(\rho)$. By the same definition, $p_s(s)$ denotes the PDF of the logarithmic density $s=\ln(\rho/\left<\rho\right>)$.

\begin{figure}[t]
\centerline{\includegraphics[width=1.0\linewidth]{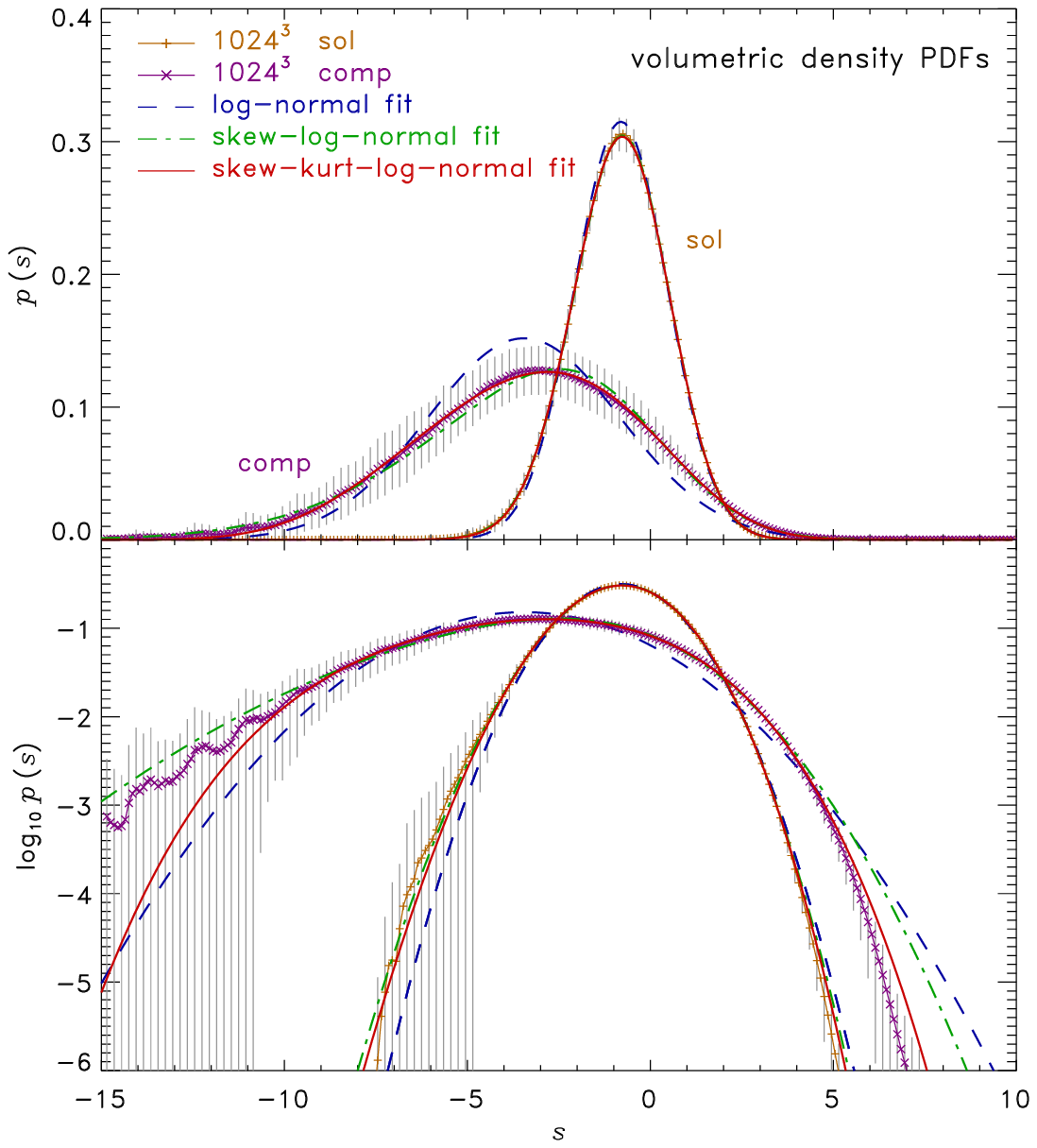}}
\caption{Volume-weighted density PDFs $p(s)$ of the logarithmic density $s=\ln(\rho/\left<\rho\right>)$ in linear scaling (\emph{top panel}), which displays the peak best, and in logarithmic scaling (\emph{bottom panel}) to depict the low- and high-density wings. The PDF obtained from compressive forcing ($1024^3\,\mathrm{comp}$) is significantly wider than the solenoidal one ($1024^3\,\mathrm{sol}$). The peak is shifted to lower values of the logarithmic density $s$, because of mass conservation, defined in eq.~(\ref{eq:mass-conservation}). The density PDF from solenoidal forcing is compatible with a Gaussian distribution. However, there are also non-Gaussian features present, which are associated with intermittency effects. These are more prominent in the density PDF obtained from compressive forcing, exhibiting statistically significant deviations from a perfect log-normal (fit using eq.~\ref{eq:log-normal} shown as dashed lines). A skewed log-normal fit (dash-dotted lines) given by eq.~(\ref{eq:skew-log-normal}) provides a better representation, but still does not fit the high-density tail of the PDF obtained for compressive forcing. Both the PDF data obtained from solenoidal and compressive forcing are best described as log-normal distributions with higher-order corrections defined in eq.~(\ref{eq:skew-kurt-log-normal}), which take into account both the non-Gaussian skewness and kurtosis of the distributions. These fits are shown as solid lines (skew-kurt-log-normal fit). The first four standardised moments defined in equations~(\ref{eq:moments}) of the distributions in $\rho$ and $s$ are summarised in Table~\ref{tab:pdfs_vol} together with the fit parameters. The grey shaded regions indicate 1-$\sigma$ error bars due to temporal fluctuations of the distributions in the regime of fully developed, supersonic turbulence. A total number of $1024^3\times81\approx10^{11}$ data points contribute to each PDF.}
\label{fig:solcomppdf}
\end{figure}

Figure~\ref{fig:solcomppdf} presents the comparison of the time-averaged volume-weighted density PDFs $p_s(s)$ obtained for solenoidal and compressive forcing. The linear plot of $p_s(s)$ (top panel) displays the peak best, whereas the logarithmic representation (bottom panel) reveals the low- and high-density wings of the distributions. Three different fits to analytic expressions (discussed below) are shown as well.

\subsection{The density PDF for solenoidal forcing} \label{sec:pdfsol}
In numerical experiments of driven supersonic isothermal turbulence with solenoidal and/or weakly compressive forcing \citep[e.g.,][]{Vazquez1994,PadoanNordlundJones1997,StoneOstrikerGammie1998,MacLow1999,NordlundPadoan1999,BoldyrevNordlundPadoan2002,LiKlessenMacLow2003,PadoanJimenezNordlundBoldyrev2004,KritsukEtAl2007,BeetzEtAl2008}, but also in decaying turbulence \citep[e.g.,][]{OstrikerGammieStone1999,Klessen2000,OstrikerStoneGammie2001,GloverMacLow2007b} it was shown that the density PDF $p_s$ is close to a log-normal distribution,
\begin{equation}
p_s\,\mathrm{d}s=\frac{1}{\sqrt{2\pi\sigma_s^2}}\,\exp\Bigg[-\frac{(s-\left<s\right>)^2}{2\sigma_s^2}\Bigg]\,\mathrm{d}s\;, \label{eq:log-normal}
\end{equation}
where the mean $\left<s\right>$ is related to the standard deviation $\sigma_s$ by $\left<s\right>=-\sigma_s^2/2$ due to the constraint of mass conservation \citep[e.g.,][]{Vazquez1994}:
\begin{equation}
\int_{-\infty}^{\infty}\exp\left(s\right)\,p_s\,\mathrm{d}s=\int_{0}^{\infty}\rho\,p_{\rho}\,\mathrm{d}\rho=\left<\rho\right>\;. \label{eq:mass-conservation}
\end{equation}
Equation~(\ref{eq:mass-conservation}) simply states that the mean density has to be recovered. This constraint together with the PDF normalisation,
\begin{equation}
\int_{-\infty}^{\infty}p_s\,\mathrm{d}s=\int_{0}^{\infty}p_{\rho}\,\mathrm{d}\rho=1 \label{eq:normalization}
\end{equation}
must always be fulfilled for any density PDF whether log-normal or non-Gaussian.

From our simulations, we obtain density PDFs in agreement with log-normal distributions for solenoidal forcing. The log-normal fit using equation~(\ref{eq:log-normal}) is shown in Figure~\ref{fig:solcomppdf} as dashed lines. However, the PDF is not perfectly log-normal, i.e., there are weak non-Gaussian contributions \citep[see also,][]{DubinskiNarayanPhillips1995}, especially affecting the wings of the distribution. The strength of these non-Gaussian features is quantified by computing higher-order moments (skewness and kurtosis) of the distributions. The first four standardised central moments \citep[see, e.g.,][]{PressEtAl1986} of a discrete dataset $\{q\}$ with $N$ elements are defined as
\begin{equation} \label{eq:moments}
\def\arraystretch{3.0}
\begin{array}{lrrl}
\displaystyle \mathrm{mean:}     \quad\quad & \displaystyle \left<q\right> & = & \displaystyle \frac{1}{N}\sum_{i=1}^N{q_i}\\
\displaystyle \mathrm{dispersion:}   \quad\quad & \displaystyle \sigma_q       & = & \displaystyle \sqrt{\left<\left(q-\left<q\right>\right)^2\right>}\\
\displaystyle \mathrm{skewness:} \quad\quad & \displaystyle \mathcal{S}_q  & = & \displaystyle \frac{\left<\left(q-\left<q\right>\right)^3\right>}{\sigma^3}\\
\displaystyle \mathrm{kurtosis:} \quad\quad & \displaystyle \mathcal{K}_q  & = & \displaystyle \frac{\left<\left(q-\left<q\right>\right)^4\right>}{\sigma^4}\;.
\end{array}
\end{equation}
Note that in our definition of the kurtosis (also called flatness), the Gaussian distribution has $\mathcal{K}=3$. We have computed the first four statistical moments of the volumetric density PDFs shown in Figure~\ref{fig:solcomppdf}. The results are summarised in Table~\ref{tab:pdfs_vol}. The 1-$\sigma$ error given for each statistical moment was obtained by averaging over 81 realisations of the turbulence as described in~\S~\ref{sec:initandpostprocess}. Both solenoidal and compressive forcing yield density PDFs with deviations from the Gaussian 3rd order (skewness $\mathcal{S}=0$) and 4th order (kurtosis $\mathcal{K}=3$) moments.

\begin{table*}[t]
\begin{center}
\caption{Statistical moments and fit parameters of the PDFs of the volumetric density $\rho$ for solenoidal and compressive forcing shown in Fig.~\ref{fig:solcomppdf}. \label{tab:pdfs_vol}}
\def\arraystretch{1.1}
\begin{tabular}{lccc}
\hline
\hline
~ & ~ & Solenoidal Forcing & Compressive Forcing\\
\hline
Standardised Moments & $\left<\rho\right>$      & $\phantom{-}1.00\pm0.00$ & $\phantom{-}1.00\pm0.00$\\
of $\rho$ & $\sigma_\rho$ & $\phantom{-}1.89\pm0.09$ & $\phantom{-}5.86\pm0.96$\\
~ & $\mathcal{S}_\rho$ & $\phantom{-}9.03\pm1.06$ & $\phantom{-}26.7\pm10.1$\\
~ & $\mathcal{K}_\rho$ & $\phantom{-}211.\pm69.8$ & $\phantom{-}1720\pm2000$\\
\hline
Standardised Moments & $\left<s\right>$      & $-0.83\pm0.05$ & $-3.40\pm0.43$\\
of $s=\ln\left(\rho/\left<\rho\right>\right)$ & $\sigma_s$ & $\phantom{-}1.32\pm0.06$ & $\phantom{-}3.04\pm0.24$\\
~ & $\mathcal{S}_s$ & $-0.10\pm0.11$ & $-0.26\pm0.20$\\
~ & $\mathcal{K}_s$ & $\phantom{-}3.03\pm0.17$ & $\phantom{-}2.91\pm0.43$\\
\hline
Skewed Log-normal Approximation & $\xi$    & $\phantom{-}0.010\pm0.050$ & $-0.048\pm0.133$\\
using equation~(\ref{eq:skew-log-normal}) & $\omega$ & $\phantom{-}1.562\pm0.035$ & $\phantom{-}4.712\pm0.193$\\
~ & $\alpha$ & $-0.911\pm0.064$ & $-2.163\pm0.173$\\
\hline
4th Order Approximation & $a_0$ & $-1.3664\pm0.0091$ & $-2.5014\pm0.0259$\\
(including skewness and kurtosis) & $a_1$ & $-0.4592\pm0.0064$ & $-0.3437\pm0.0132$\\
using equation~(\ref{eq:skew-kurt-log-normal}) & $a_2$ & $-0.3067\pm0.0052$ & $-0.0831\pm0.0030$\\
~ & $a_3$ & $-0.0073\pm0.0011$ & $-0.0065\pm0.0011$\\
~ & $a_4$ & $-0.0002\pm0.0005$ & $-0.0004\pm0.0001$\\
\hline
\end{tabular}
\end{center}
\end{table*}

\subsection{The density PDF for compressive forcing} \label{sec:pdfcomp}
Contrary to the solenoidal case, the PDF obtained for compressive forcing is not at all well fitted with the perfect log-normal functional form (dashed line in Figure~\ref{fig:solcomppdf} for compressive forcing). Due to the constraints of mass conservation (eq.~\ref{eq:mass-conservation}) and normalisation (eq.~\ref{eq:normalization}), the peak position and its amplitude cannot be reproduced simultaneously. The skewness and kurtosis for the compressive forcing case are also listed in Table~\ref{tab:pdfs_vol}. Non-Gaussian values of skewness and kurtosis, i.e., higher-order moments require modifications to the analytic expression of the log-normal PDF given by equation~(\ref{eq:log-normal}). A first step of modification is to allow for a finite skewness, which is possible with a skewed log-normal distribution \citep{Azzalini1985}
\begin{equation}
p(s)=\frac{1}{\pi\,\omega}\,\exp{\left[-\frac{(s-\xi)^2}{2\omega^2}\right]}\,\int_{-\infty}^{(s-\xi)\alpha/\omega}\exp{\left(-\frac{t^2}{2}\right)}\,\mathrm{d}t\;, \label{eq:skew-log-normal}
\end{equation}
where $\alpha$, $\xi$ and $\omega$ are fit parameters. Defining $\delta=\alpha/\sqrt{1+\alpha^2}$, the first four standardised central moments of the distribution are linked to the parameters $\alpha$, $\xi$ and $\omega$, such that

\begin{equation} \label{eq:skewlognormalmoments}
\def\arraystretch{3.0}
\begin{array}{lrrl}
\displaystyle \mathrm{mean:}     \quad\quad & \displaystyle \left<s\right> & = & \displaystyle \xi+\omega\,\delta\sqrt{2/\pi}\\
\displaystyle \mathrm{dispersion:}   \quad\quad & \displaystyle \sigma_s       & = & \displaystyle \omega\left(1-2\delta^2/\pi\right)^{1/2}\\
\displaystyle \mathrm{skewness:} \quad\quad & \displaystyle \mathcal{S}_s  & = & \displaystyle \frac{4-\pi}{2}\frac{(\delta\sqrt{2/\pi})^3}{(1-2\delta^2/\pi)^{3/2}}\\
\displaystyle \mathrm{kurtosis:} \quad\quad & \displaystyle \mathcal{K}_s  & = & \displaystyle \frac{2(\pi-3)(\delta\sqrt{2/\pi})^4}{(1-2\delta^2/\pi)^2}\;.
\end{array}
\end{equation}

Skewed log-normal fits are added to Figure~\ref{fig:solcomppdf} as dash-dotted lines and the corresponding fit parameters are given in Table~\ref{tab:pdfs_vol}. However, for a skewed log-normal distribution, the kurtosis is a function of the skewness, since the skewness and kurtosis in equations~(\ref{eq:skewlognormalmoments}) both depend on the same parameter $\delta$ only.

Better agreement between an analytic functional form and the measured PDF can be obtained, if the actual kurtosis of the data is taken into account as an independent parameter in the analytical approach. The fundamental derivation of a standard Gaussian distribution is given by
\begin{equation}
\ln{p(s)} = a_0+a_1s+a_2s^2\;,
\end{equation}
where one parameter is constrained by the normalisation and the two remaining ones are determined by the mean and the dispersion. We can extend this to a modified Gaussian-like distribution by including higher-order moments:
\begin{equation}
p(s)=\exp\left[a_0+a_1s+a_2s^2+a_3s^3+a_4s^4+\mathcal{O}(s^5)\right]\;. \label{eq:skew-kurt-log-normal}
\end{equation}
Here, the expansion is stopped at the 4th moment. One parameter is again given by the normalisation, and the remaining four parameters are related to the mean, dispersion, skewness and kurtosis. Fits obtained with this formula are included in Figure~\ref{fig:solcomppdf} as solid lines. The fit parameters are listed in Table~\ref{tab:pdfs_vol}. This new functional form is in good agreement with the data from solenoidal and compressive forcing, fitting both the peak and the wings very well. They follow the constraints of mass conservation and normalisation given by equations~(\ref{eq:mass-conservation}) and~(\ref{eq:normalization}). We have computed the first four moments of the fitted function and find very good agreement with the first four moments of the actual PDFs.

The fitted parameters $a_3$ and $a_4$, which represent the higher-order terms tend to zero compared to the standard Gaussian parameters $a_0$, $a_1$ and $a_2$ (see Table~\ref{tab:pdfs_vol}). This means that the higher-order corrections to the standard Gaussian are small. However, we point out that they are absolutely necessary to obtain a good analytic representation of the PDF data, given the fact that equations~(\ref{eq:mass-conservation}) and~(\ref{eq:normalization}) must always be fulfilled and that the analytic PDF should return the correct values of the numerically computed moments of the measured distributions.

In the various independent numerical simulations mentioned above, the density PDFs were close to log-normal distributions as in our solenoidal and compressive forcing cases. However, most of these studies also report considerable deviations from Gaussian PDFs, which affected mainly the low- and high-density wings of their distributions. These deviations can be associated with rare events caused by strong intermittent fluctuations during head-on collisions of strong shocks and oscillations in very low-density rarefaction waves \citep[e.g.,][]{PassotVazquez1998,KritsukEtAl2007}. The pronounced deviations from the log-normal shape of the density PDF for compressively driven turbulence were also discussed by \citet{SchmidtEtAl2009}. Even stronger deviations from log-normal PDFs were reported in strongly self-gravitating turbulent systems \citep[e.g.,][]{Klessen2000,FederrathGloverKlessenSchmidt2008,KainulainenEtAl2009}.

Intermittency is furthermore inferred from observations, affecting the wings of molecular line profiles \citep{FalgaronePhillips1990}, and the statistics of centroid velocity increments \citep{HilyBlantFalgaronePety2008}. \citet{GoodmanPinedaSchnee2009} measured column density PDFs using dust extinction and emission, as well as molecular lines of gas in the \object{Perseus MC}. Using dust extinction maps, \citet{LombardiAlvesLada2006} obtained the column density PDF for the \object{Pipe nebula}. The PDFs found in these studies roughly follow log-normal distributions. However, deviations from perfect log-normal distributions are clearly present in the density PDFs obtained in these studies. They typically exhibit non-Gaussian features. For instance, \citet{LombardiAlvesLada2006} had to apply combinations of multiple Gaussian distributions to obtain good agreement with the measured PDF data.

\subsection{Density--Mach number correlation and signatures of intermittency in the density PDFs}
As discussed by \citet{PassotVazquez1998}, a Gaussian distribution in the logarithm of the density, i.e., a log-normal distribution in $\rho$ is expected for supersonic, isothermal turbulent flows. The fundamental assumption behind this model is that density fluctuations are built up by a hierarchical process. The local density $\rho(\vect{r},t)$ at a given position $\vect{r}$ is determined by a Markov process, i.e., by the product $\rho(t_n)=\delta(t_{n-1})\rho(t_{n-1})=\dots=\delta(t_0)\rho(t_0)$ of a large number of \emph{independent} random fluctuations $\delta(t_n)>0$ in time \citep{Vazquez1994}. If these fluctuations were indeed independent, the quantity $s=\ln(\rho/\left<\rho\right>)$ would be determined by the sum of this large number of local fluctuations and the distribution in $s$ becomes a Gaussian distribution according to the central limit theorem. Since the equations~(\ref{eq:hydro1}) and~(\ref{eq:hydro2}) are invariant under the transformation $s \to s+s_0$ for an arbitrary constant $s_0$, the random variable $s(t_n)$ should be independent of the local Mach number, and independent of the density at previous times $t_{n-1},t_{n-2},\dots,t_{0}$. As pointed out by \citet{Vazquez1994} and \citet{PassotVazquez1998}, this independence breaks down in strong shocks and density extrema, because $s_0$ cannot be arbitrarily high due to mass conservation, and an upper boundary $s_+$ exists. In consequence, if $s_+$ is reached locally, the density cannot increase anymore by a subsequent fluctuation, and the next density is \emph{not} independent of the previous timestep, causing the fundamental assumption to break down. This also applies to strong rarefaction waves, because creating shocks always produces strongly rarefied regions outside the shock.

When the fundamental assumption breaks down, density and velocity statistics are expected to become correlated \citep{Vazquez1994,PassotVazquez1998,KritsukEtAl2007}. Since in isothermal gas, the sound speed is constant, this translates directly into Mach number--density correlations. The average local Mach number $M\!=\!v/c_\mathrm{s}$ may therefore exhibit some dependence on the average local density. For instance, it is intuitively clear that head-on collisions of strong shocks produce very high density peaks. In the stagnation point of the flow, the local velocity and consequently the local Mach number will almost drop to zero. The time evolution of the maximum and minimum density in Figure~\ref{fig:evol} shows these intermittent fluctuations \citep[see also, ][]{PorterPouquetWoodward1992, KritsukEtAl2007}. The intermittent phenomenon corresponds to the situation explained above, for which $s_+$ might have been reached, and some dependence of the Mach number on density is expected.

In real molecular clouds, the maximum densities are similarly bounded, and cannot reach infinitely high values, either. This is--unlike the finite resolution constraints in simulations--because the gas becomes optically thick at a certain density ($\rho\gtrsim10^{-14}\,\mathrm{g}\,\mathrm{cm}^{-3}$), and cannot cool efficiently anymore \citep[e.g.,][and references therein]{Larson1969,Penston1969,Larson2005,JappsenEtAl2005}. The gas is not close to isothermal anymore in this regime, and adiabatic compression induced by turbulent motions remain finite in real molecular clouds. Thus, the reason for the breakdown of the density--Mach number independence is different in simulations and observations, but it might still be fundamental for the deviations from a log-normal PDF. Moreover, the existence of a characteristic scale may lead to a breakdown of the hierarchical model, and thus to a breakdown of the fundamental assumption. The scale at which supersonic turbulence becomes subsonic is such a scale. This scale is called the sonic scale, and is discussed later in \S~\ref{sec:sonicscale}.

\begin{figure*}[t]
\centerline{
\includegraphics[width=0.5\linewidth]{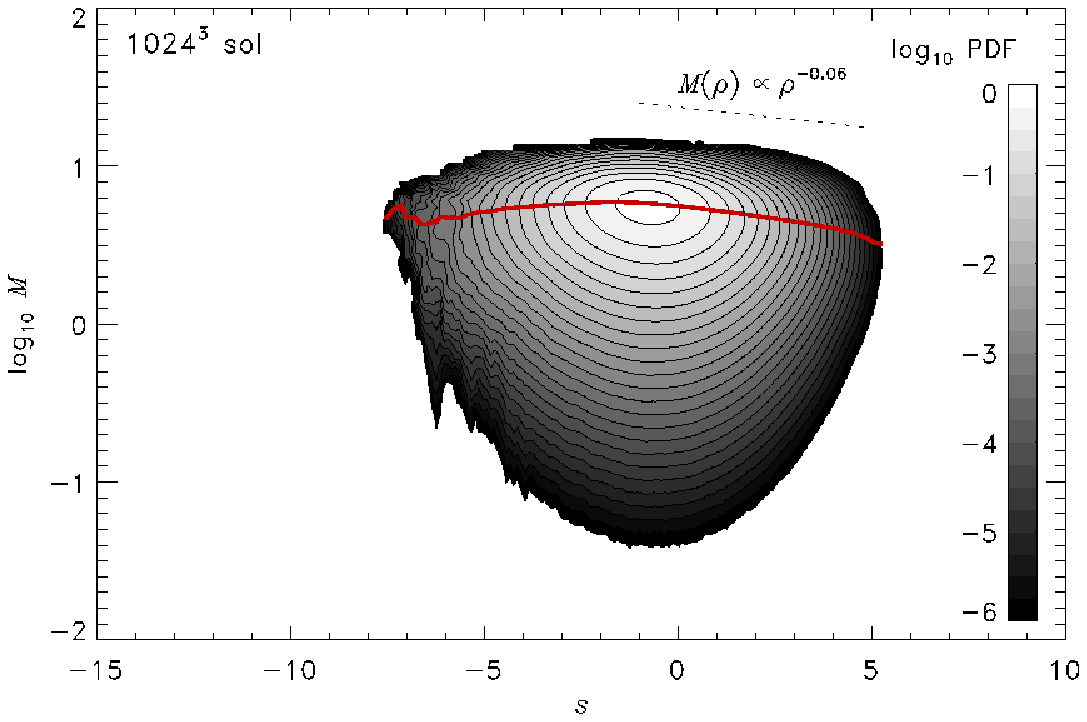}
\includegraphics[width=0.5\linewidth]{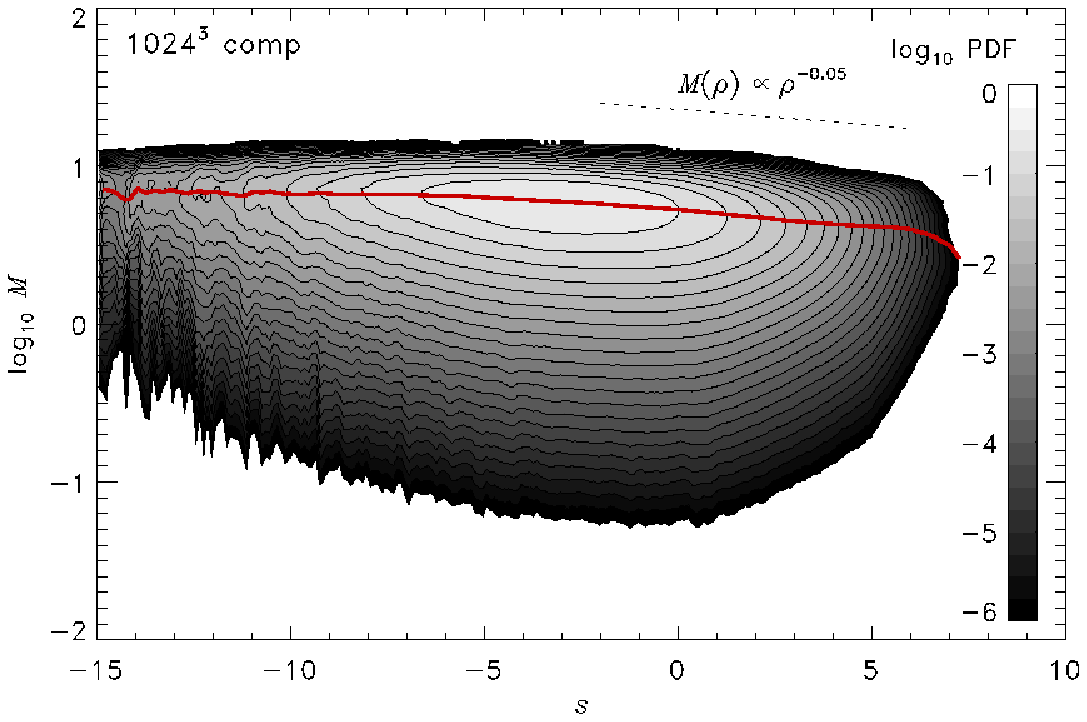}
}
\caption{Volume-weighted correlation PDFs of local Mach number $M$ versus logarithmic density $s$ for solenoidal (\emph{left}) and compressive forcing (\emph{right}). Adjacent contour levels are spaced by $0.25\,\mathrm{dex}$ in probability density. Density and Mach number exhibit a very weak, but non-zero correlation in both forcing cases, which provides an explanation for the non-Gaussian features in the density PDFs of Fig.~\ref{fig:solcomppdf} \citep{Vazquez1994,PassotVazquez1998,KritsukEtAl2007}. The two solid lines, intersecting the maxima of both distributions show the mean Mach number as a function of the logarithmic density $s=\ln(\rho/\left<\rho\right>)$. The tendency for high-density gas having lower Mach numbers on average is indicated as power laws in the high-density parts of the distributions. This suggests that the Mach number $M(\rho)\propto\rho^{-0.06}$ for solenoidal and $M(\rho)\propto\rho^{-0.05}$ for compressive forcing.}
\label{fig:2dpdfs}
\end{figure*}

We have computed the probability distributions for Mach number--density correlations. Figure~\ref{fig:2dpdfs} shows the volume-weighted correlation PDFs of local Mach number $M$ versus density $\rho$. Although the correlation between density and Mach number is weak as expected for isothermal turbulence \citep{Vazquez1994,PassotVazquez1998}, these two quantities are not entirely uncorrelated, which may explain the deviations from perfect log-normal distributions. There is a weak trend for high-density regions to exhibit lower Mach numbers on average. Power-law estimates for densities above the mean logarithmic density indicate Mach number--density correlations of the form $M(\rho)\propto\rho^{-0.06}$ for solenoidal and $M(\rho)\propto\rho^{-0.05}$ for compressive forcing. A similar power law exponent can be obtained from \citet[][Fig.~4]{KritsukEtAl2007}.

% \citet{PinedaCaselliGoodman2008} provided velocity dispersions computed from $^{12}$CO lines and from $^{13}$CO lines in six subregions of the \object{Perseus MC}. It it worth pointing out that the velocity dispersions derived from the optically thick $^{12}$CO are always larger than from the optically thin $^{13}$CO. On average, the $^{12}$CO velocity dispersion is 23\% larger than the $^{13}$CO velocity dispersion in the Perseus MC. We interpret this fact as a consequence of low-density gas typically exhibiting larger velocity dispersions. In computing statistical moments like the velocity dispersion, low-density material gets a higher relative weight for optically thick tracers like $^{12}$CO. This is because $^{12}$CO becomes optically thick faster and saturates at high column densities. In contrast, for optically thin tracers, each column is almost equally weighted. Since the velocity dispersions derived for the Perseus MC are always larger in the optically thick $^{12}$CO compared to the optically thin $^{13}$CO, it follows that the low-density gas in the Perseus MC exhibits larger velocity dispersions than the high-density gas on average. This is in qualitative agreement with the idea that high-density gas builds up in colliding turbulent gas flows, where the density is enhanced and the Mach number decreases on average in a roughly isothermal medium. Moreover, it agrees well with our findings from Figure~\ref{fig:2dpdfs}, which shows that high-density gas exhibits smaller velocity dispersions on average.

\subsection{Numerical resolution dependence of the density PDFs} \label{sec:pdfresol}
\begin{figure}[t]
\centerline{\includegraphics[width=1.0\linewidth]{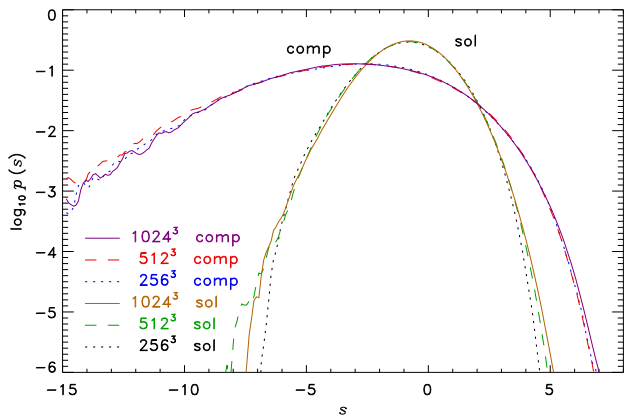}}
\caption{Density PDFs at numerical resolutions of $256^3$, $512^3$ and $1024^3$ grid cells. The PDFs show very good overall convergence, especially around the peaks. Table~\ref{tab:resolpdf} shows that the standard deviations are converged with numerical resolution. The high-density tails, however, are not converged even at a numerical resolution of $1024^3$ grid points, indicating a systematic shift to higher densities with resolution. This limitation is shared among all turbulence simulations \citep[see also,][]{HennebelleAudit2007,KitsionasEtAl2009,PriceFederrath2010}. The low-density wings are subject to strong temporal fluctuations due to intermittent bursts caused by head-on collisions of shocks followed by strong rarefaction waves \citep[e.g.,][]{KritsukEtAl2007}. The intermittency causes deviations from a perfect Gaussian distribution and accounts for non-Gaussian higher-order moments (skewness and kurtosis) in the distributions.}
\label{fig:resolpdf}
\end{figure}

The high-density tails of the PDFs in Figure~\ref{fig:solcomppdf} are not perfectly fit, even when the skewness and kurtosis are taken into account. This is partly due to non-zero 5th, 6th and higher-order moments in the distributions, and partly because our numerical resolution is insufficient to sample the high-density tail perfectly. Figure~\ref{fig:resolpdf} shows that even at a numerical resolution of $1024^3$ grid points, the high-density tails are not converged in both solenoidal and compressive forcing and tend to underestimate high densities. This limitation is shared among all turbulence simulations \citep[see, for instance, the turbulence comparison project by][]{KitsionasEtAl2009}, since the strongest and most intermittent fluctuations building up in the tails will always be truncated due to limited numerical resolution \citep[see also][]{HennebelleAudit2007,KowalLazarianBeresnyak2007,PriceFederrath2010}. However, the peak and the standard deviation of the PDFs are reproduced quite accurately at a resolution of $256^3$. Table~\ref{tab:resolpdf} shows the values of the linear standard deviation $\sigma_\rho$ and logarithmic standard deviation $\sigma_s$ for numerical resolutions of $256^3$, $512^3$ and $1024^3$. There appears to be no strong systematic dependence of the standard deviations on the numerical resolution for resolutions above $256^3$. The statistical fluctuations are the dominant source of uncertainty in the derived values of the standard deviations. It should be noted however that we have tested only the case of an RMS Mach number of about 5-6 here. There might be a stronger resolution dependence for higher Mach numbers, due to the stronger shocks produced in higher Mach number turbulence, which should be tested in a separate study.

\begin{table}[t]
\begin{center}
\caption{Standard deviations of the density PDFs as a function of numerical resolution for solenoidal and compressive forcing shown in Fig.~\ref{fig:resolpdf}. \label{tab:resolpdf}}
\def\arraystretch{1.1}
\begin{tabular}{@{\hspace{0.3cm}}l@{\hspace{0.3cm}}c@{\hspace{0.3cm}}c@{\hspace{0.3cm}}c@{\hspace{0.3cm}}c@{\hspace{0.3cm}}}
\hline
\hline
Grid Res. & \multicolumn{2}{c}{Solenoidal Forcing} & \multicolumn{2}{c}{Compressive Forcing}\\
\hline
~ & $\sigma_\rho$ & $\sigma_s$ & $\sigma_\rho$ & $\sigma_s$\\
$\phantom{1}256^3$~\mydotfill  & $1.79\pm0.08$ & $1.36\pm0.07$ & $5.66\pm0.79$ & $3.09\pm0.21$\\
$\phantom{1}512^3$~\mydotfill  & $1.89\pm0.10$ & $1.35\pm0.05$ & $5.59\pm0.67$ & $3.15\pm0.34$\\
$1024^3$~\mydotfill            & $1.89\pm0.09$ & $1.32\pm0.06$ & $5.86\pm0.96$ & $3.04\pm0.24$\\
\hline
\end{tabular}
\end{center}
\end{table}

\subsection{The column density PDFs and comparison with observations} \label{sec:columndensitypdfs}
\begin{figure}[t]
\centerline{\includegraphics[width=1.0\linewidth]{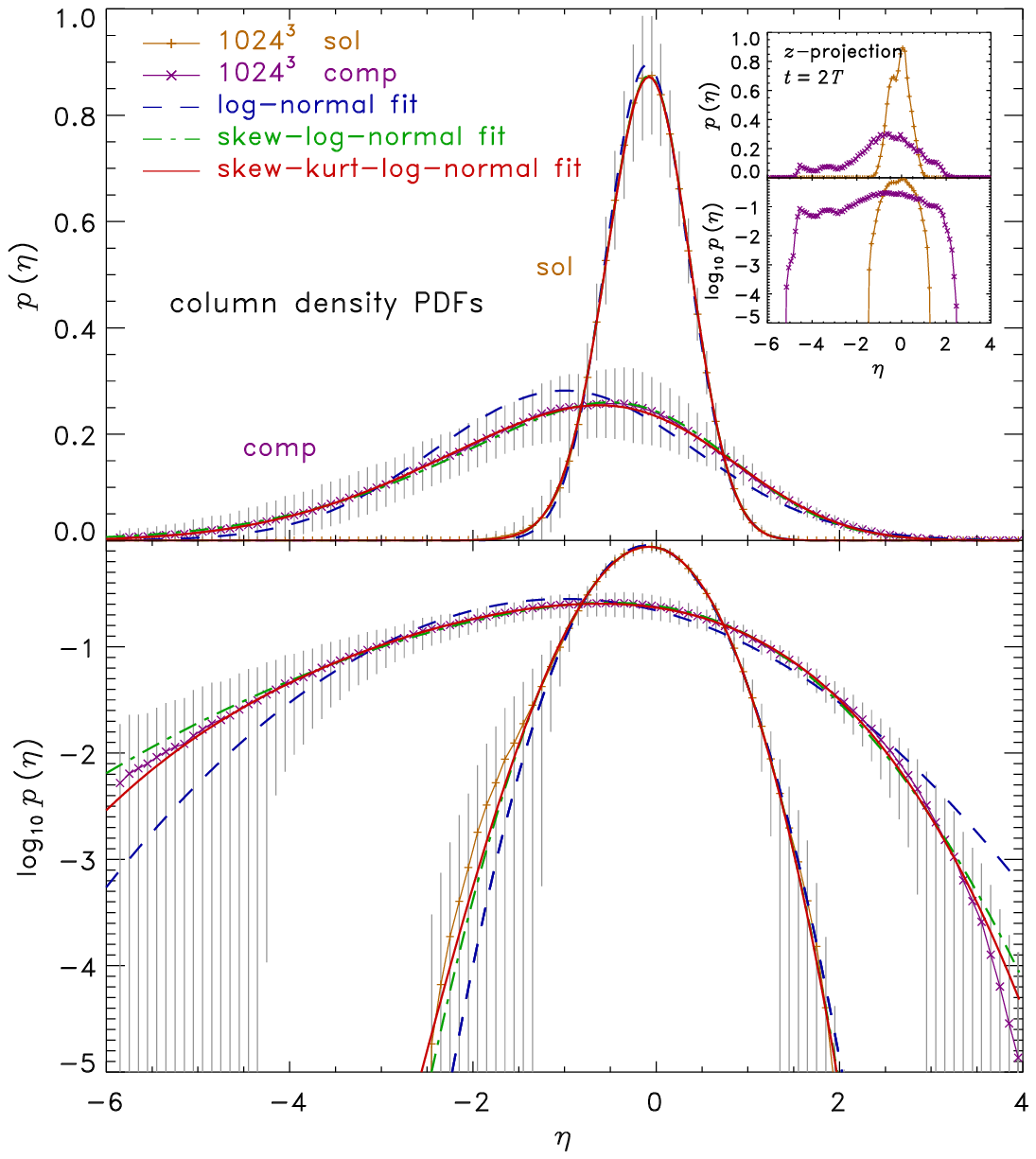}}
\caption{Same as Figure~\ref{fig:solcomppdf}, but the time- and projection-averaged logarithmic \emph{column} density PDFs of $\eta=\ln (\Sigma_i/\left<\Sigma_i\right>)$ are shown. $\Sigma_i$ and $\left<\Sigma_i\right>$ denote the column density and the mean column density integrated along the $i=x,\,y,\,z$ principal axes respectively. As for the volumetric PDFs of Fig.~\ref{fig:solcomppdf}, the standard deviation of the column density PDF obtained from compressive forcing is roughly three times larger than from solenoidal forcing (see Table~\ref{tab:pdfs_coldens}). The inset in the upper right corner shows the PDFs of column density computed in $z$-projection at a fixed time $t=2\,T$, corresponding to the snapshots shown in Figure~\ref{fig:snapshots}. The density dispersions computed for these instantaneous PDFs are $\sigma_\Sigma=0.49$ and $\sigma_\eta=0.45$ for solenoidal forcing, and $\sigma_\Sigma=1.34$ and $\sigma_\eta=1.56$ for compressive forcing. Although these distributions are quite noisy, the influence of the forcing is still clearly discernible. Thus, by studying instantaneous column density PDFs, which are accessible to observations, one should be able to distinguish solenoidal from compressive forcing.}
\label{fig:solcomppdfcoldens}
\end{figure}

The strong difference between the statistics of the solenoidal and compressive forcing cases seen in the PDFs of the volumetric density shown in Figure~\ref{fig:solcomppdf} is reflected by the corresponding column density PDFs. The time-averaged and projection-averaged column density PDFs are shown in Figure~\ref{fig:solcomppdfcoldens}. Analogous to Table~\ref{tab:pdfs_vol} for the volumetric density PDFs, we summarise the statistical quantities and fit parameters for the column density PDFs in Table~\ref{tab:pdfs_coldens}. The main results and conclusions obtained for the volumetric density distributions also hold for the column density distributions. Compressive forcing yields a column density standard deviation roughly three times larger than solenoidal forcing. The relative difference between solenoidal and compressive forcing is thus roughly the same for the volumetric and the column density distributions. However, the absolute values are lower for the column density distributions compared to the volumetric density distributions. The reason for this is that by computing projections of the volumetric density fields, density fluctuations are effectively averaged out by integration along the line-of-sight, and as a consequence, the column density dispersions become smaller compared to the corresponding volumetric density dispersions.

\begin{table*}[t]
\begin{center}
\caption{Same as Table~\ref{tab:pdfs_vol}, but for the PDFs of the column density $\Sigma$ shown in Fig.~\ref{fig:solcomppdfcoldens}. \label{tab:pdfs_coldens}}
\def\arraystretch{1.1}
\begin{tabular}{lccc}
\hline
\hline
~ & ~ & Solenoidal Forcing & Compressive Forcing\\
\hline
Standardised Moments & $\left<\Sigma\right>$      & $\phantom{-}1.00\pm0.00$ & $\phantom{-}1.00\pm0.00$\\
of $\Sigma$ & $\sigma_\Sigma$ & $\phantom{-}0.47\pm0.05$ & $\phantom{-}1.74\pm0.43$\\
~ & $\mathcal{S}_\Sigma$ & $\phantom{-}1.38\pm0.38$ & $\phantom{-}4.57\pm1.44$\\
~ & $\mathcal{K}_\Sigma$ & $\phantom{-}6.32\pm2.20$ & $\phantom{-}36.8\pm24.3$\\
\hline
Standardised Moments & $\left<\eta\right>$      & $-0.10\pm0.02$ & $-1.00\pm0.33$\\
of $\eta=\ln\left(\Sigma/\left<\Sigma\right>\right)$ & $\sigma_\eta$ & $\phantom{-}0.46\pm0.06$ & $\phantom{-}1.51\pm0.28$\\
~ & $\mathcal{S}_\eta$ & $-0.04\pm0.30$ & $-0.17\pm0.29$\\
~ & $\mathcal{K}_\eta$ & $\phantom{-}2.97\pm0.40$ & $\phantom{-}2.69\pm0.45$\\
\hline
Skewed Log-normal Approximation & $\xi$    & $\phantom{-}0.180\pm0.088$ & $\phantom{-}0.717\pm0.102$\\
using equation~(\ref{eq:skew-log-normal}) & $\omega$ & $\phantom{-}0.539\pm0.058$ & $\phantom{-}2.392\pm0.160$\\
~ & $\alpha$ & $-0.878\pm0.342$ & $-2.371\pm0.300$\\
\hline
4th Order Approximation & $a_0$ & $-0.1524\pm0.0451$ & $-1.4547\pm0.0532$\\
(including skewness and kurtosis) & $a_1$ & $-0.3900\pm0.1080$ & $-0.2902\pm0.0417$\\
using equation~(\ref{eq:skew-kurt-log-normal}) & $a_2$ & $-2.4643\pm0.1994$ & $-0.2669\pm0.0259$\\
~ & $a_3$ & $-0.1748\pm0.1469$ & $-0.0370\pm0.0080$\\
~ & $a_4$ & $\phantom{-}0.0204\pm0.1239$ & $-0.0035\pm0.0024$\\
\hline
\end{tabular}
\end{center}
\end{table*}

The small inset in the upper right corner of Figure~\ref{fig:solcomppdfcoldens} additionally shows the column density PDFs computed along the $z$-axis at one single time $t=2\,T$ corresponding to the map shown in Figure~\ref{fig:snapshots}. This figure shows the effect of studying one realisation only, without time- and/or projection-averaging. This is interesting to consider, because observations can only measure column density distributions at one single time. Improving the statistical significance would only be possible by studying multiple fields and averaging in space rather than in time invoking the ergodic theorem as suggested by \citet{GoodmanPinedaSchnee2009}. However, even by studying one turbulent realisation only, the difference between solenoidal and compressive forcing is recovered from the dispersions of the distributions. We therefore expect that using observations of column density PDFs, one can distinguish purely solenoidal from purely compressive forcing by measuring the dispersion of the column density PDF.

\citet{GoodmanPinedaSchnee2009} provided measurements of the column density PDFs in the \object{Perseus MC} obtained with three different methods: dust extinction, dust emission, and $^{13}$CO gas emission. Although systematic differences were found between the three methods, they conclude that in general, the measured column density PDFs are close to, but not perfect log-normal distributions, which is consistent with our results. They furthermore provided the column density PDFs and the column density dispersions for six subregions in the Perseus MC. The difference between the dispersions measured for these subregions is not as large as the difference between purely solenoidal and purely compressive forcing. The largest difference in the column density dispersions among the six subregions found by \citet{GoodmanPinedaSchnee2009} is only about 50\% relative to the average column density dispersion measured in the Perseus MC. This indicates that both purely solenoidal and purely compressive forcing are very unlikely to occur in nature. On the other hand, a varying mixture of solenoidal and compressive modes close to the natural mixture of 2:1 can easily explain the 50\% difference in density dispersion measured among the different regions. In particular, the Shell region \citep{RidgeEtAl2006}, which surrounds the B star \object{HD 278942} exhibits the largest density dispersion among all the subregions studied by \citet{GoodmanPinedaSchnee2009}, although its velocity dispersion is rather small compared to the others. This indicates that turbulent motions may be driven compressively rather than solenoidally within the Shell region. \citet{GoodmanPinedaSchnee2009} indeed mentioned that the gas in the Shell is dominated by an ''obvious driver'', skewing the column density distribution towards lower values compared to the other regions. Due to the constraints of mass conservation (eq.~\ref{eq:mass-conservation}) and normalisation (eq.~\ref{eq:normalization}), both the peak position and the peak value of the PDF skew to lower values, if the density dispersion increases (see Figure~\ref{fig:solcomppdfcoldens}). Taken together, this suggests that the Shell in the Perseus MC represents an example of strongly compressive turbulence forcing rather than purely solenoidal forcing.

\subsection{The forcing dependence of the density dispersion--Mach number relation} \label{sec:density_mach_correlations}
In \citet{FederrathKlessenSchmidt2008}, we investigated the density dispersion--Mach number relation \citep{PadoanNordlundJones1997,PassotVazquez1998}\footnote{Note that \citet{PassotVazquez1998} suffers from a number of typographical errors as a result of last-minute change of notation. Please see \citet[][footnote~5]{MacLowEtAl2005} for a number of corrections.},
\begin{equation}
\frac{\sigma_\rho}{\left<\rho\right>}=b\mathcal{M}\;. \label{eq:sigrhoofmach}
\end{equation}
This relation was also investigated in \citet[][Fig.~11]{KowalLazarianBeresnyak2007}, indicating that the standard deviation of turbulent density fluctuations, $\sigma_\rho$ is directly proportional to the sonic Mach number in the supersonic regime. It must be noted, however, that it was only directly tested for rather low RMS Mach numbers, $\mathcal{M}\lesssim2.5$ \citep{KowalLazarianBeresnyak2007} and $\mathcal{M}\lesssim3.5$ \citep{PassotVazquez1998}, compared to typical Mach numbers in molecular clouds. If additionally a log-normal PDF, equation~(\ref{eq:log-normal}) is assumed, then equation~(\ref{eq:sigrhoofmach}) can be expressed as
\begin{equation}
\sigma_s^2=\ln\big(1+b^2\mathcal{M}^2\big)\;, \label{eq:sigsofmach}
\end{equation}
with the \emph{same} parameter $b$ \citep{PadoanNordlundJones1997,FederrathKlessenSchmidt2008}. 

We begin our discussion of the forcing dependence of the density dispersion--Mach number relation with a problem raised by \citet{MacLowEtAl2005} and \citet{GloverMacLow2007b}. \citet{MacLowEtAl2005} and \citet{GloverMacLow2007b} claimed that the density dispersion--Mach number relation found by \citet{PassotVazquez1998}, $\sigma_s=b\mathcal{M}$ (which is a Taylor expansion of eq.~\ref{eq:sigsofmach} for small RMS Mach numbers), with $b\approx1$ did not at all fit their results for pressure and density PDFs, while equation~(\ref{eq:sigsofmach}) with $b\approx0.5$ \citep{PadoanNordlundJones1997} provided a much better representation of their data. The main difference in the density dispersion--Mach number relations by \citet{PadoanNordlundJones1997} and \citet{PassotVazquez1998} is the proportionality constant $b$. It is $b\approx0.5$ and $b\approx1$ in \citet{PadoanNordlundJones1997} and \citet{PassotVazquez1998}, respectively. Our forcing analysis provides the solution to this apparent difference, which lies at the heart of the disagreement of the PDF data analysed in \citet{MacLowEtAl2005} and \citet{GloverMacLow2007b} with the model by \citet{PassotVazquez1998}. \citet{PassotVazquez1998} used 1D models. In 1D, only compressive forcing is possible, because no transverse waves can exist. In contrast, \citet{MacLowEtAl2005} and \citet{GloverMacLow2007b} used a mixture of solenoidal and compressive forcing in 3D. In this section, we show that the parameter $b$ in both equations~(\ref{eq:sigrhoofmach}) and~(\ref{eq:sigsofmach}) is a function of the forcing parameter $\zeta$. Indeed, using the relation $\sigma_s=b\mathcal{M}$ analysed in \citet{PassotVazquez1998}, but with a lower proportionality constant ($b=0.5$ in contrast to $b=1$) gives a very good representation of the PDF data in \citet[][Fig.~8]{MacLowEtAl2005}. Thus, an investigation of the parameters that control $b$ seems necessary and important.

% There are two important points to make, why they did not find good agreement of their data with the model by \citet{PassotVazquez1998}. First, \citet{MacLowEtAl2005} and \citet{GloverMacLow2007b} used $\sigma_s=b\mathcal{M}$. This equation is similar to equation~(\ref{eq:sigrhoofmach}), but with the \emph{linear} density dispersion, $\sigma_\rho$ replaced by the \emph{logarithmic} density dispersion, $\sigma_s$. This representation of equation~(\ref{eq:sigrhoofmach}) is valid for low RMS Mach numbers as shown by the numerical models of \citet{PassotVazquez1998} up to rms Mach 3.3. Replacing $\sigma_\rho$ with $\sigma_s$ in equation~(\ref{eq:sigrhoofmach}) represents a Taylor expansion of equation~(\ref{eq:sigsofmach}) for small RMS Mach numbers. Secondly, and more importantly, as shown here the parameter $b$ in both equations~(\ref{eq:sigrhoofmach}) and~(\ref{eq:sigsofmach}) is a function of the forcing parameter $\zeta$. \citet{MacLowEtAl2005} and \citet{GloverMacLow2007b} both used a mixture of modes for their forcing, while \citet{PassotVazquez1998} modelled purely compressive forcing. The implications of this important difference in the forcing are discussed below.

Moreover, relations~(\ref{eq:sigrhoofmach}) and~(\ref{eq:sigsofmach}) are key ingredients for the analytical models of the stellar initial mass function by \citet{PadoanNordlund2002} and \citet{HennebelleChabrier2008}, as well as for the star formation rate model by \citet{KrumholzMcKee2005} and \citet{KrumholzMcKeeTumlinson2009} and for the star formation efficiency model by \citet{Elmegreen2008}. In all these models, $b$ is assumed to be $0.5$, which is an empirical result from magnetohydrodynamical simulations by \citet{PadoanNordlundJones1997}. On the other hand, \citet{PassotVazquez1998} found $b\!\approx\!1$ from 1D hydrodynamical simulations. \citet{FederrathKlessenSchmidt2008} resolved this disagreement between \citet{PadoanNordlundJones1997} and \citet{PassotVazquez1998} by showing that $b$ is a function of the ratio $\zeta\in[0,1]$ of compressive to solenoidal modes of the turbulence forcing. However, \citet{FederrathKlessenSchmidt2008} only tested the two limiting cases of purely solenoidal forcing ($\zeta=1$) and purely compressive forcing ($\zeta=0$). They approximated the regime of mixtures with a heuristic model, which had a linear dependence of $b$ on $\zeta$:
% \footnote{Note that \citet{PassotVazquez1998} suffers from a number of typographical errors as a result of last-minute change of notation. Please see \citet[][footnote~5]{MacLowEtAl2005} for a number of corrections. We emphasise in this context that each of the five equations expressing the direct proportionality of $\sigma_s$ with RMS Mach number $\tilde{M}$: $\sigma_s=\beta\tilde{M}$ within the whole subsection ''B. The case $\gamma\!=\!1$'' in \citet{PassotVazquez1998} represent Taylor expansions of the logarithmic expression $\sigma_s=[\ln(1+\beta^2\tilde{M}^2)]^{1/2}$ (our eq.~\ref{eq:sigsofmach}) for \emph{small} RMS Mach numbers. The fundamental proportionality between the density dispersion of the \emph{linear} density $\sigma_\rho$ and the RMS Mach number for arbitrarily large RMS Mach numbers from which the latter two expressions can be derived by assuming the functional form of the log-normal PDF given by eq.~(\ref{eq:log-normal}): $\sigma_\rho=\beta\tilde{M}$ (our eq.~\ref{eq:sigrhoofmach}) was motivated using the isothermal shock jump conditions and given in \citet[][eq.~5]{PadoanNordlundJones1997}. Eq.~5 in \citet[][]{MacLowEtAl2005} is therefore only expected to hold for small RMS Mach numbers with purely compressive forcing represented by $\beta=1$ in the notation used in this footnote.}
\begin{equation}
\tilde{b}=1+\left(\frac{1}{D}-1\right)\zeta=
\left\{
\def\arraystretch{1.5}
\begin{array}{lr}
1-\frac{2}{3}\zeta, & \textnormal{for $\;D=3$}\;\phantom{.}\\
1-\frac{1}{2}\zeta, & \textnormal{for $\;D=2$}\;\phantom{.}\\
1,                  & \textnormal{for $\;D=1$}\;.
\end{array}
\right.
\label{eq:b}
\end{equation}

\begin{figure}[t]
\centerline{\includegraphics[width=1.0\linewidth]{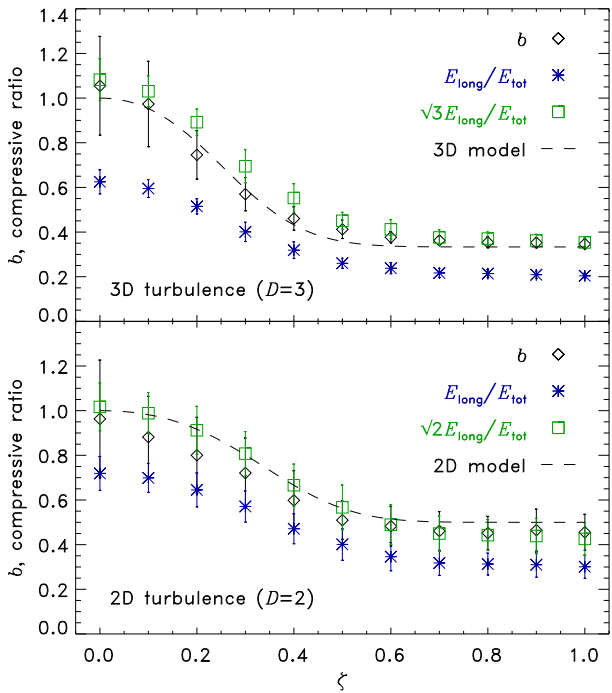}}
\caption{\emph{Diamonds:} The proportionality parameter $b$ in the density dispersion--Mach number relation, eq.~(\ref{eq:sigrhoofmach}), computed as $b=\sigma_\rho/\left(\left<\rho\right>\mathcal{M}\right)$ for eleven 3D models at numerical resolution of $256^3$ grid cells (\emph{top panel}) and eleven 2D models at numerical resolution of $1024^2$ grid cells (\emph{bottom panel}), ranging from purely compressive forcing ($\zeta=0$) to purely solenoidal forcing ($\zeta=1$). The parameter $b$ decreases smoothly from $b\approx1$ for compressive forcing to $b\approx1/3$ in 3D and $b\approx1/2$ in 2D for solenoidal forcing. \emph{Stars:} Ratio $\left<\Psi\right>=E_\mathrm{long}/E_\mathrm{tot}$ of longitudinal to total power in the velocity power spectrum (see~\S~\ref{sec:vspectra}). This quantity provides a measure for the relative amount of compression induced by the turbulent velocity field, and appears to be correlated with the standard deviation of the density PDF. \emph{Squares:} Same as stars, but multiplied by the geometrical factor $\!\sqrt{D}$ with $D=3$ for the three-dimensional case and $D=2$ for the two-dimensional case. The quantity $\!\sqrt{D}\,\left<\Psi\right>$ provides a good numerical estimate of the PDF parameter $b$. The dashed lines show model fits using equation~(\ref{eq:b_esti}) for $D=3$ (\emph{top panel}) and $D=2$ (\emph{bottom panel}).}
\label{fig:zeta}
\end{figure}

Here, we refine this model based on eleven additional simulations with $\zeta=[0,1]$ separated by $\Delta\zeta=0.1$ for RMS Mach numbers of 5 in 2D and 3D. These simulations allow us to test eleven different mixtures of forcing controlled by the parameter $\zeta$ (see eq.~\ref{eq:forcing_ratio}). The models were run at a numerical resolution of $256^3$ and $1024^2$ grid points in 3D and 2D, respectively. We use a lower resolution in 3D, because using our standard resolution of $1024^3$ would be too computationally intensive. However, as shown in \S~\ref{sec:pdfresol}, the standard deviation of the density is fairly well reproduced at $256^3$, as is the RMS Mach number $\mathcal{M}$ \citep[][Fig.~2]{FederrathKlessenSchmidt2009}, which allows a reasonably accurate determination of $b$. The results are plotted in Figure~\ref{fig:zeta} as diamonds for 3D (top panel) and 2D (bottom panel). We used equation~(\ref{eq:sigrhoofmach}) to measure $b$, because unlike equation~(\ref{eq:sigsofmach}), this version of the standard deviation--Mach number relation does not rest on the assumption of a log-normal PDF. In fact, if equation~(\ref{eq:sigsofmach}) was used to derive $b$ for models with $\zeta<0.5$, $b$ would be overestimated significantly (by up to an order of magnitude for $\zeta=0$), because the deviations from the perfect log-normal distribution are stronger for $\zeta<0.5$ (cf.~\S~\ref{sec:pdfcomp}; see also \citet{SchmidtEtAl2009}).

Figure~\ref{fig:zeta} shows that the dependence of $b$ on $\zeta$ is non-linear. For 3D turbulence the parameter $b$ increases smoothly from $b\approx1/3$ for $\zeta=1$ to $b\approx1$ for $\zeta=0$, and for 2D turbulence from $b\approx1/2$ for $\zeta=1$ to $b\approx1$ for $\zeta=0$. However, there is an apparent break at $\zeta\approx0.5$, which represents the natural forcing mixture used in many previous studies. For $\zeta\gtrsim0.5$ the $b$-parameter remains close to the value obtained for purely solenoidal forcing, i.e.~$b\approx0.3-0.4$ in 3D and $b\approx0.5$ in 2D. The flat part of the data in Figure~\ref{fig:zeta} for $\zeta>0.5$ explains why in previous studies with a natural forcing mixture \citep[e.g.,][]{MacLowEtAl1998,KlessenHeitschMacLow2000,LiKlessenMacLow2003,KritsukEtAl2007,GloverFederrathMacLowKlessen2009}, the turbulence statistics were close to the purely solenoidal forcing case \citep[e.g.,][]{PadoanNordlundJones1997,StoneOstrikerGammie1998,BoldyrevNordlundPadoan2002,PadoanNordlund2002,KowalLazarianBeresnyak2007,LemasterStone2008,BurkhartEtAl2009}. In contrast, $b$ increases much more strongly for $\zeta\lesssim0.5$, until it reaches $b\approx1$ for purely compressive forcing \citep[e.g.,][]{PassotVazquez1998,FederrathKlessenSchmidt2008,SchmidtEtAl2009}.

Equation~(\ref{eq:b}) thus needs to be refined to account for the non-linear dependence of $b$ on the forcing. Moreover, equation~(\ref{eq:b}) was based on the analytic expression of the forcing parameter $\zeta$ (cf.~\S~\ref{sec:forcing}). However, the numerical estimate of $b$ depends on how well the code can actually induce compression through the build-up of divergence in the velocity field. Thus, different codes can produce slightly different values of $b$ for the same forcing parameter $\zeta$. This is because of the varying efficiency of codes to convert the energy provided by a given forcing into actual velocity fluctuations \citep[e.g.][]{KitsionasEtAl2009,PriceFederrath2010}. To construct a refined model for $b$ that does not directly rest on the analytic forcing parameter $\zeta$ and that accounts for the non-linear dependence on the forcing, we recall that $b$ is a normalised measure of compression. Compression is caused by converging flows and shocks, which have a finite magnitude of velocity divergence. A normalised measure of compression is thus also provided by dividing the power in longitudinal modes of the velocity field by the total power of all modes in the velocity field,
\begin{equation}
\left<\Psi\right>=\frac{E_\mathrm{long}}{E_\mathrm{tot}}\;.
\end{equation}
% A scale-dependent version of this ratio is given in equation~(\ref{eq:Eratio}) later in \S~\ref{sec:spectra}. To obtain the average of that ratio, we integrate equation~(\ref{eq:Eratio}) over all wavenumbers $k$, and call it $\left<\Psi\right>=E_\mathrm{long}/E_\mathrm{tot}$. 
We therefore expect a dependence of $b$ on $\left<\Psi\right>$.

Figure~\ref{fig:zeta} shows $\left<\Psi\right>$ as a function of $\zeta$ (plotted as stars) for 3D and 2D turbulence. It is indeed correlated with $b$, however, $\left<\Psi\right>$ is less than $b$ by a factor of roughly $\!\sqrt{3}$ in 3D and $\!\sqrt{2}$ in 2D. The squares in Figure~\ref{fig:zeta} show $\!\sqrt{3}\,\left<\Psi\right>$ in 3D and $\!\sqrt{2}\,\left<\Psi\right>$ in 2D, which seems to provide a good estimate of $b$. The factor $\!\sqrt{3}$ is a geometrical factor for 3D turbulence (the diagonal in a cube of size unity). It is $\!\sqrt{2}$ in 2D turbulence (the diagonal in a square of size unity), and $\!\sqrt{1}$ in 1D. The latter in particular is trivial, because in 1D only longitudinal modes can exist, and thus $\!\sqrt{1}\,\left<\Psi\right>=1$ for any value of $\zeta$ (cf.~Fig.~\ref{fig:forcing_ratio}). The larger geometrical factors in 2D and 3D account for the fact that the longitudinal velocity fluctuations, which induce compression occupy only one of the available spatial directions (two in 2D and three in 3D) on average. For the general case of supersonic turbulence in $D=1,2$ and 3 dimensions, these ideas lead to
\begin{equation} \label{eq:b_refined}
\tilde{b}=\sqrt{D}\left<\Psi\right>\;,
\end{equation}
which is solely based on the ratio of the power in longitudinal modes in the velocity field to the total power of all modes in the velocity field, $\left<\Psi\right>$.

In addition to the refined model based on the compressive ratio $\left<\Psi\right>$ in equation~(\ref{eq:b_refined}), we provide a fit function for $b$ based on the forcing parameter $\zeta$. The dashed lines in Figure~\ref{fig:zeta} show
\begin{equation} \label{eq:b_esti}
\tilde{b}(\zeta)=\frac{1}{D}+\frac{D-1}{D}\,\left(\frac{F_\mathrm{long}(\zeta)}{F_\mathrm{tot}(\zeta)}\right)^3\;.
\end{equation}
The forcing ratio $F_\mathrm{long}/F_\mathrm{tot}$ is given by equation~\ref{eq:forcing_ratio}. The first summand in equation~(\ref{eq:b_esti}) is the expected ratio of longitudinal modes (compression) in a supersonic turbulent medium for a purely solenoidal forcing, i.e. a forcing that does not directly induce compression. The second summand is the contribution to the compression directly induced by the forcing. The model equation~(\ref{eq:b_esti}) is similar to equation~(\ref{eq:b}), but with a non-linear dependence of $b$ on the forcing parameter $\zeta$.

We suggest that the dependence of $b$ on the forcing solves a puzzle reported by \citet{PinedaCaselliGoodman2008}. They provided measurements of velocity dispersions and $^{12}$CO excitation temperatures for the six subregions in the \object{Perseus MC}. The molecular excitation temperatures serve as a guide for the actual gas temperature, from which the sound speed can be estimated. From these values, the local RMS Mach numbers are computed as the ratio of the local velocity dispersion to the local sound speed. \citet{GoodmanPinedaSchnee2009} and \citet{PinedaCaselliGoodman2008} pointed out that there is clearly \emph{no} correlation of the form suggested by equation~(\ref{eq:sigsofmach}) for a \emph{fixed} parameter $b$ across the investigated subregions in the Perseus MC. For instance, the Shell region exhibits an intermediate to small velocity dispersion derived from $^{12}$CO and $^{13}$CO observations, while its density dispersion is the largest in the Perseus MC. This provides additional support to our suggestion that the Shell in Perseus is dominated by compressive turbulence forcing for which $b$ takes a higher value compared to solenoidal forcing. The apparent lack of density dispersion--Mach number correlation reported by \citet{PinedaCaselliGoodman2008} and \citet{GoodmanPinedaSchnee2009} for a fixed parameter $b$ can thus be explained, because $b$ is in fact \emph{not} fixed across different subregions in the Perseus MC.

We plan to measure $b$ in different regions of the ISM in future studies. However, the main problem in a quantitative analysis of equation~(\ref{eq:sigrhoofmach}) with observational data is that the column density dispersion is typically smaller than the 3D density dispersion (compare Tab.~\ref{tab:pdfs_vol} and Tab.~\ref{tab:pdfs_coldens}). The relation between the column density PDF and the volumetric density PDF is non-trivial and depends on whether the column density tracer is optically thin or optically thick and on the scale of the turbulence driving. However, \citet{BruntFederrathPrice2010a} developed a promising technique to estimate the 3D density variance from 2D observations with an accuracy of about 10\%.

% Using position-position-velocity data of optically thin molecular lines alongside column density data may provide a way to recover the volumetric density PDF and its dispersion from observations, making a quantitative comparison possible. This needs to be investigated in more detail in a future study.

\section{Intermittency} \label{sec:cvis}
Intermittency manifests itself in
\begin{itemize}
\item[\emph{i})] non-Gaussian (often exponential) wings of PDFs of quantities involving density and/or velocity, its derivatives (e.g., vorticity) and combinations of density and velocity (e.g., $\rho^{1/2}v$ and $\rho^{1/3}v$ as discussed in Appendix~\ref{app:spectra_sqrtrho_rho3}),
\item[\emph{ii})] anomalous scaling of the higher-order structure functions of the velocity field \citep[e.g.,][]{AnselmetEtAl1984} and centroid velocity increments \citep{LisEtAl1996,HilyBlantFalgaronePety2008}, and
\item[\emph{iii})] coherent structures of intense vorticity ($\nabla\times\vect{v}$) \citep[see][for results of incompressible turbulence]{VincentMeneguzzi1991,MoisyJimenez2004}, and of strong shocks and rarefaction waves ($\nabla\cdot\vect{v}$).
\end{itemize}

Filamentary coherent structures of vorticity (intermittency item~\emph{iii}) are indeed observed in our two supersonic models. In Figure~\ref{fig:snapshots} (middle panel), we show the projected vorticity for solenoidal and compressive forcing, respectively. Most of the filaments of high vorticity coincide with the positions of shocks and therefore also with high density and negative divergence in the velocity field (Figure~\ref{fig:snapshots}, bottom panel). This is furthermore inferred from observations of the Ursa Majoris Cloud by \citet{FalgaroneEtAl1994} and is consistent with the results of weakly compressible decaying turbulence experiments by \citet{PorterPouquetWoodward1992ThCFD} and \citet{PorterPouquetWoodward1994}, who concluded that intense vorticity is typically associated with intermittency.

\subsection{The probability distribution of centroid velocity increments}
Since there is evidence of filamentary coherent structures in the vorticity (intermittency item~\emph{iii}) of our models, and because there is additional evidence of non-Gaussian tails in the density PDFs (intermittency item~\emph{i}) discussed in~\S~\ref{sec:pdfs}, we now proceed to examine the PDFs and the scaling of centroid velocity increments (intermittency item~\emph{ii}) to assess the strength of the intermittency. We compare centroid velocity increments (CVIs) for solenoidal and compressive forcing and discuss the interpretation of observations based on that comparison. Following the analysis by \citet{LisEtAl1996}, who discuss CVIs computed for the turbulence simulation by \citet{PorterPouquetWoodward1994}, and following the CVI analysis of the \object{Polaris Flare} and of the \object{Taurus MC} by \citet{HilyBlantFalgaronePety2008}, the centroid velocity increment is defined as
\begin{equation} \label{eq:cvi}
\delta C_{\ell}(\vect{r}) = \left< C(\vect{r})-C(\vect{r}+\vect{\ell}) \right>\;,
\end{equation}
where the angle average $\left<\phantom{C}\right>$ is computed over all possible directions of the vector $\vect{\ell}$ in the plane perpendicular to the line-of-sight. Thus, $\delta C_{\ell}(\vect{r})$ only depends on the norm of the lag vector $\ell=\left|\vect{\ell}\right|$, which separates two points $\vect{r}=(x,y)$ and $\vect{r}+\vect{\ell}$ in the plane of the sky $(x,y)$. The normalised centroid velocity, $C(\vect{r})$ in equation~(\ref{eq:cvi}) is defined as
\begin{equation} \label{eq:cv}
C(\vect{r}) = \frac{\int \rho(\vect{r},z)\,v_z(\vect{r},z)\,\mathrm{d}z}{\int\rho(\vect{r},z)\,\mathrm{d}z}\;.
\end{equation}
The variable $v_z(\vect{r},z)$ denotes the line-of-sight velocity in $z$-direction. We have however computed $C(\vect{r})$ separately along each of the three principal lines-of-sight $x$, $y$ and $z$ of our Cartesian domain in order to examine the effects of varying the projection. Also note that we have computed normalised centroid velocities \citep{LazarianEsquivel2003}, since we want to compare to \citet{HilyBlantFalgaronePety2008}. Another point to mention here is that the centroid velocities, $C(\vect{r})$ are typically computed using an intensity weighting instead of a density weighting. This is because the gas density cannot be measured directly, whereas the emission intensity is accessible to observations. By using density weighting we implicitly assume optically thin emission. For optically thick emission, uniform weighting would be more appropriate \citep{LisEtAl1996}.

\begin{figure*}[t]
\centerline{
\includegraphics[width=0.5\linewidth]{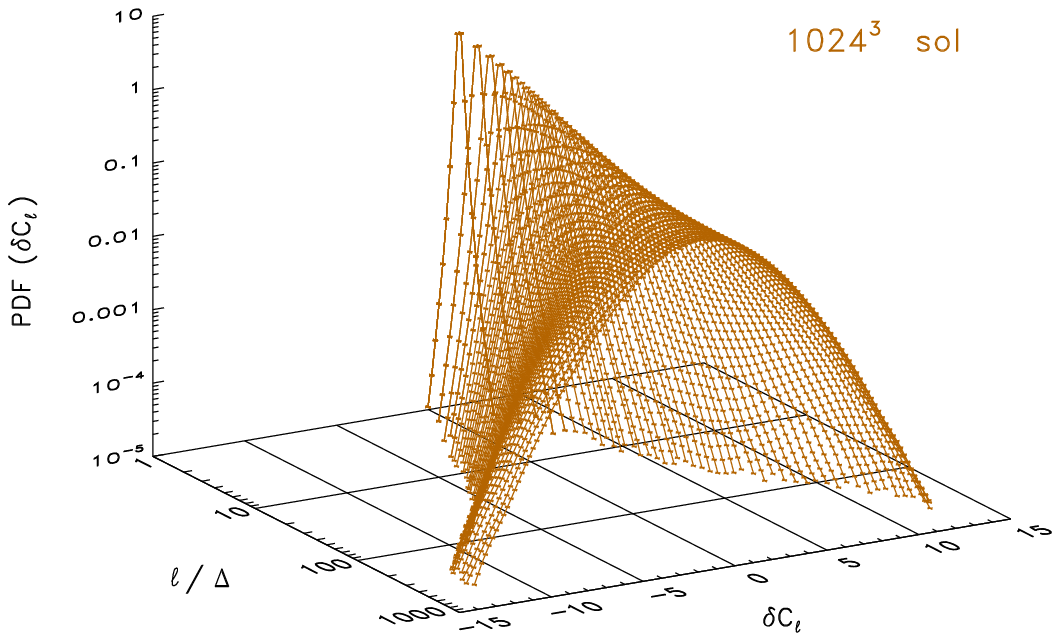}
\includegraphics[width=0.5\linewidth]{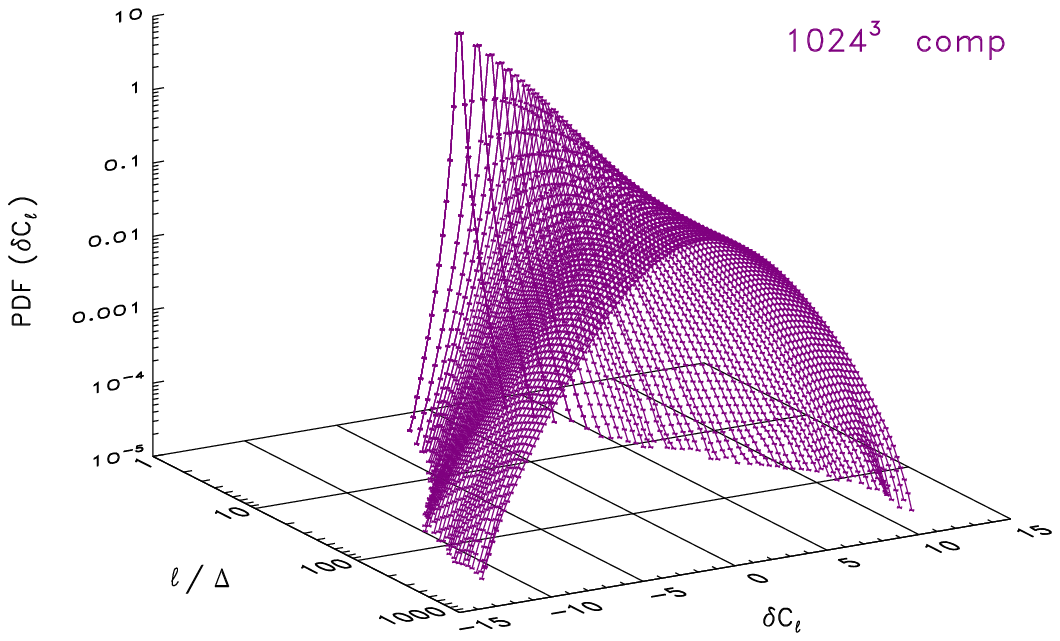}
}
\caption{PDFs of centroid velocity increments, computed using equations~(\ref{eq:cvi}) and~(\ref{eq:cv}) are shown as a function of lag $\ell$ in units of grid cells $\Delta=L/1024$ for solenoidal forcing (\emph{left}) and compressive forcing (\emph{right}). The PDFs are very close to Gaussian distributions for long lags, whereas for short lags, they develop exponential tails, which is a manifestation of intermittency \citep[e.g.,][and references therein]{HilyBlantFalgaronePety2008}.}
\label{fig:cvipdfs}
\end{figure*}

Figure~\ref{fig:cvipdfs} shows the PDFs of $\delta C_\ell(\vect{r})$ computed for varying lag $\ell$ in units of the numerical cell size $\Delta=L/1024$. They should be compared to \citet[][Fig.~4-6]{HilyBlantFalgaronePety2008}. The PDFs are mainly Gaussian for large lags, whereas for smaller separations, they develop exponential tails, indicating intermittent behaviour. This result is consistent with the numerical simulation analysed by \citet{LisEtAl1996}, and with observations of the \object{$\rho$ Oph Cloud}, the \object{Orion B} and the Polaris Flare by \citet{LisEtAl1998}, \citet{MieschScaloBally1999} and \citet{HilyBlantFalgaronePety2008}, respectively.

\begin{figure}[t]
\centerline{\includegraphics[width=1.0\linewidth]{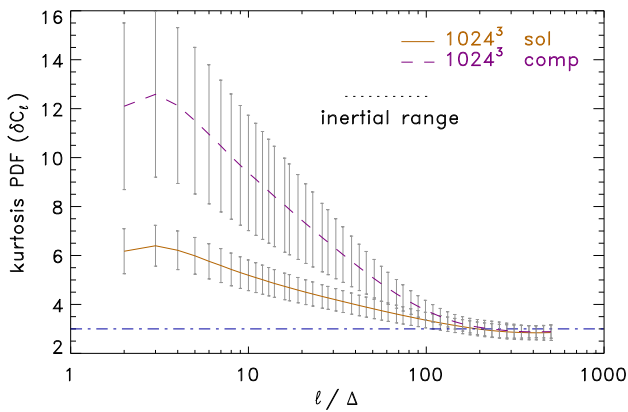}}
\caption{Kurtosis $\mathcal{K}$ of the PDFs of centroid velocity increments shown in Fig.~\ref{fig:cvipdfs} as a function of the lag $\ell$ in units of grid cells $\Delta=L/1024$ for solenoidal and compressive forcing. Note that a kurtosis value of $3$ (horizontal dot-dashed line) corresponds to the value for a Gaussian distribution. Non-Gaussian values of the kurtosis are obtained for $\ell \lesssim 100 \Delta$. The error bars contain both snapshot-to-snapshot variations as well as the variations between centroid velocity increments computed by integration along the $x$, $y$ and $z$ axes. This figure can be compared to observations of the \object{Polaris Flare} and \object{Taurus MC} \citep[see Fig.~7 of][]{HilyBlantFalgaronePety2008}.}
\label{fig:cvipdfskurt}
\end{figure}

Following the analysis by \citet{HilyBlantFalgaronePety2008}, we computed the kurtosis $\mathcal{K}$ of the PDFs of CVIs using the definition in equations~(\ref{eq:moments}). Note that $\mathcal{K}=3$ corresponds to a Gaussian distribution, and $\mathcal{K}=6$ corresponds to an exponential function. The kurtosis of the CVI PDFs is shown in Figure~\ref{fig:cvipdfskurt} as a function of spatial lag $\ell$, and can be directly compared to \citet[][Fig.~7]{HilyBlantFalgaronePety2008}. Both forcing types exhibit nearly Gaussian values of the kurtosis at lags $\ell\gtrsim100\Delta$. On the other hand, for $\ell\lesssim100\Delta$, both forcing types produce non-Gaussian PDFs. Solenoidal forcing approaches the exponential value $\mathcal{K}=6$ for $\ell\lesssim10\Delta$. Compressive forcing yields exponential values already for lags $\ell\approx40\Delta$, while solenoidal forcing has $\mathcal{K}\approx4$ on these scales. This indicates stronger intermittency in the case of compressive forcing. For $\ell\lesssim30\Delta$, compressive forcing yields even super-exponential values of $\mathcal{K}$. For both solenoidal and compressive forcings, we show later in~\S~\ref{sec:spectra} that $\ell\lesssim30\Delta$ is in the dissipation range for numerical turbulence. Compressive velocity modes dominate in this regime (see Fig.~\ref{fig:spectratio}), which may result artificially in extreme intermittency. For $\ell\approx30\Delta$, compressive forcing gives $\mathcal{K}=6.0\pm1.0$, which is roughly 35\% larger than the Polaris Flare observations at their resolution limit. The solenoidal case on the other hand gives $\mathcal{K}\approx4.3\pm0.5$, which is in very good agreement with the IRAM and KOSMA data discussed by \citet[][Fig.~7]{HilyBlantFalgaronePety2008}. Depending on the actual lag used for the comparison, both solenoidal and compressive forcing seem to be consistent with the observations. However, it should be noted that the lags cannot be easily compared for the real clouds and the simulations, because simulated and observed fields have different spatial resolution. Moreover, the simulated fields have periodic boundaries, while the true fields don't. Nevertheless, the similarity of the observed and the numerically simulated CVIs indicates that turbulence intermittency plays an important role in both our simulations and in real molecular clouds.

The Polaris Flare has a very low star formation rate and is therefore appropriate for studying the statistics of interstellar supersonic turbulence without contamination by internal energy sources. In contrast, the \object{Taurus MC} is actively forming stars. Against our expectations, the Taurus MC data display very \emph{weak} intermittent behaviour and the kurtosis remains at the Gaussian values $\mathcal{K}\approx3$ in \citet[][Fig.~7]{HilyBlantFalgaronePety2008}. However, the Taurus field studied by \citet{HilyBlantFalgaronePety2008} is located far from star-forming regions in a translucent part of the Taurus MC (E.~Falgarone 2009, private communication). This may explain why the Taurus field displays only very weak intermittency. It would be interesting to repeat the analysis of centroid velocity increments for regions of confirmed star formation, including regions with winds, outflows and ionisation feedback from young stellar objects to see whether these regions indeed display stronger intermittency.

\subsection{The structure function scaling of centroid velocity increments} \label{sec:cvisfs}
In this section, we discuss the scaling of the $p\,$th order structure function of CVIs, defined as
\begin{equation} \label{eq:cvisf}
\mathrm{CVISF}_p(\ell) = \left<\left|\delta C_\ell(\vect{r})\right|^p\right>_\vect{r}\;.
\end{equation}

We have averaged over a large enough sample of independent increments $\delta C_{\ell}(\vect{r})$ that increasing the sample size produced no change in the value of $\mathrm{CVISF}_p(\ell)$ for $p\leq6$, which is demonstrated in Appendix~\ref{app:cviconvergence}. Figure~\ref{fig:cvis} shows the CVI structure functions for solenoidal and compressive forcing. The CVI structure functions were fit to power laws of the form
\begin{equation} \label{eq:cvisf_scaling}
\mathrm{CVISF}_p(\ell) \propto \ell^{\,\zeta_p}
\end{equation}
within the inertial range\footnote{By its formal definition for \emph{incompressible} turbulence studies \citep[e.g.,][]{Frisch1995}, the inertial range is the range of scales for which the turbulence statistics are not directly influenced by the forcing acting on scales larger than the inertial range, and not directly influenced by the viscosity acting on scales smaller than the inertial range. The inertial range is typically very small in numerical experiments, because of the high numerical viscosity caused by the discretisation scheme, given the resolutions achievable with current computer technology (see also~\S~\ref{sec:spectra} and Appendix~\ref{app:spectra}).}, defined equivalently to the study in \citet{FederrathKlessenSchmidt2009}. The value for each power-law exponent is indicated in Figure~\ref{fig:cvis} and summarised in Table~\ref{tab:cvis}.

\begin{figure*}[t]
\centerline{
\includegraphics[width=0.5\linewidth]{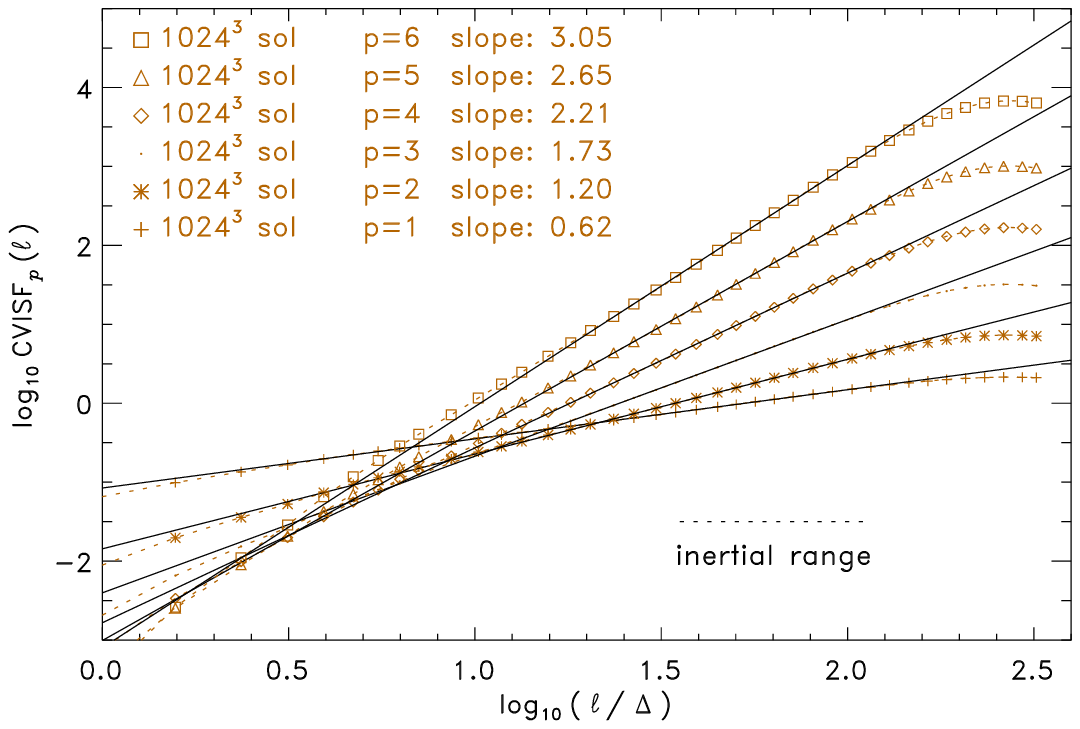}
\includegraphics[width=0.5\linewidth]{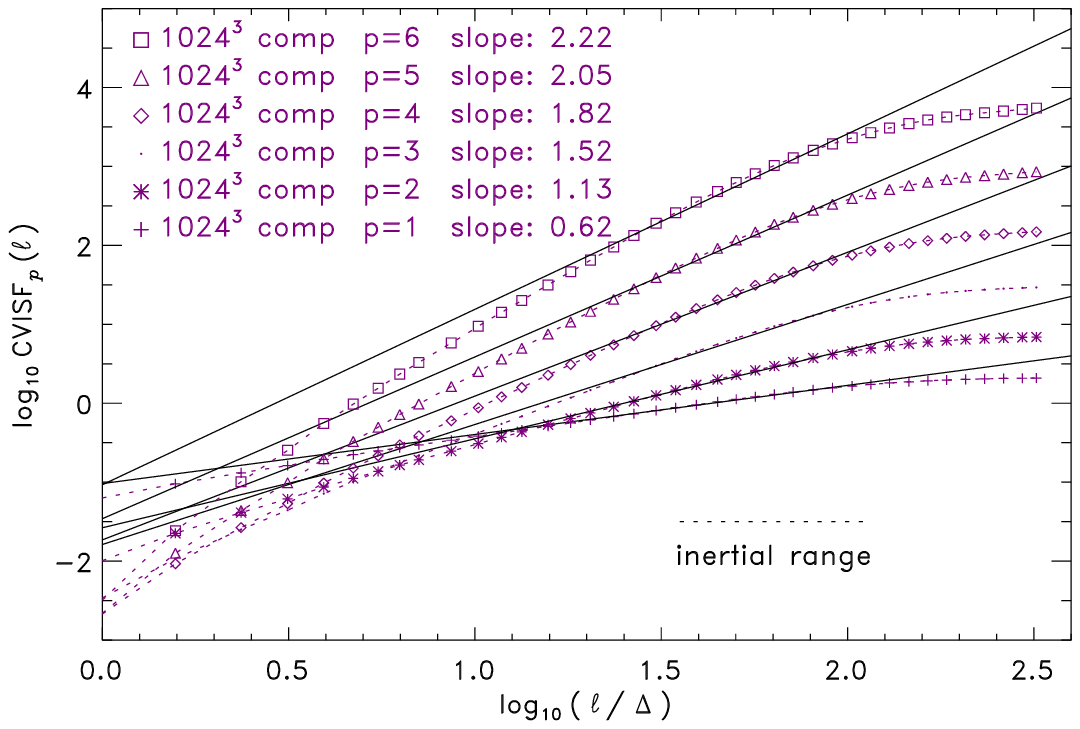}
}
\caption{Scaling of the structure functions of centroid velocity increments defined in equation~(\ref{eq:cvisf}) for solenoidal forcing (\emph{left}) and compressive forcing (\emph{right}) up to the 6th order. Scaling exponents obtained using power-law fits following equation~(\ref{eq:cvisf_scaling}) within the inertial range are indicated in the figures and summarised in Tab.~\ref{tab:cvis}.}
\label{fig:cvis}
\end{figure*}

\begin{table*}[t]
\begin{center}
\caption{Scaling of the structure functions of centroid velocity increments. \label{tab:cvis}}
\def\arraystretch{1.1}
\begin{tabular}{lcccccc}
\hline
\hline
Absolute Scaling Exponents~\mydotfill & $\zeta_1$ & $\zeta_2$ & $\zeta_3$ & $\zeta_4$ & $\zeta_5$ & $\zeta_6$\\
\hline
CVI SFs ($1024^3$ sol)\hspace{2cm} & $0.62$ & $1.20$ & $1.73$ & $2.21$ & $2.65$ & $3.05$\\
CVI SFs ($1024^3$ comp) & $0.62$ & $1.13$ & $1.52$ & $1.82$ & $2.05$ & $2.22$\\
\hline
Relative Scaling Exponents~\mydotfill & $\widetilde{\zeta}_1$ & $\widetilde{\zeta}_2$ & $\widetilde{\zeta}_3$ & $\widetilde{\zeta}_4$ & $\widetilde{\zeta}_5$ & $\widetilde{\zeta}_6$\\
\hline
CVI SFs using ESS$^{\,a}$ ($1024^3$ sol) & $0.36$ & $0.70$ & $1.00$ & $1.27$ & $1.51$ & $1.72$\\
CVI SFs using ESS$^{\,a}$ ($1024^3$ comp) & $0.38$ & $0.72$ & $1.00$ & $1.23$ & $1.41$ & $1.56$\\
\object{Polaris Flare}$^{\,b}$ & $0.37$ & $0.70$ & $1.00$ & $1.27$ & $1.53$ & $1.77$\\
\object{Polaris Flare}$^{\,c}$ & $0.38$ & $0.71$ & $1.00$ & $1.28$ & $1.54$ & $1.80$\\
Intermittency Model SL94$^{\,d}$ & $0.36$ & $0.70$ & $1.00$ & $1.28$ & $1.54$ & $1.78$\\
Intermittency Model B02$^{\,e}$ & $0.42$ & $0.74$ & $1.00$ & $1.21$ & $1.40$ & $1.56$\\
\hline
\end{tabular}
\end{center}
$^a$ Using extended self-similarity (ESS) \citep{BenziEtAl1993}.\\
$^b$ Measurement of CVI structure functions by \citet{HilyBlantFalgaronePety2008}.\\
$^c$ Measurement of CVI structure functions by \citet{HilyBlantFalgaronePety2008} using $^{12}$CO(2--1) data by \citet{BenschStutzkiOssenkopf2001}.\\
$^d$ Intermittency model $\widetilde{\zeta}_p=p/9+\mathcal{C}\left(1-\left(1-2/(3\mathcal{C})\right)^{p/3}\right)$ defined in eq.~(\ref{eq:intermittencymodel}) using a fractal co-dimension $\mathcal{C}=2$ \citep{SheLeveque1994}, which corresponds to filamentary structures ($\mathcal{D}=1$).\\
$^e$ Same as $^d$, but for co-dimension $\mathcal{C}=1$ \citep{Boldyrev2002,BoldyrevNordlundPadoan2002} corresponding to sheet-like structures ($\mathcal{D}=2$).
\end{table*}

\begin{figure*}[t]
\centerline{
\includegraphics[width=0.5\linewidth]{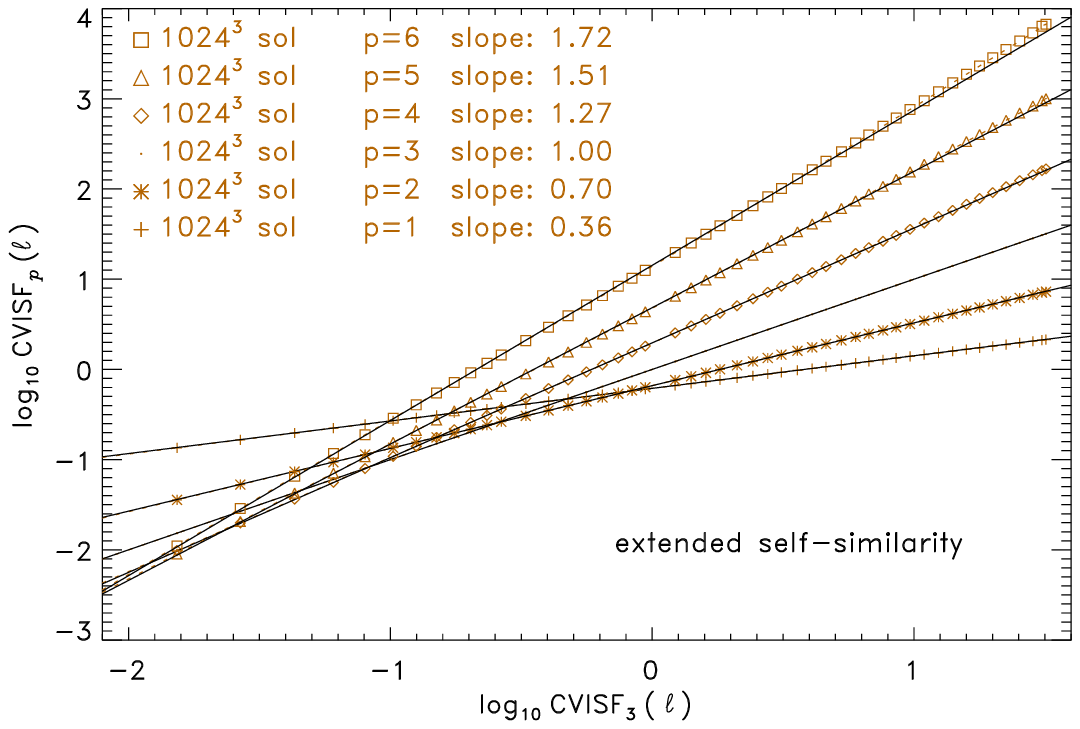}
\includegraphics[width=0.5\linewidth]{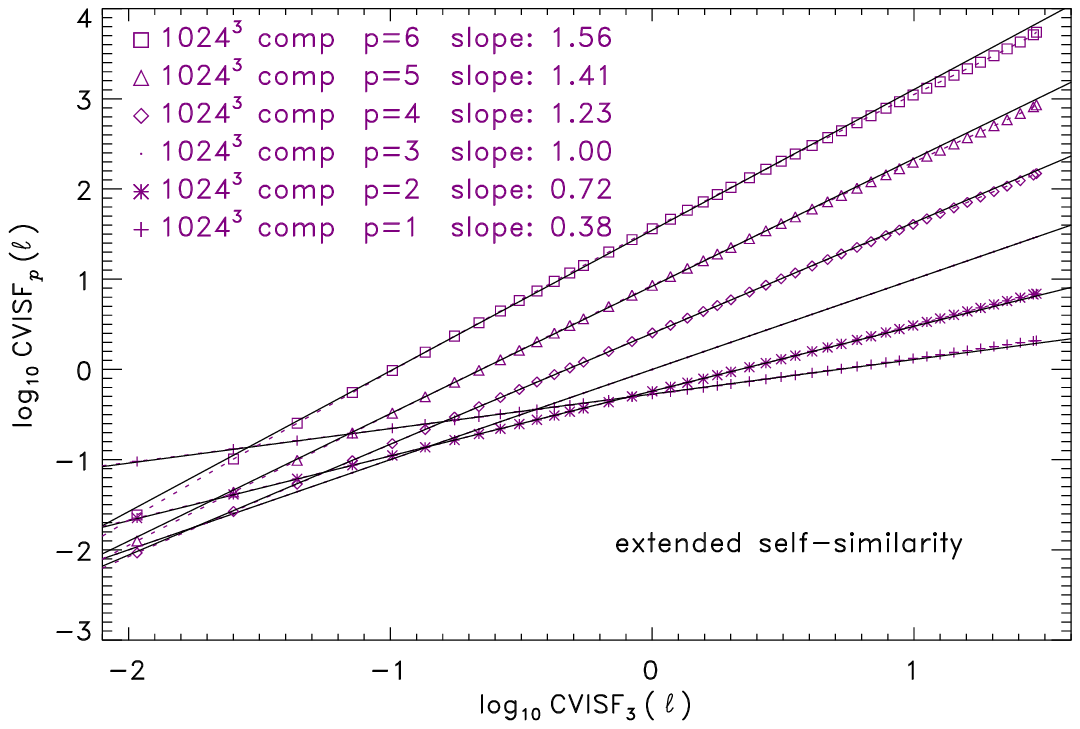}
}
\caption{Same as Fig.~\ref{fig:cvis}, but using the extended self-similarity hypothesis \citep{BenziEtAl1993}, allowing for a direct comparison of the scaling exponents of centroid velocity increments with the study by \citet{HilyBlantFalgaronePety2008} for the \object{Polaris Flare} and \object{Taurus MC} (see Tab.~\ref{tab:cvis}).}
\label{fig:cvisess}
\end{figure*}

For a direct comparison of CVI structure functions with the study by \citet[][Fig.~8]{HilyBlantFalgaronePety2008}, we apply the extended self-similarity (ESS) hypothesis \citep{BenziEtAl1993}, which states that the inertial range scaling may be extended beyond the inertial range, such that power-law fits can be applied over a larger dynamic range. The ESS hypothesis is used by plotting the $p\,$th order $\mathrm{CVISF}_p(\ell)$ against the 3rd order $\mathrm{CVISF}_3(\ell)$ \citep{BenziEtAl1993}. These plots are shown in Figure~\ref{fig:cvisess}. Indeed, the scaling range is drastically increased using ESS. All ESS data points are consistent with a single power law for each CVI structure function order $p\leq6$. We summarise the scaling exponents with and without using the ESS hypothesis in Table~\ref{tab:cvis}.

Table~\ref{tab:cvis} furthermore provides the ESS scaling exponents obtained for the \object{Polaris Flare} \citep[][Tab.~3]{HilyBlantFalgaronePety2008}, as well as the scaling exponents obtained from intermittency models of the structure function scaling exponents
\begin{equation} \label{eq:intermittencymodel}
\widetilde{\zeta}_p\equiv\frac{\zeta_p}{\zeta_3}=\frac{p}{9}+\mathcal{C}\left(1-\left(1-\frac{2}{3\mathcal{C}}\right)^{p/3}\right)
\end{equation}
by \citet{SheLeveque1994} ($\mathcal{C}=2$) and \citet{Boldyrev2002} ($\mathcal{C}=1$). In these models, the fractal co-dimension $\mathcal{C}$ is related to the fractal dimension of the most intermittent structures $\mathcal{D}$ by $\mathcal{C}=3-\mathcal{D}$. The \citet{SheLeveque1994} model assumes 1D vortex filaments as the most intermittent structures ($\mathcal{D}=1$), whereas the \citet{Boldyrev2002} model assumes sheet-like structures with $\mathcal{D}=2$.

For solenoidal forcing, the scaling of the CVI structure functions using ESS is very similar to the \citet{SheLeveque1994} model. This model is appropriate for \emph{incompressible} turbulence, for which the most intermittent structures are expected to be filaments \citep[][$\mathcal{D}=1$]{SheLeveque1994}. Interestingly, their model seems to be consistent with the measurements in the Polaris Flare by \citet{HilyBlantFalgaronePety2008} and with our solenoidal forcing case. In contrast, the scaling exponents derived for compressive forcing are better consistent with the intermittency model by \citet[][$\mathcal{D}=2$]{Boldyrev2002}. This direct comparison indicates that turbulence in the Polaris Flare observed by \citet{HilyBlantFalgaronePety2008} behaves like solenoidally forced turbulence. However, it does not imply that turbulence in the Polaris Flare is close to incompressible, since our numerical models are clearly supersonic in the inertial range (see~\S~\ref{sec:spectra}). It rather means that CVI scaling is different from the absolute scaling exponents following from the intermittency models by \citet{SheLeveque1994} and \citet{Boldyrev2002}. This is mainly because of two reasons: First, these models do not account for density fluctuations \citep[see however][]{SchmidtFederrathKlessen2008}, and second, CVIs are 2D projections of the 3D turbulence. The statistics derived from CVIs is a convolution of density and velocity statistics projected onto a 2D plane. As shown by \citet{OssenkopfEtAl2006} and \citet{EsquivelEtAl2007}, CVI statistics differ significantly from pure velocity statistics, if the ratio of density dispersion to mean density is high. This is usually the case in supersonic flows, and is also the case for both our numerical experiments (see Table~\ref{tab:pdfs_vol}). It explains the difference between the structure functions derived from the pure velocity statistics compared to convolved velocity--density statistics \citep{SchmidtFederrathKlessen2008}. The deviations from the \citet{Kolmogorov1941c} scaling ($\widetilde{\zeta}_p=p/3$) for the 3D data analysed in \citet{SchmidtFederrathKlessen2008} are significantly larger than those derived via CVI in 2D, revealing a significant loss in the signatures of intermittency in the projected CVI data \citep[see also][for a discussion of projection effects]{BruntEtAl2003,BruntMacLow2004}. This also means that direct tests of the theoretical models will be very difficult to achieve, unless a means of relating the CVI-based moments to the 3D moments is developed. Moreover, the fractal dimension of structures changes in a non-trivial way upon projection \citep{StutzkiEtAl1998,SanchezEtAl2005,FederrathKlessenSchmidt2009}, which severely limits the comparison of CVI statistics with the 3D intermittency models by \citet{SheLeveque1994}, \citet{Boldyrev2002} and \citet{SchmidtFederrathKlessen2008}.

Nevertheless, a \emph{direct} comparison of CVI structure function scaling obtained in numerical experiments and observations can provide useful information to distinguish between different parameters of the turbulence, as for instance different turbulence forcings.

\section{Principal component analysis} \label{sec:pca}
Principal component analysis (PCA) is a multivariate tool \citep{MurtaghHeck1987} introduced by \citet{HeyerSchloerb1997} for measuring the scaling of interstellar turbulence. It has been used for studying the structure and scaling in several molecular cloud regions, simulations and synthetic images \citep{BruntHeyer2002a,BruntHeyer2002b,BruntEtAl2003,HeyerBrunt2004,HeyerWilliamsBrunt2006}. PCA can be used to characterise structure on different scales. For best comparison with observations, we choose to work in position-position-velocity (PPV) space. Since our simulation data are typically stored in position-position-position (PPP) space, we transformed our PPP cubes into PPV space prior to PCA. As for the CVIs discussed in the previous section, we use the approximation of optically thin radiative transfer to derive radiation intensity. This means that we essentially assume that the emission is proportional to the gas density. The PPV data therefore represent a simulated measured intensity $T(x_i,y_i,v_{z,j}) \equiv T_{ij}$ at spatial position $\vect{r}_i=(x_i,y_i)$ and spectral position $v_{z,j}$. The indices $i$ and $j$ thus represent the spatial and spectral coordinates respectively. A detailed description of the PCA technique is given by \citet{HeyerSchloerb1997} and \citet{BruntHeyer2002a}. The most important steps necessary to derive the characteristic length scales and corresponding velocity scales using PCA are described below. First, the covariance matrix
\begin{equation}
U_{jk} = \frac{1}{N_x N_y}\,T_{ij}\,T_{ik}
\end{equation}
is constructed by summation over all spatial points $N_x$ and $N_y$. Solving the eigenvalue equation
\begin{equation}
\underline{U}\,\vect{u}^{(l)} = \lambda^{(l)}\,\vect{u}^{(l)}
\end{equation}
yields the $l\,$th eigenvalue $\lambda^{(l)}$ and the $l\,$th eigenvector $\vect{u}^{(l)}$ of the covariance matrix. The subsequent projection
\begin{equation}
I_i^{(l)} = T_{ik}^{\phantom{(l)}}u_k^{(l)}
\end{equation}
onto the eigenvectors yields the $l\,$th eigenimage $I_i^{(l)}$. Autocorrelation functions (ACFs) are then computed for each of the eigenimages and eigenvectors. The spatial scale on which the two-dimensional ACF of the $l\,$th eigenimage falls off by $1/e$ defines the $l\,$th characteristic spatial scale. Following the same procedure, the corresponding characteristic velocity scale is determined from the ACF of the $l\,$th eigenvector, which contains the spectral information.

\begin{figure*}[t]
\centerline{
\includegraphics[width=0.5\linewidth]{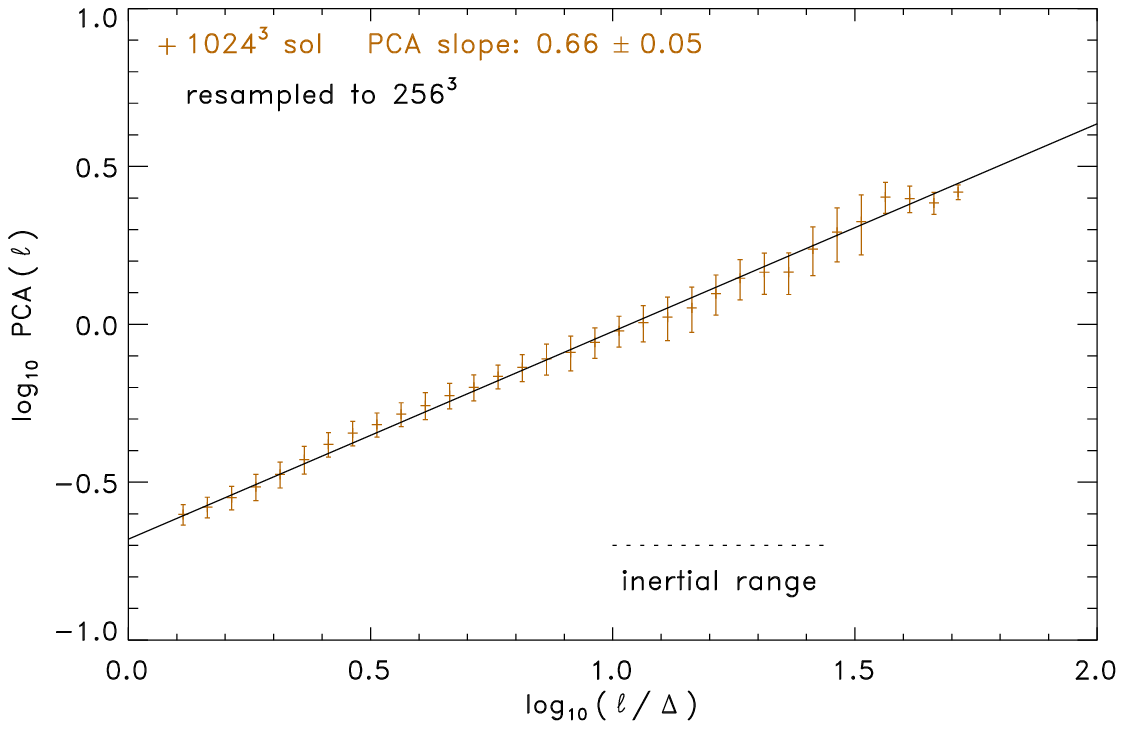}
\includegraphics[width=0.5\linewidth]{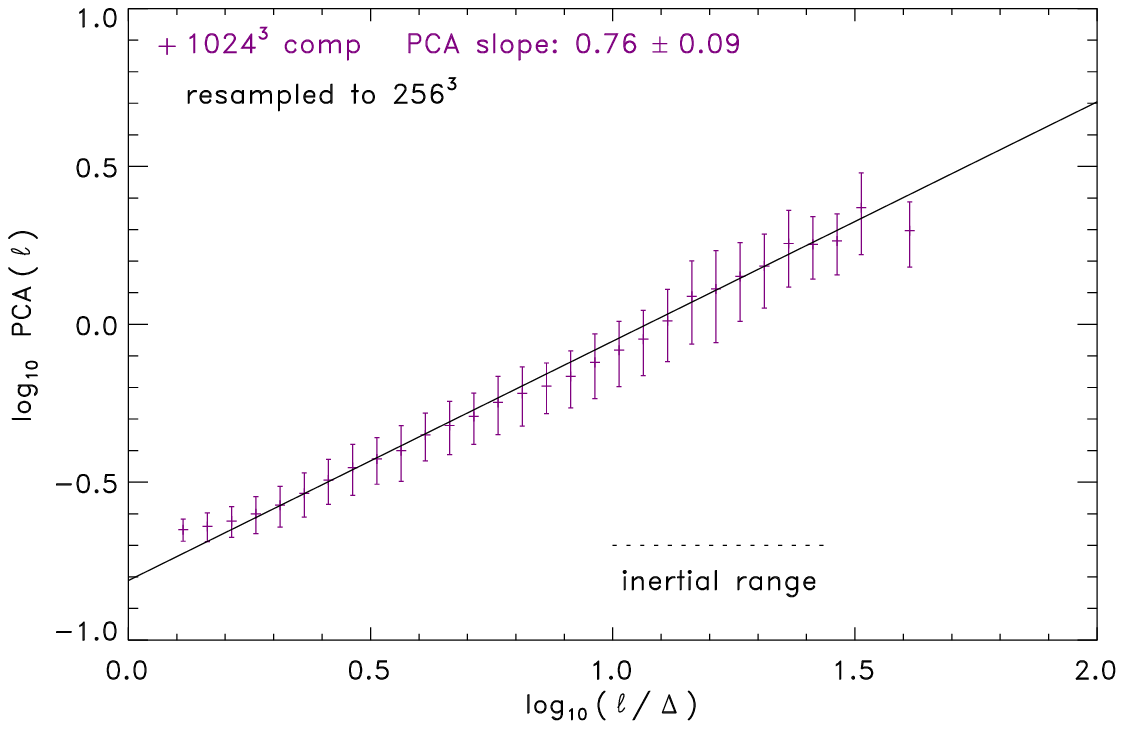}
}
\caption{Principal component analysis (PCA) for solenoidal (\emph{left}) and compressive forcing (\emph{right}). The PCA slopes obtained for solenoidal and compressive forcings are summarised and compared with observations by \citet{HeyerWilliamsBrunt2006} in Table~\ref{tab:pca}. The error bars contain the contribution from temporal variations and from three different projections along the $x$, $y$ and $z$-axes. The data were re-sampled from $1024^3$ to $256^3$ grid points prior to PCA. The re-sampling speeds up the PCA and has virtually no effect on the inertial range scaling \citep[see e.g.,][]{PadoanEtAl2006,FederrathKlessenSchmidt2009}.}
\label{fig:pcas}
\end{figure*}

Figure~\ref{fig:pcas} shows our time-~and projection-averaged set of spatial and velocity scales obtained with PCA. We have fitted power laws to the PCA data, which yielded PCA scaling exponents $\alpha_\mathrm{PCA}$ for solenoidal and compressive forcing respectively. For solenoidal forcing we find $\alpha_\mathrm{PCA}=0.66\pm0.05$ and for compressive forcing we find $\alpha_\mathrm{PCA}=0.76\pm0.09$ (see Table~\ref{tab:pca}). The different PCA slopes $\alpha_\mathrm{PCA}$ derived for solenoidal and compressive forcing suggest that by using PCA, differences in the mixture of transverse and longitudinal modes of the velocity field can be detected. However, the difference between solenoidal and compressive forcing is only at the 1-$\sigma$ level.

\begin{table*}[t]
\begin{center}
\caption{Comparison of measured PCA scaling slopes. \label{tab:pca}}
\def\arraystretch{1.1}
\begin{tabular}{lccccc}
\hline
\hline
\hspace{3.5cm} & $1024^3$ sol & $1024^3$ comp & \object{Rosette Zone I}$^{\,a}$ & \object{Rosette Zone II}$^{\,b}$ & \object{G216-2.5}$^{\,c}$\\
\hline
$\alpha_\mathrm{\,PCA}$~\mydotfill & $0.66\pm0.05$ & $0.76\pm0.09$ & ~ & ~ & ~\\
% \hline
$\alpha_\mathrm{\,PCA}$ from $^{12}$CO(1--0)~\mydotfill & ~ & ~ & $0.79\pm0.06$ & $0.66\pm0.06$ & $0.63\pm0.04$\\
% \hline
$\alpha_\mathrm{\,PCA}$ from $^{13}$CO(1--0)~\mydotfill & ~ & ~ & $0.86\pm0.09$ & $0.67\pm0.12$ & $0.56\pm0.02$\\
\hline
\end{tabular}
\end{center}
$^a$ PCA by \citet{HeyerWilliamsBrunt2006} of the interior of an HII region in the \object{Rosette MC}.\\
$^b$ Same as $^a$, but exterior of the HII region.\\
$^c$ PCA by \citet{HeyerWilliamsBrunt2006} for \object{G216-2.5} (\object{Maddalenas's Cloud}).
\end{table*}

\citet{HeyerWilliamsBrunt2006} applied PCA to the \object{Rosette MC} and to \object{G216-2.5} (\object{Maddalenas's Cloud}). These two clouds are quite different in dynamical and evolutionary state, although they exhibit roughly the same turbulence Mach number. \citet{HeyerWilliamsBrunt2006} measured the Mach number $\mathcal{M}_{1\mathrm{pc}}\!\approx\!4\!-\!5$ on a scale of $1\,\mathrm{pc}$ for both clouds. The Rosette MC exhibits confirmed massive star formation, whereas G216-2.5 has a low star formation rate, similar to the Polaris Flare discussed in the previous section. \citet{HeyerWilliamsBrunt2006} measured PCA slopes for both clouds and additionally provided the PCA slopes in two distinct subregions of the Rosette MC. The first subregion is inside the HII region (Zone I) surrounding the massive star cluster \object{NGC 2244}\footnote{The formation of the \object{star cluster XA} in the Rosette MC was likely triggered by the accumulation of material in the expanding shell surrounding the OB star cluster \object{NGC 2244} \citep{WangEtAl2008,WangEtAl2009}. This emphasises the importance of expanding HII regions in triggering subsequent star formation by compression of gas in expanding shells \citep{ElmegreenLada1977}.}, while the other subregion is outside of this HII region (Zone II). The measured PCA slopes obtained from $^{12}$CO and $^{13}$CO observations are summarised in Table~\ref{tab:pca} together with our estimates for solenoidal and compressive forcing. The PCA scaling exponent for solenoidal forcing is very close to the PCA scaling exponents derived from the $^{12}$CO observations in the G216-2.5 ($\alpha_\mathrm{PCA}=0.63\pm0.04$) and in Zone II of the Rosette MC ($\alpha_\mathrm{PCA}=0.66\pm0.06$). In contrast, the PCA slope derived from $^{12}$CO observations in Zone I of the Rosette MC ($\alpha_\mathrm{PCA}=0.79\pm0.06$) is better consistent with our compressive forcing case. This indicates that Zone I contains more kinetic energy in compressive modes than Zone II and G216-2.5. The corresponding $^{13}$CO observations reported in \citet{HeyerWilliamsBrunt2006} yield slightly larger differences between the PCA scaling exponents derived for Zone I on the one hand, and Zone II and G216-2.5 on the other hand (see also Table~\ref{tab:pca}). This supports the idea that Zone I in the Rosette MC, and Zone II as well as G216-2.5 contain quite different amounts of compressive modes in the velocity field, which may be the result of different turbulence forcing mechanisms, similar to the differences obtained in purely solenoidal and compressive forcings.

\section{Fourier spectra} \label{sec:spectra}

\subsection{Velocity Fourier spectra} \label{sec:vspectra}
Fourier spectra of the velocity field $E(k)$ are typically used to distinguish between \citet{Kolmogorov1941c} turbulence, $E(k)\propto k^{-5/3}$ and \citet{Burgers1948} turbulence, $E(k)\propto k^{-2}$. For highly compressible, isothermal, supersonic, turbulent flow, it has been shown that the inertial range scaling is close to Burgers turbulence. For instance, \citet{KritsukEtAl2007} found $E(k)\propto k^{-1.95}$ and \citet{SchmidtEtAl2009} obtained $E(k)\propto k^{-1.87}$ from high-resolution numerical simulations.

The Fourier spectrum of a quantity provides a measure of the scale dependence of this quantity. Velocity Fourier spectra are thus defined as
\begin{equation} \label{eq:ft_e}
E(k)\,\mathrm{d}k = \frac{1}{2}\int\widehat{\vect{v}}\cdot\widehat{\vect{v}}^{*}\,4\pi k^2\,\mathrm{d}k\;,
\end{equation}
where $\widehat{\vect{v}}$ denotes the Fourier transform of the velocity field \citep[e.g.,][]{Frisch1995}. The total Fourier spectrum can be separated into transverse ($\vect{k} \perp \widehat{\vect{v}}$) and longitudinal ($\vect{k} \parallel \widehat{\vect{v}}$) parts by applying a Helmholtz decomposition. Note that integrating the transverse energy spectrum yields the kinetic energy in transverse (rotational) modes, while integration of the longitudinal energy spectrum yields the kinetic energy in longitudinal (compressible) modes. Furthermore, by integrating the velocity spectrum from $k_1$ to $k_2$, one obtains the kinetic energy content on length scales corresponding to the wavenumber interval $[k_1,k_2]$. Since the mean velocity is zero in our simulations, integration of the total velocity Fourier spectrum $E(k)$ over all wavenumbers yields the total variance of velocity fluctuations $\sigma_v^2$:
\begin{equation}
\int_1^{k_\mathrm{c}}\,E(k)\,\mathrm{d}k = \frac{1}{2}\,\sigma_v^2\;.
\end{equation}
The upper bound of the integral is the cutoff wavenumber $k_\mathrm{c}=N$ for a cubic dataset with $N^3$ data points. Thus, $k_\mathrm{c}=1024$ for our standard resolution of $1024^3$ grid cells.

\begin{figure*}[t]
\centerline{
\includegraphics[width=0.5\linewidth]{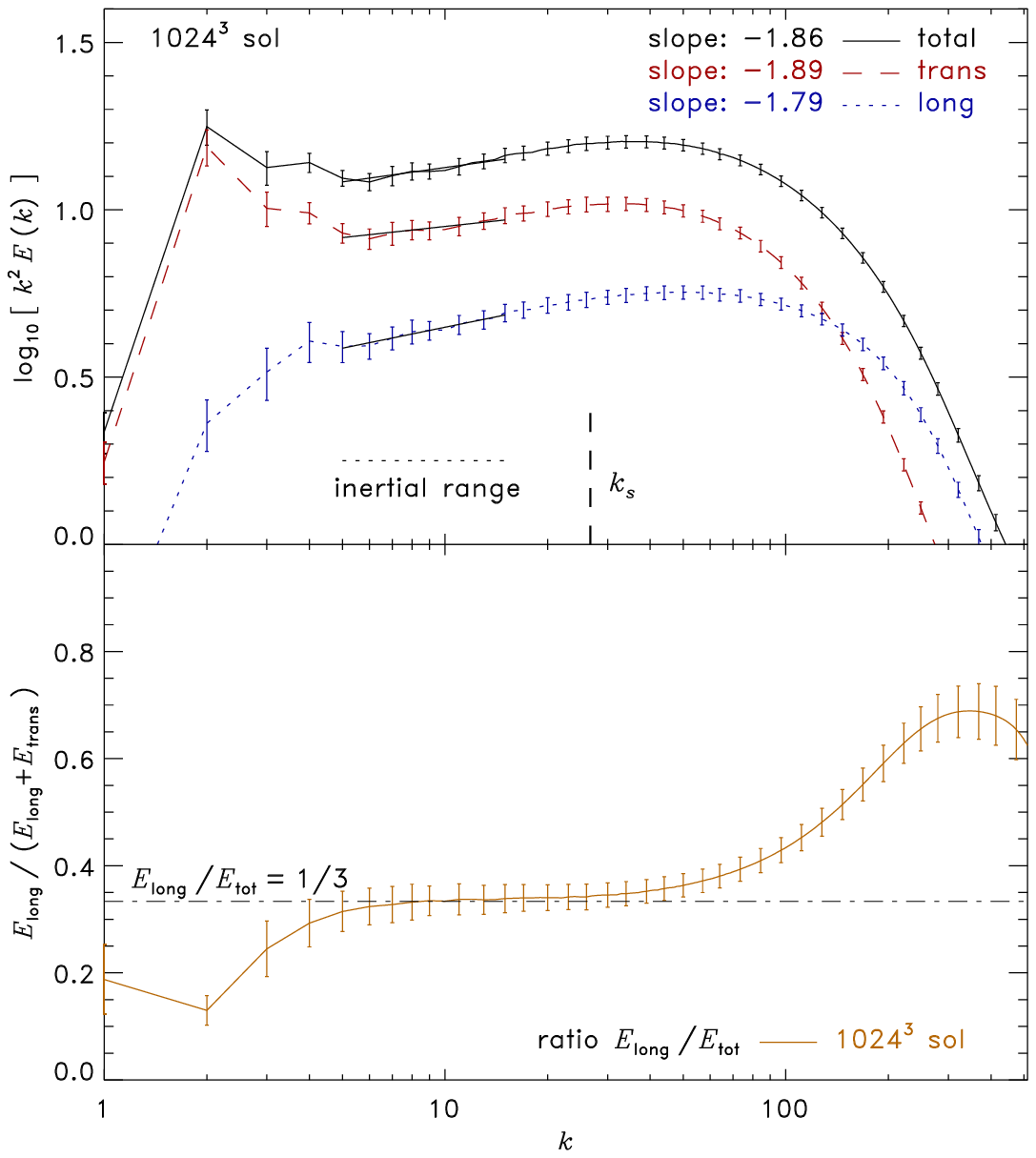}
\includegraphics[width=0.5\linewidth]{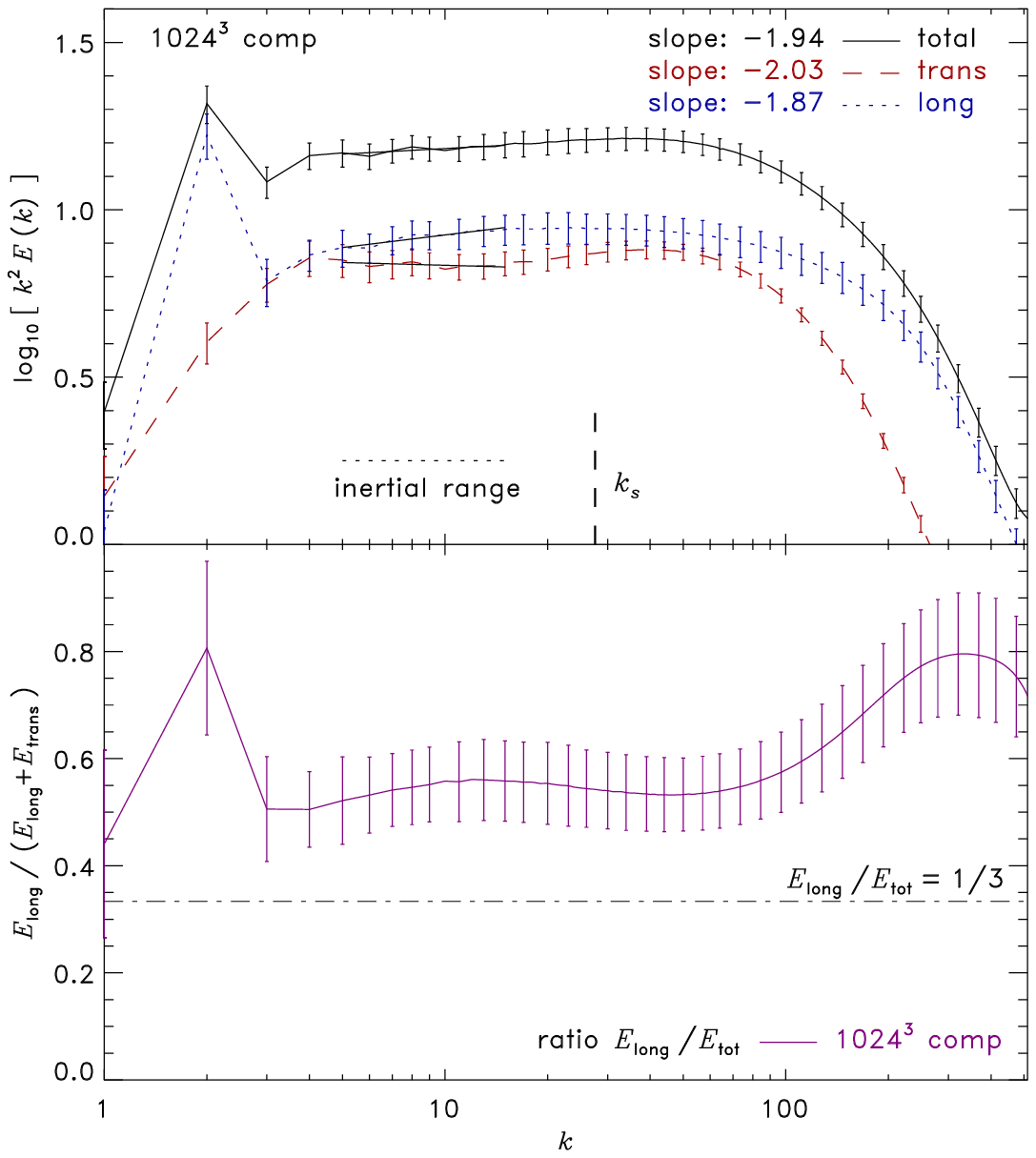}
}
\caption{\emph{Top panels:} Total, transverse (rotational) and longitudinal (compressible) velocity Fourier spectra $E(k)$ defined in equation~(\ref{eq:ft_e}) and  compensated by $k^2$ for solenoidal (\emph{left}) and compressive forcing (\emph{right}). Error bars indicate temporal variations, which account for an uncertainty of roughly $\pm0.05$ of all scaling slopes reported for the inertial range $5 \lesssim k \lesssim 15$. The inferred inertial range scaling exponents for both solenoidal and compressive forcing are consistent with independent numerical simulations and with observations of the size--linewidth relation (see text). Note that the transverse part, $E_\mathrm{trans}$ falls off more steeply than the longitudinal part, $E_\mathrm{long}$ for both forcing types in the inertial range. \emph{Bottom panels:} Ratio of the energy in longitudinal velocity modes $E_\mathrm{long}$ to the total energy in velocity modes $E_\mathrm{tot}=E_\mathrm{trans}+E_\mathrm{long}$. For solenoidal forcing, we obtain $E_\mathrm{long}/E_\mathrm{tot}\!\approx\!1/3$ in the inertial range (horizontal dash-dotted line), because compression can only occur in one of the three spatial dimensions on average \citep{ElmegreenScalo2004,FederrathKlessenSchmidt2008}. For compressive forcing, this ratio is roughly $1/2$, which corresponds to an equipartition of longitudinal and transverse velocity modes. Note however that compressive forcing can compress the gas in all three spatial dimensions \emph{directly}, whereas solenoidal forcing can only induce compression \emph{indirectly} through the velocity field \citep{FederrathKlessenSchmidt2008}. The excess of longitudinal modes at high wavenumbers $k\gtrsim40$ stems from numerical dissipation, which is more effectively dissipating transverse than longitudinal modes on small scales due to the discretisation onto a grid. This suggests that roughly $30$ grid cells are needed to accurately resolve a vortex, while a shock is typically resolved with roughly $3$ grid cells using the piecewise parabolic method \citep{ColellaWoodward1984}. However, for a numerical resolution of $1024^3$ grid cells, we find that wavenumbers $k\lesssim40$ are almost unaffected by the discretisation and by the parameters of the numerical scheme (see Appendix~\ref{app:spectra}).}
\label{fig:spectratio}
\end{figure*}

In Figure~\ref{fig:spectratio} we show the total velocity Fourier spectra $E(k)$ as defined in equation~(\ref{eq:ft_e}) together with its decomposition into transverse $E_\mathrm{trans}$ and longitudinal $E_\mathrm{long}$ parts for solenoidal and compressive forcing respectively. The prominent signature of the different forcings on the main driving scale, $k=2$ is clearly noticeable: Solenoidal forcing excites mostly transverse modes, whereas compressive forcing excites mostly longitudinal modes in the velocity field at $k=2$. However, the forcing has direct influence only for $1<k<3$ (see~\S~\ref{sec:forcing}). Further down the cascade, the turbulent flow develops its own statistics as a result of non-linear interactions in the inertial range $5\lesssim k\lesssim15$. We emphasise that this scaling range was chosen very carefully, since turbulence simulations will only provide a small inertial range even at resolutions of $1024^3$ grid cells \citep[see, e.g.,][]{KleinEtAl2007,LemasterStone2009}. This is mainly caused by the bottleneck phenomenon \citep[e.g.,][]{PorterPouquetWoodward1994,DoblerEtAl2003,HaugenBrandenburg2004,SchmidtHillebrandtNiemeyer2006,KritsukEtAl2007}, which may slightly affect the Fourier spectra in the dissipation range. However, the bottleneck phenomenon had no significant impact on the turbulence statistics in our numerical study for wavenumbers $k\lesssim40$. This is demonstrated in Appendix~\ref{app:spectra}, where we present the resolution dependence of the Fourier spectra and the dependence on parameters of the PPM numerical scheme. We conclude that the statistical quantities derived for wavenumbers $k\lesssim40$ are not significantly affected by the numerical scheme or limited resolution applied in the present study.

We apply power-law fits to the inertial range data with the resulting slopes indicated in Figure~\ref{fig:spectratio} (top panels). Both solenoidal and compressive forcing yield slopes consistent with size--linewidth relations inferred from observations \citep[e.g.,][]{Larson1981,Myers1983,PeraultFalgaronePuget1986,SolomonEtAl1987,FalgaronePugetPerault1992,MieschBally1994,OssenkopfMacLow2002,PadoanEtAl2003,HeyerBrunt2004,PadoanEtAl2006,OssenkopfKripsStutzki2008b,HeyerEtAl2009}, and with the results of independent numerical simulations \citep[e.g.,][]{KlessenHeitschMacLow2000,BoldyrevNordlundPadoan2002,PadoanJimenezNordlundBoldyrev2004,KritsukEtAl2007,SchmidtEtAl2009}. Note that size--linewidth relations of the form $\sigma_v \propto l^\gamma$ with scaling exponents $\gamma=0.4\!-\!0.5$ correspond to Fourier spectra $E(k)\propto k^{-\beta}$ with scaling exponents in the range $\beta=1.8\!-\!2.0$, because $\gamma=(\beta-1)/2$. However, it must be emphasised that the relation between scaling exponents obtained from observational maps of centroid velocities (as discussed in \S~\ref{sec:cvisfs}) and 3D velocity fields from simulations is non-trivial, because of projection-smoothing and intensity-weighting. Projection-smoothing increases the scaling exponents of the 2D projection of a 3D field such that $\gamma_\mathrm{2D} = \gamma_\mathrm{3D} + 1/2$ \citep[e.g.,][]{StutzkiEtAl1998,BruntMacLow2004}. However, \citet{BruntMacLow2004} showed that the effect of projection-smoothing is compensated statistically (but not identically) by intensity-weighting of observed centroid velocity maps. Thus, our measurements of velocity scaling seem consistent with observations.

It is important to note that the transverse parts $E_\mathrm{trans}(k)$ fall off more steeply than the longitudinal parts $E_\mathrm{long}(k)$ for both forcing types within the inertial range. For solenoidal forcing, we find $E_\mathrm{trans}(k) \propto k^{-1.89}$ and $E_\mathrm{long}(k) \propto k^{-1.79}$, and for compressive forcing, $E_\mathrm{trans}(k) \propto k^{-2.03}$ and $E_\mathrm{long}(k) \propto k^{-1.87}$. This result indicates that longitudinal modes can survive down to small scales, such that compression may not be neglected anywhere in the turbulent cascade. \citet[][Fig.~9,$\,$10]{LemasterStone2009} obtain $E_\mathrm{trans}(k) \propto k^{-2.0}$ and $E_\mathrm{long}(k) \propto k^{-1.8}$ for their hydrodynamical model with solenoidal forcing at a resolution of $1024^3$ grid points in the Athena code. This is consistent with our findings for the scale dependence of the transverse and longitudinal parts and shows that the kinetic energy in longitudinal modes must not be neglected within the inertial range.

In order to quantify the relative importance of compression over rotation in the turbulent motions, we present plots of the ratio
\begin{equation} \label{eq:Eratio}
\Psi(k)=\frac{E_\mathrm{long}(k)}{E_\mathrm{long}(k)+E_\mathrm{trans}(k)}=\frac{E_\mathrm{long}(k)}{E_\mathrm{tot}(k)}
\end{equation}
in the bottom panels of Figure~\ref{fig:spectratio}. Solenoidal forcing yields $\Psi\!\approx\!1/3$ in the inertial range. We emphasise that the ratio $\Psi\!\approx\!1/3$ was expected from the fact that compression can only occur in one of the three available spatial dimensions on average in the case of supersonic flow driven by a purely solenoidal force field \citep{ElmegreenScalo2004,FederrathKlessenSchmidt2008}. This is the fundamental idea on which the heuristic model of the density dispersion--Mach number relation given by equation~(\ref{eq:b}) was based. For compressive forcing, we find $\Psi\!\approx\!1/2$ in the inertial range as a result of the direct compression induced by compressive forcing. Thus, solenoidal and compressive forcing produce quite similar amounts of compressive modes in the velocity field ($\Psi\!\approx\!1/3$ versus $\Psi\!\approx\!1/2$). This means that even fully compressive forcing can behave very similar to solenoidal forcing in the inertial range, as far as pure velocity statistics are concerned. However, we show in the next section that the density statistics are very different in the inertial range. The same is true for combined density--velocity statistics (see Appendix~\ref{app:spectra_sqrtrho_rho3}).

We also note here that the rise of $\Psi$ at $k \gtrsim 40$ for both forcing types is a numerical effect, which comes from the discretisation of the velocity field onto a grid with finite resolution. This shows that energy in rotational modes cannot be accounted for accurately if vortices are smaller than roughly $30$ grid cells in each direction, whereas longitudinal modes (i.e. shocks) may still be well resolved. As a result, the transverse kinetic energy is underestimated for $k\gtrsim40$ up to the resolution limit $k_\mathrm{c}=1024$. However, the plateau of almost constant $\Psi$ for $k\lesssim40$ indicates that the discretisation had no significant influence on scales with wavenumbers $k\lesssim40$. The effect of underestimating the transverse kinetic energy due to the discretisation of fluid variables is also observed in the ZEUS-3D simulations by \citet[][Fig.~2]{PavlovskiSmithMacLow2006} for wavenumbers $k\gtrsim10$ at numerical resolution of $256^3$ grid cells. In Appendix~\ref{app:spectra}, we furthermore demonstrate that our results for the Fourier spectra are not affected by the specific choice of parameters of the numerical scheme for wavenumbers $k\lesssim40$.

\subsection{Logarithmic density Fourier spectra}
In analogy to the velocity Fourier spectra $E(k)$, we define logarithmic density fluctuation spectra
\begin{equation} \label{eq:ft_s}
S(k)\,\mathrm{d}k = \int(\widehat{s-\left<s\right>})\,(\widehat{s-\left<s\right>})^{*}\,4\pi k^2\,\mathrm{d}k\;.
\end{equation}
We subtract the mean logarithmic density prior to the Fourier transformation such that $S(k)$ is a measure of density fluctuations as a function of scale. Therefore, integrating $S(k)$ over all scales yields the square of the logarithmic density dispersion $\sigma_s$
\begin{equation} \label{eq:ft_s_integral}
\int_1^{k_\mathrm{c}}\,S(k)\,\mathrm{d}k = \sigma_s^2\;.
\end{equation}
Furthermore, integrating $S(k)$ over the wavenumber range $[k_1,k_2]$ yields the typical density fluctuations on length scales corresponding to this range of scales.

\begin{figure*}[t]
\centerline{
\includegraphics[width=0.5\linewidth]{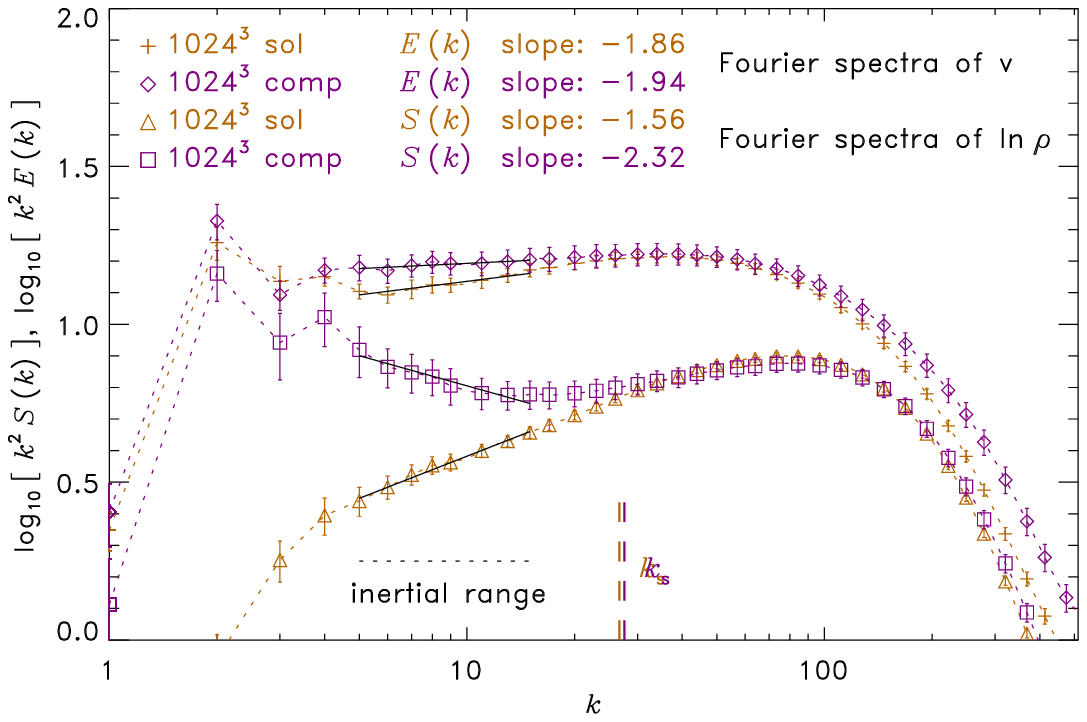}
\includegraphics[width=0.5\linewidth]{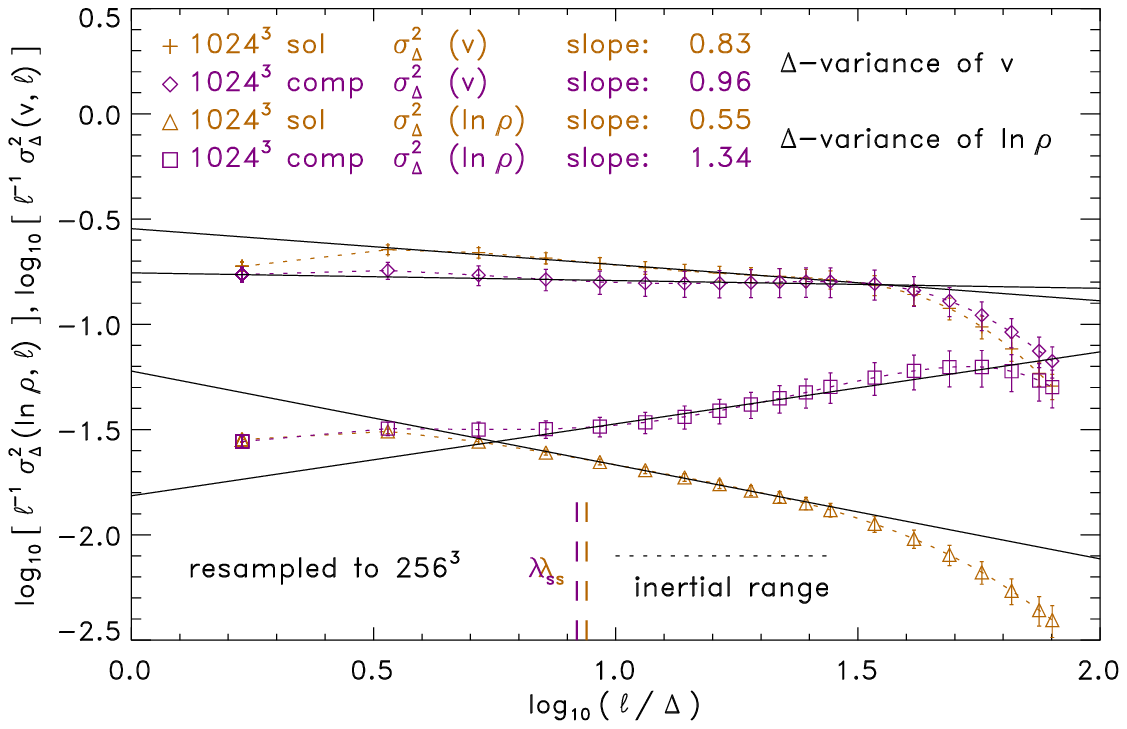}
}
\caption{\emph{Left panel:} Fourier spectra of the velocity, $E(k)$ defined in eq.~(\ref{eq:ft_e}) (crosses and diamonds) and Fourier spectra of the logarithmic density fluctuations, $S(k)$ defined in eq.~(\ref{eq:ft_s}) (triangles and squares) for solenoidal and compressive forcing, respectively. Both $E(k)$ and $S(k)$ are compensated by $k^2$ allowing for a better determination of the inertial range scaling. The density fluctuation power spectra differ significantly in the inertial range $5 \lesssim k \lesssim 15$ with $S(k)\propto k^{-1.56}$ for solenoidal and $S(k)\propto k^{-2.32}$ for compressive forcing. The scale on which the density fluctuation spectra from solenoidal and compressive forcing cross each other and where the slope obtained in compressive forcing breaks and approaches the shallower slope of the solenoidal forcing case roughly coincides with the sonic wavenumber $k_\mathrm{s}$ (vertical dashed lines) defined in eq.~(\ref{eq:ks}). \emph{Right panel:} Same as left panel, but instead of using Fourier spectra to determine the inertial range scaling, we use the $\Delta$-variance method to derive the scaling slopes in physical space. Note that the scaling slopes $\alpha$ obtained with the $\Delta$-variance technique are related to the slopes $\beta$ of the Fourier spectra by $\beta=\alpha+1$ \citep{StutzkiEtAl1998}. Error bars denote 1-$\sigma$ temporal fluctuations.}
\label{fig:spect}
\end{figure*}

Figure~\ref{fig:spect} (left) shows the logarithmic density fluctuation spectra $S(k)$ together with the total velocity Fourier spectra $E(k)$ in one plot. In contrast to the scaling of the velocity $E(k)$, the scaling of $S(k)\propto k^{-\beta}$ is significantly different for solenoidal ($\beta=1.56\pm0.05$) and compressive forcing ($\beta=2.32\pm0.09$) in the inertial range.

% The strong difference between density Fourier spectra of solenoidal and compressive forcing was also seen in the $\rho$ spectra--without taking the logarithm of the density prior to Fourier transformation--in \citet[][Fig.~3]{FederrathKlessenSchmidt2009}, which leads to fractal dimension estimates of $D\approx2.6$ for solenoidal and $D\approx2.3$ for compressive forcing.

\section{$\Delta$-variance of the velocity and density} \label{sec:deltavar}
The $\Delta$-variance technique provides a complementary method for measuring the scaling exponent of Fourier spectra in the physical domain using a wavelet transformation \citep{StutzkiEtAl1998}. We apply the $\Delta$-variance to our simulation data using the tool developed and provided by \citet{OssenkopfKripsStutzki2008a}. This tool implements an improved version of the original $\Delta$-variance \citep{StutzkiEtAl1998,BenschStutzkiOssenkopf2001}. The $\Delta$-variance measures the amount of structure on a given length scale $\ell$ by filtering the dataset $q(\vect{r})$ with an up-down-function $\bigodot\!_\ell$ (typically a French-hat or Mexican-hat filter) of size $\ell$, and computing the variance of the filtered dataset. The $\Delta$-variance is defined as
\begin{equation}
\sigma_\Delta^2(\ell) = \left<\left(q(\vect{r})\ast\bigodot\!\frac{\!}{\!}_\ell(\vect{r})\right)^2\right>_{\!\vect{r}}\,,
\end{equation}
where the average is computed over all data points at positions $\vect{r}=(x,y,z)$. The operator $\ast$ stands for the convolution. We use the original French-hat filter with a diameter ratio of $3.0$ as in previous studies using the $\Delta$-variance technique \citep[e.g.,][]{StutzkiEtAl1998,MacLowOssenkopf2000,OssenkopfKlessenHeitsch2001,OssenkopfMacLow2002,OssenkopfEtAl2006}.

Figure~\ref{fig:spect} (right panel) shows that the inertial range scaling obtained with the $\Delta$-variance technique is in very good agreement with the scaling measured in the Fourier spectra. Note that the scaling exponents $\beta$ of Fourier spectra are ideally related to the scaling exponents $\alpha$ of the $\Delta$-variance by $\alpha=\beta-1$ \citep{StutzkiEtAl1998}. The small deviations from this analytical relation are caused by the finite size of the dataset, the re-sampling procedure prior to the $\Delta$-variance analysis applied here and the choice of the filter function \citep{OssenkopfKripsStutzki2008a}. However, these deviations are on the order of 4\% and therefore smaller than the average snapshot-to-snapshot variations.

For the $\Delta$-variance of the velocity field, $\sigma_\Delta^2(v,\ell)\propto\ell^\alpha$, we find scaling exponents $\alpha=0.83\pm0.05$ for solenoidal forcing and $\alpha=0.96\pm0.05$ for compressive forcing. This translates into size-linewidth relations $\sigma_\Delta(v,\ell)\propto\ell^\gamma$ with scaling exponents $\gamma=\alpha/2$. Thus, we find $\gamma=0.42\pm0.03$ for solenoidal forcing and $\gamma=0.48\pm0.03$ for compressive forcing. \citet{OssenkopfMacLow2002} found a common power-law slope $\gamma=0.5\pm0.04$ for the \object{Polaris Flare}, ranging over three orders of magnitude in length scale from about $50\,\mathrm{pc}$ down to roughly $0.05\,\mathrm{pc}$. This scaling exponent is roughly consistent with both our solenoidal and compressive forcing data, but slightly better consistent with compressive forcing. Note that the centroid velocity analysis by \citet{OssenkopfMacLow2002} is also subject to the combined effects of projection-smoothing and intensity-weighting discussed in \citet{BruntMacLow2004} and discussed in~\S~\ref{sec:vspectra}. Thus, the comparison of 3D scaling of the velocity with 2D observations should always be made with the caution that projection-smoothing and intensity-weighting roughly cancel each other out in a statistical sense \citep{BruntMacLow2004}.

We are not aware of any observational study considering the scaling of \emph{logarithmic} intensity. The use of logarithmic density is useful in isothermal simulations, because the equations of hydrodynamics, equations~(\ref{eq:hydro1}) and~(\ref{eq:hydro2}), are invariant under transformations in $s=\ln(\rho/\langle\rho\rangle)$. In observations however, the intensity, $T$ is measured instead of the density, but the intensity can be transformed into $s'=\ln(T/\langle T\rangle)$, which gives a normalised quantity similar to $s=\ln(\rho/\langle\rho\rangle)$. This enables a straightforward comparison of simulation and observational data (yet with the limitations listed in~\S~\ref{sec:limitations}). It is also interesting to look at logarithmic density and intensity scaling, because this scaling parameter is used in analytic models of the mass distribution of cores and stars by \citet{HennebelleChabrier2008,HennebelleChabrier2009}.

Unlike a \emph{logarithmic} scaling analysis, the scaling of the \emph{linear} integrated intensity, $\sigma_\Delta(\rho,\ell)\propto\ell^\gamma$ was analysed by \citet{StutzkiEtAl1998} and \citet{BenschStutzkiOssenkopf2001}. They found $\gamma\approx0.5\!-\!0.9$ for the Polaris Flare, in good agreement with the scaling exponent $\gamma=0.8\pm0.1$ obtained from solenoidal forcing in \citet{FederrathKlessenSchmidt2009}. In contrast, the scaling exponent obtained for compressive forcing is significantly higher ($\gamma=1.4\pm0.3$). \citet{BenschStutzkiOssenkopf2001} measured scaling exponents $\gamma\approx1.0\!-\!1.5$ in small-scale maps ($\ell\lesssim0.1\,\mathrm{pc}$) of the Polaris Flare and in \object{Perseus/NGC1333}, which are consistent with our estimates for compressive forcing \citep[][Tab.~1]{FederrathKlessenSchmidt2009}. Since both solenoidal and compressive forcings display strong intermittency at short lags (see Fig.~\ref{fig:cvipdfskurt}), intermittency appears to be primarily measurable on scales smaller than the turbulence injection scale. Taking together the results by \citet{BenschStutzkiOssenkopf2001} with ours for solenoidal and compressive forcing indicates that interstellar turbulence is driven primarily on large scales, potentially with a significant amount of compressive modes present on the forcing scale \citep[see also][]{BruntHeyerMacLow2009}.

\section{The sonic scale} \label{sec:sonicscale}
The velocity Fourier spectra $E(k)$ discussed in~\S~\ref{sec:vspectra} can be described as power laws $E(k)\propto k^{-\beta}$ with negative power-law exponents, $\beta>1$. This means that the typical velocity fluctuations are decreasing when going to smaller scales. The value of the integral $\int_{k_1}^{k_\mathrm{c}} E(k)\,\mathrm{d}k$ over a finite range of wavenumbers with $k_1$ as the lower bound and the cutoff wavenumber $k_\mathrm{c}$ as the upper bound therefore becomes smaller with increasing $k_1$. Thus, the turbulent flow is expected to change from a supersonic  to a subsonic flow on a certain length scale. This scale separates the supersonic regime on large scales, where the velocity fluctuations are supersonic from the subsonic regime, which is located on smaller scales, where the typical velocity fluctuations are small compared to the thermal motions of the gas. This transition scale is called the \emph{sonic scale} $\lambda_\mathrm{s}$. Following \citet{SchmidtEtAl2009}, the corresponding \emph{sonic wavenumber} $k_\mathrm{s}$ in Fourier space is defined by solving the equation
\begin{equation} \label{eq:ks}
\int_{k_\mathrm{s}}^{k_\mathrm{c}} E(k)\,\mathrm{d}k \simeq \frac{1}{2}\,c_\mathrm{s}^2
\end{equation}
implicitly for $k_\mathrm{s}$. The sonic scale is thus defined as the scale on which the mean square velocity fluctuations become comparable to the mean square of the sound speed.

We solved equation~(\ref{eq:ks}) for the sonic wavenumbers $k_\mathrm{s}$ for both the solenoidal and compressive forcing cases. The sonic wavenumbers for solenoidal and compressive forcings are indicted in Figure~\ref{fig:spect} (left) as vertical dashed lines. We find $k_\mathrm{s}=26$ for solenoidal forcing and $k_\mathrm{s}=27$ for compressive forcing. The corresponding sonic scales $\lambda_\mathrm{s}$ are also indicated in Figure~\ref{fig:spect} (right) as vertical dashed lines.

The Fourier spectra $S(k)$ shown in Figure~\ref{fig:spect} (left) and the corresponding $\Delta$-variance curves shown in Figure~\ref{fig:spect} (right) for solenoidal and compressive forcing cross each other roughly at the sonic wavenumber and on the sonic scale, respectively. For compressive forcing $S(k)$ is significantly steeper on scales larger than the sonic scale ($k\lesssim k_\mathrm{s}$) compared to scales $k\gtrsim k_\mathrm{s}$. $S(k)$ for compressive forcing approaches the shallower slope of $S(k)$ for solenoidal forcing at $k\approx k_\mathrm{s}$. For $k\gtrsim k_\mathrm{s}$ there are neither significant differences between the density spectra $S(k)$ nor the velocity spectra $E(k)$ for solenoidal and compressive forcings.

The strong break in the logarithmic density fluctuation spectra $S(k)$ for compressive forcing around $k_\mathrm{s}$ appears to be linked to the transition from supersonic motions on large scales to subsonic motions on scales smaller than the sonic scale. In order to quantify this, we estimated the typical density fluctuations on supersonic scales ($k<k_\mathrm{s}$) by evaluating $\sigma^2_s(k\!<\!k_\mathrm{s})=\int_1^{k_\mathrm{s}}S(k)\,\mathrm{d}k$. We obtain $\sigma_s(k\!<\!k_\mathrm{s})\!\approx\!1.22$ for solenoidal and $\sigma_s(k\!<\!k_\mathrm{s})\!\approx\!3.05$ for compressive forcing, which is on the order of the logarithmic density dispersions $\sigma_s$ found from the density PDFs (see Table~\ref{tab:pdfs_vol}). This means that most of the power in density fluctuations is located on scales larger than the sonic scale. In contrast, on scales smaller than the sonic scale the typical density fluctuations can be estimated by solving $\sigma^2_s(k\!>\!k_\mathrm{s})=\int_{k_\mathrm{s}}^{k_\mathrm{c}}S(k)\,\mathrm{d}k$. We obtain $\sigma_s(k\!>\!k_\mathrm{s})\!\approx\!0.45$ for both types of forcing. This shows that density fluctuations on scales below the sonic scale are small compared to the typical density fluctuations in the supersonic regime at $k<k_\mathrm{s}$ \citep[see also][]{VazquezBallesterosKlessen2003}. Moreover, Figure~\ref{fig:spect} shows that the typical logarithmic density fluctuations are similar for both solenoidal and compressive forcings on scales smaller than the sonic scale. Note that the sum of logarithmic density fluctuations on all scales is
\begin{equation}
\left[\sigma^2_s(k\!<\!k_\mathrm{s})+\sigma^2_s(k\!>\!k_\mathrm{s})\right]^{1/2}\approx1.30
\end{equation}
for solenoidal forcing and
\begin{equation}
\left[\sigma^2_s(k\!<\!k_\mathrm{s})+\sigma^2_s(k\!>\!k_\mathrm{s})\right]^{1/2}\approx3.08
\end{equation}
for compressive forcing. As expected from equation~(\ref{eq:ft_s_integral}), these values are in excellent agreement with the total logarithmic density dispersions $\sigma_s$, obtained from the density PDFs shown in Table~\ref{tab:pdfs_vol}.

\begin{figure*}[t]
\begin{center}
\begin{tabular}{rr}
\includegraphics[width=0.45\linewidth]{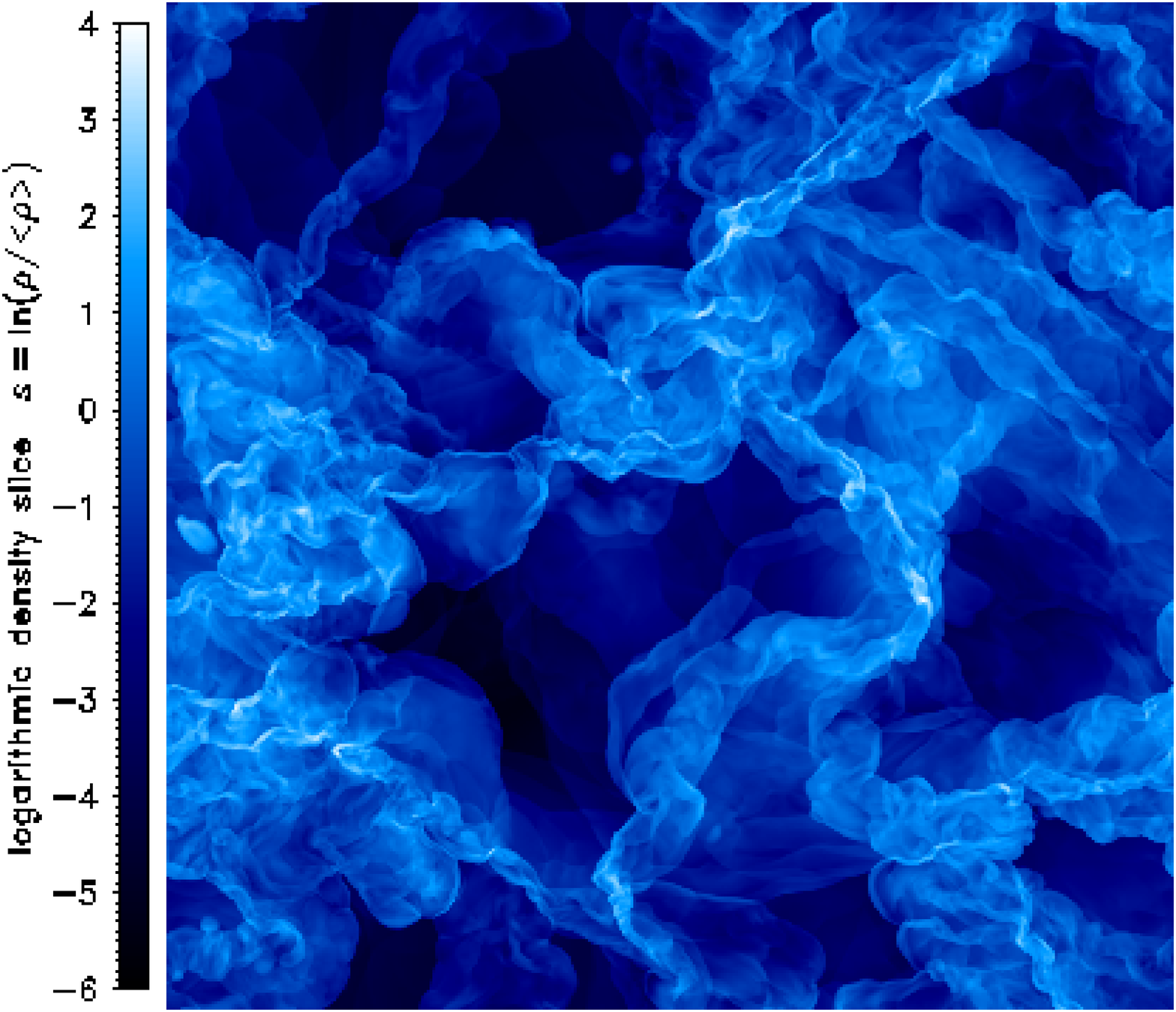} &
\includegraphics[width=0.45\linewidth]{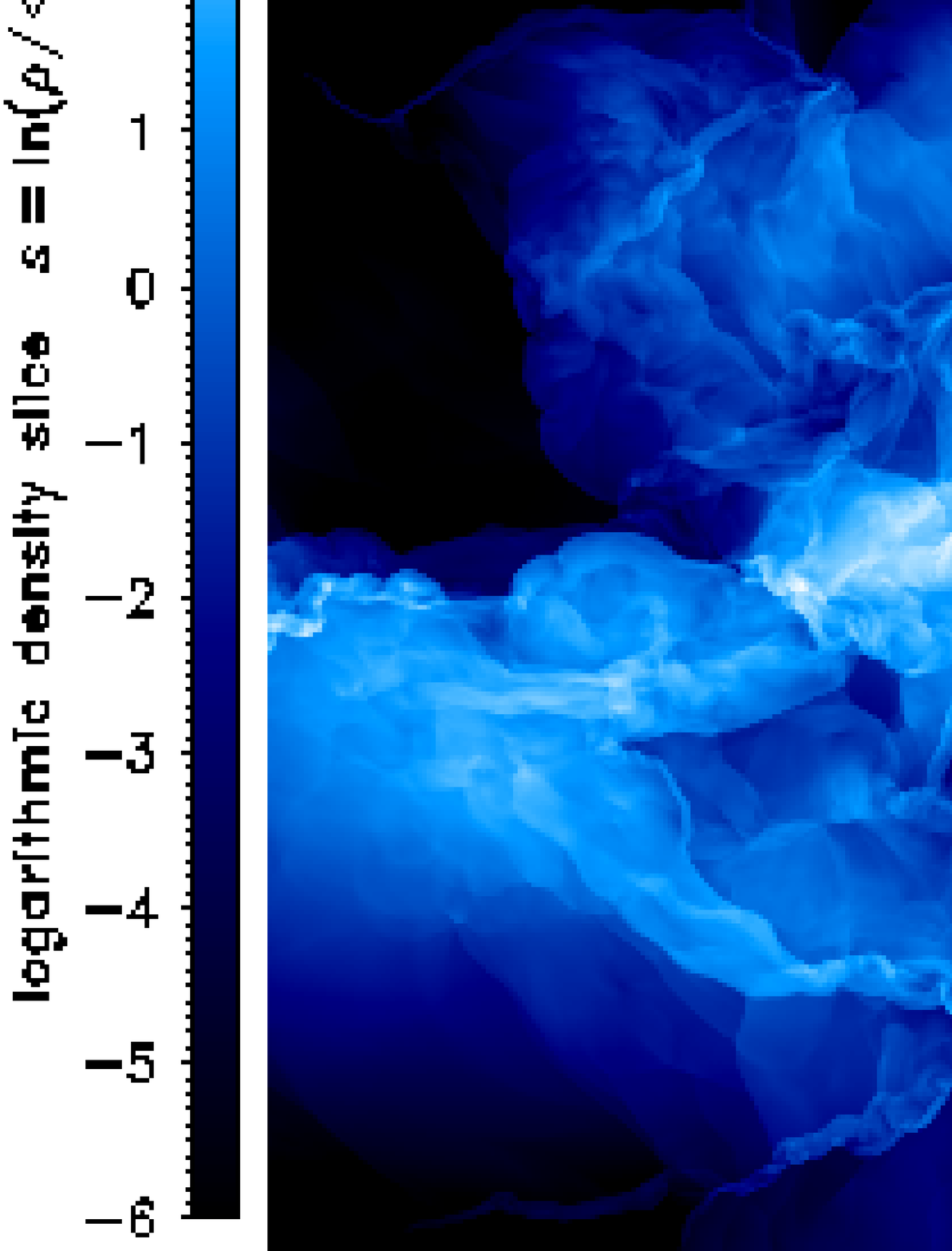} \\
\includegraphics[width=0.45\linewidth]{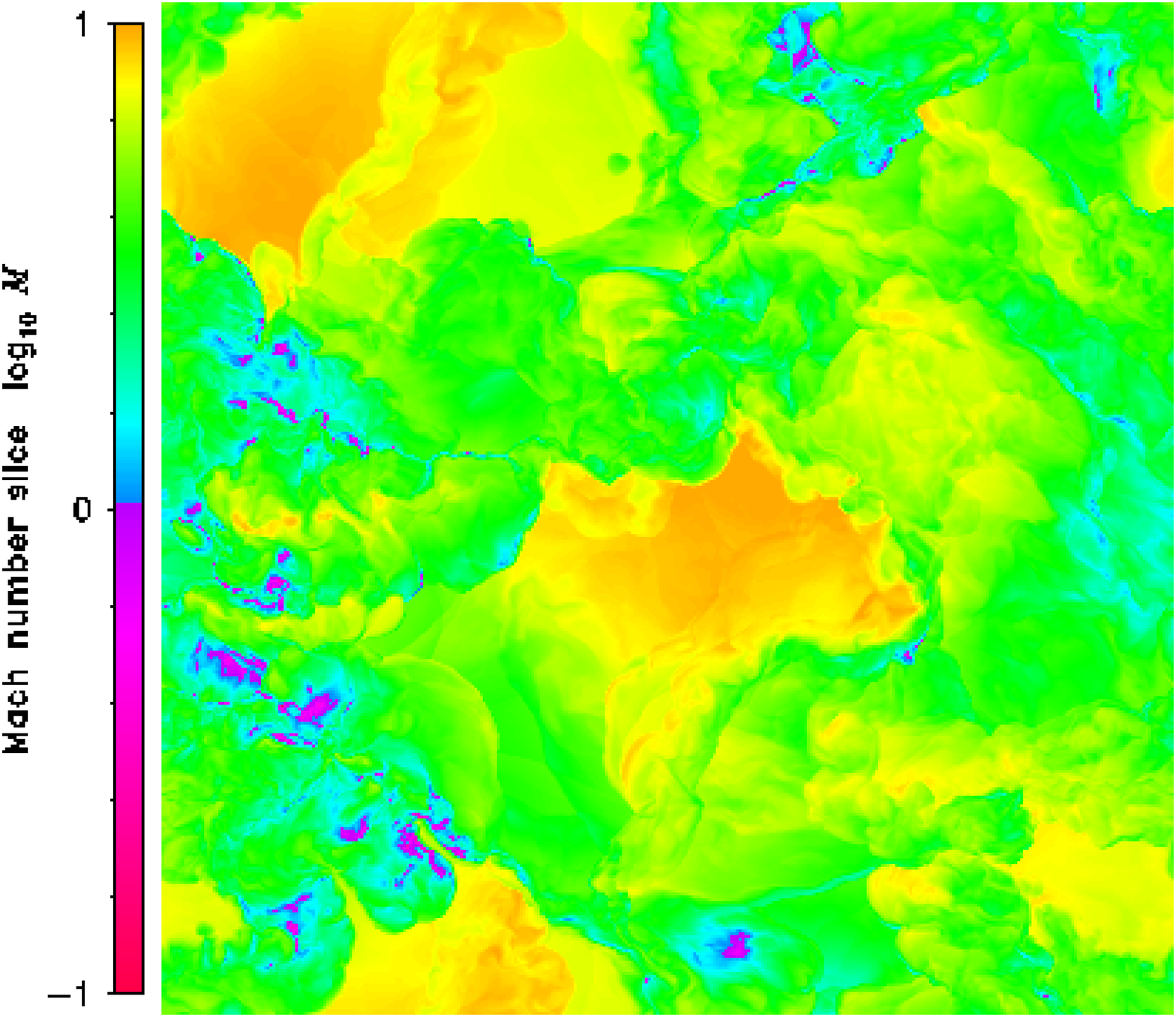} &
\includegraphics[width=0.45\linewidth]{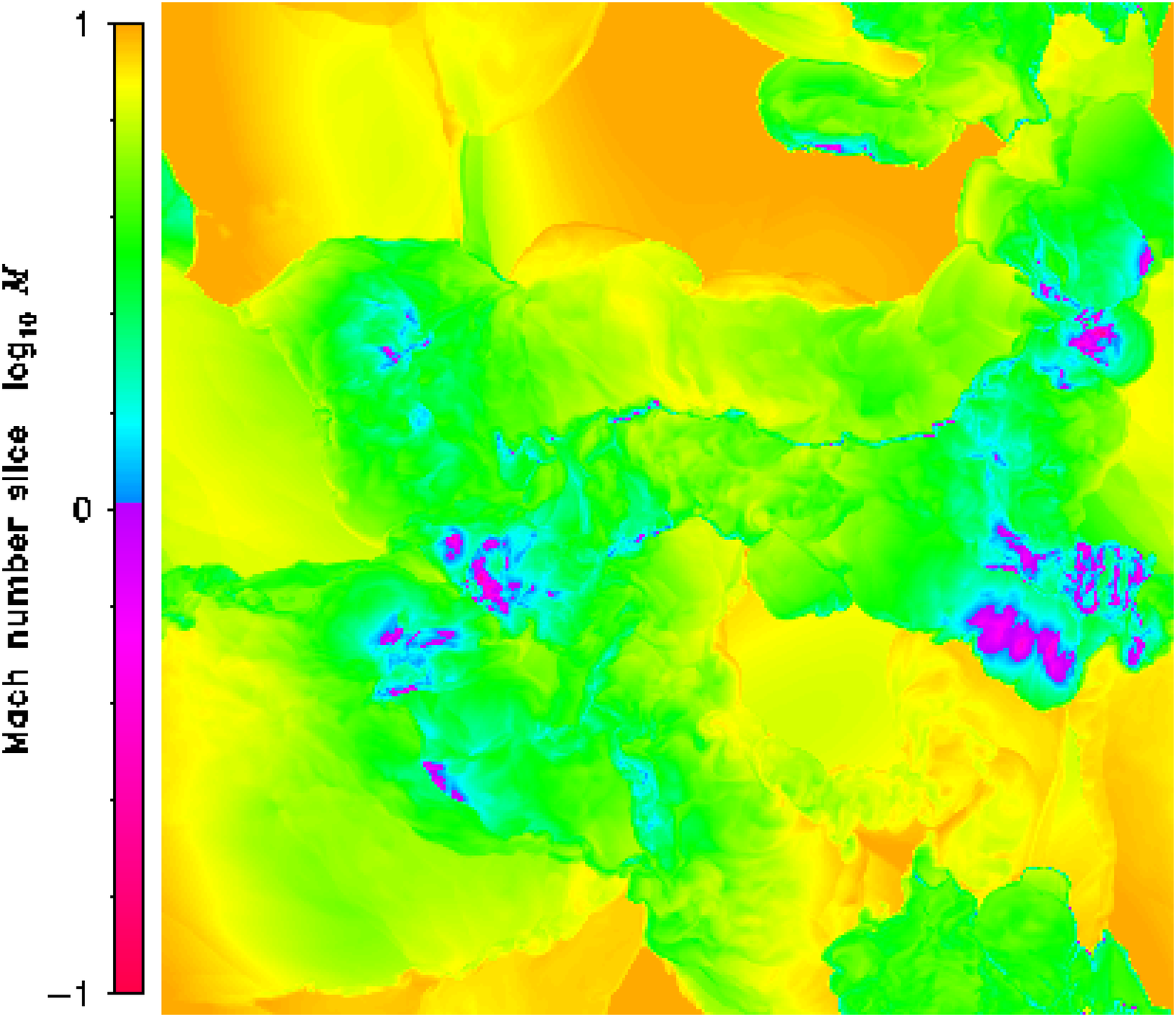}
\end{tabular}
\end{center}
\caption{$z$-slices through the local density (\emph{top panels}) and Mach number fields (\emph{bottom panels}) at $z=0$ and $t=2\,T$ for solenoidal forcing (\emph{left}), and compressive forcing (\emph{right}). Regions with subsonic velocity dispersions ($\mathrm{Mach}\!<\!1$) are distinguished from regions with supersonic velocity dispersions ($\mathrm{Mach}\!>\!1$) in the colour scheme. The correlation between density and Mach number is quite weak. However, as shown in Fig.~\ref{fig:2dpdfs}, high-density regions exhibit lower Mach numbers on average. Thus, dense cores might naturally exhibit transonic to subsonic velocity dispersions, because their sizes are expected to be comparable to the sonic scale. The sonic scale may be the transition scale to coherent cores \citep[e.g.,][]{GoodmanEtAl1998}. Although many of these `cores' here are transient, some of them are dense enough to become gravitationally bound, and accumulate enough mass to decouple from the overall supersonic turbulent flow. \emph{See the A\&A online version for a movie, showing the time evolution of this figure.}}
\label{fig:slices}
\end{figure*}

A spatial representation of the structures exhibiting subsonic velocity dispersions is shown in Figure~\ref{fig:slices} (bottom panel). These structures are identified in slices through the local Mach number $M$ as regions with $M\lesssim1$. Figure~\ref{fig:slices} (top panel) displays the corresponding density slices. The density--Mach number correlations are quite weak, as expected for isothermal turbulence (cf.~\S~\ref{sec:density_mach_correlations}). However, Figure~\ref{fig:2dpdfs} shows that high-density regions exhibit lower Mach numbers on average. In real molecular clouds, the sonic scale is expected to be located on length scales $\lambda_\mathrm{s}\approx 0.1\,\mathrm{pc}$ within factors of a few \citep[e.g.,][]{FalgaronePugetPerault1992,BarrancoGoodman1998,GoodmanEtAl1998,SchneeEtAl2007}. For instance, \citet{HeyerWilliamsBrunt2006} found $\lambda_\mathrm{s}\approx0.3\!-\!0.4\,\mathrm{pc}$ for the \object{Rosette MC} and $\lambda_\mathrm{s}\approx0.1\!-\!0.2\,\mathrm{pc}$ for \object{G216-2.5}. Furthermore, the sonic scale may be associated with the transition to coherent cores \citep{GoodmanEtAl1998,BallesterosKlessenVazquez2003,KlessenEtAl2005}. Recent simulations of turbulent core formation by \citet{SmithClarkBonnell2009} also suggest that star-forming cores typically exhibit transonic to subsonic velocity dispersions. This can be understood if cores form close to the sonic scale in a globally supersonic turbulent medium. Figure~\ref{fig:slices} suggests that regions with subsonic velocity dispersions have different shapes and sizes for both solenoidal and compressive forcings. The movie (\emph{available in the A\&A online version}) shows that these structures are transient objects, forming and dissolving in the turbulent flow \citep[e.g., see also][]{VazquezEtAl2005ApJ}. If we had included self-gravity in the present study, some of these regions would have likely collapsed gravitationally, because turbulent support becomes insufficient in some of these subsonic cores \citep[e.g.,][]{MacLowKlessen2004}.

\section{Limitations} \label{sec:limitations}
As a result of the simplicity of the hydrodynamic simulations presented in this paper, comparisons with observational data are limited and should be considered with caution. These limitations are listed below:
\begin{itemize} 
\item We assume an isothermal equation of state, so our models are strictly speaking only applicable to molecular gas of low enough density to be optically thin to dust cooling. Variations in the equation of state can lead to changes in the density statistics \citep[e.g.,][]{PassotVazquez1998,LiKlessenMacLow2003,AuditHennebelle2009}. The results of the present study apply primarily to the dense interstellar molecular gas for which an isothermal equation of state is an adequate approximation \citep{WolfireEtAl1995,Ferriere2001,PavlovskiSmithMacLow2006,GloverFederrathMacLowKlessen2009}.
\item The numerical resolution of our simulations is limited. As shown in Figure~\ref{fig:resolpdf}, the high-density tails of the PDFs systematically shift to higher densities \citep[see also][]{HennebelleAudit2007,KitsionasEtAl2009,GloverFederrathMacLowKlessen2009,PriceFederrath2010}. However, the mean and the dispersions are well converged at the numerical resolutions of $256^3$, $512^3$ and $1024^3$ grid points used in this study. The inertial scaling range is very small even at resolutions of $1024^3$ grid cells. However, the systematic difference in the inertial range scaling between resolutions of $512^3$ and $1024^3$ grid points is less than 3\% (see Appendix~\ref{app:spectra}), which is less than the typical temporal variations between different realisations of the turbulent velocity and density fields.
\item Our simulations adopt periodic boundary conditions. This implies that our simulations can only be representative of a subpart of a molecular cloud, for which we study turbulence statistics with high-resolution numerical experiments. However, we cannot take account of the boundary effects in real molecular clouds. Simulations of large-scale colliding flows \citep[e.g.,][]{HeitschEtAl2006,VazquezEtAl2006,HennebelleEtAl2008,BanerjeeEtAl2009} are more suitable for studying the boundary effects during the formation of molecular clouds.
\item We only analysed driven turbulence. However, there is ongoing debate about whether turbulence is driven or decaying \citep[e.g.,][]{StoneOstrikerGammie1998,MacLow1999,LemasterStone2008,OffnerKleinMcKee2008}. We are aware of the possibility that turbulence may in fact be excited on scales larger than the size of molecular clouds \citep[e.g.,][]{BruntHeyerMacLow2009}, but may be globally decaying (if not replenished by a mechanism acting on galactic scales). As discussed in \S~\ref{sec:forcing}, this large-scale decay can however act as an effective turbulence forcing on smaller scales, because kinetic energy is transported from large to small scales through the turbulence cascade.
\item Centroid velocity and principal component analysis were applied to PPV cubes constructed from the simulated velocity and density fields assuming optically thin radiation transfer to estimate the intensity of emission lines. This approximation will of course not hold for optically thick tracers. A full radiative transfer calculation taking account of the level population \citep[e.g.,][]{KetoEtAl2004,SteinackerBacmannHenning2006,PinteEtAl2009,HauschildtBaron2009,BaronHauschildtChen2009} of self-consistently formed and evolved chemical tracer molecules \citep[e.g.,][]{GloverMacLow2007a,GloverMacLow2007b,GloverFederrathMacLowKlessen2009} would be needed to advance on this issue.
\item We neglected magnetic fields. In order to test the role of magnetic fields in star formation \citep[e.g.,][]{CrutcherHakobianTroland2009,LunttilaEtAl2008}, we would have to include the effects of magnetic fields and ambipolar diffusion. For instance, the IMF model by \citet{PadoanNordlund2002} requires magnetic fields to explain the present-day mass function, while it is still not clear whether magnetic fields are dynamically important for typical molecular clouds. However, \citet{HeyerEtAl2008} showed that magnetohydrodynamic turbulence in the \object{Taurus MC} may lead to an alignment of flows along the field lines.
\item The present study did not include the effects of self-gravity, because we specifically focus on the pure turbulence statistics obtained in solenoidal and compressive forcings. In a follow-up study, we will include self-gravity and sink particles \citep[e.g.,][]{BateBonnellPrice1995,KrumholzMcKeeKlein2004,JappsenEtAl2005,FederrathBanerjeeClarkKlessen2010} to study the influence of the different forcings on the mass distributions of sink particles. First results indicate that the sink particle formation rate is at least one order of magnitude higher for compressive forcing compared to solenoidal forcing. \citet{VazquezBallesterosKlessen2003} argue that the star formation efficiency is mainly controlled by the RMS Mach number and the sonic scale of the turbulence (cf.~\S~\ref{sec:sonicscale}). However, our preliminary results of simulations including self-gravity show that the star formation efficiency measured at a given time (i.e., the star formation rate) is much higher for compressive forcing than for solenoidal forcing with the \emph{same} RMS Mach number \emph{and} sonic scale. This provides additional support to our main conclusion that the type of forcing must be taken into account in any theory of turbulence-regulated star formation. This needs to be investigated in future, high-resolution numerical experiments including self-gravity and sink particles.
\end{itemize}

\section{Summary and conclusions} \label{sec:conclusions}
We presented high-resolution hydrodynamical simulations of driven isothermal supersonic turbulence, which showed that the structural characteristics of turbulence forcing significantly affect the density and velocity statistics of turbulent gas \citep[see also][]{SchmidtEtAl2009}. We compared solenoidal (divergence-free) forcing with compressive (curl-free) turbulence forcing. Five different analysis techniques were used to compare our simulation data with existing observational data reported in the literature: probability density functions (PDFs), centroid velocity increments, principal component analysis, Fourier spectrum functions, and $\Delta$-variances. We find that different regions in the turbulent ISM exhibit turbulence statistics consistent with different combinations of solenoidal and compressive forcing. Varying the forcing parameter $\zeta\in[0,1]$ in equation~(\ref{eq:forcing_ratio}), we showed that a continuum of turbulence statistics exists between the two limiting cases of purely solenoidal ($\zeta=1$) and purely compressive forcing ($\zeta=0$). For $\zeta>0.5$, turbulence behaves almost like in the case of purely solenoidal forcing, while for $\zeta<0.5$, turbulence is highly sensitive to changes in $\zeta$ (cf.~\ref{fig:zeta}). Note that $\zeta=0.5$ represents the natural forcing mixture used in many previous turbulence simulations. Because the behaviour of all forcing mixtures with $\zeta>0.5$ is similar to that of purely solenoidal turbulence with $\zeta=1$ (see \ref{fig:zeta}), turbulence statistics is biased towards finding solenoidal-like values. However, observations of regions around massive stars that drive swept-up shells into the surrounding medium (e.g., the shell in the \object{Perseus MC} and in the \object{Rosette MC}) seem better consistent with models of mainly compressive forcing ($\zeta<0.5$). Note that expanding HII regions around massive stars, and supernova explosions typically create such swept-up shells, which are considered to be important drivers of interstellar turbulence \citep{MacLowKlessen2004,BreitschwerdtEtAl2009}\footnote{See also \citet{TamburroEtAl2009} for an observational study.}. A detailed list of our results is provided below:

\begin{enumerate}
\item The standard deviation (dispersion) of the probability distribution function (PDF) of the gas density is roughly three times larger for compressive forcing than for solenoidal forcing. This holds for both the 3D density distributions (Figure~\ref{fig:solcomppdf} and Table~\ref{tab:pdfs_vol}) and the 2D column density distributions (Figure~\ref{fig:solcomppdfcoldens} and Table~\ref{tab:pdfs_coldens}). We extended the density dispersion--Mach number relations, equation~(\ref{eq:sigrhoofmach}) and~(\ref{eq:sigsofmach}) originally investigated by \citet{PadoanNordlundJones1997} and \citet{PassotVazquez1998}. Based on the varying degree of compression obtained by solenoidal and compressive forcing, we developed a heuristic model for the proportionally constant $b$ in the density dispersion--Mach number relation, which takes account of the forcing parameter $\zeta$ \citep{FederrathKlessenSchmidt2008}. In the case of compressive forcing the proportionality constant $b$ is close to $b\approx1$, which confirms the result by \citet{PassotVazquez1998}. In contrast, solenoidal forcing yields $b\approx1/3$, which is in excellent agreement with recent independent high-resolution numerical simulations using solenoidal forcing \citep[e.g.,][]{BeetzEtAl2008}.

\item A parameter study of eleven models with varying forcing parameter $\zeta=[0,1]$, separated by $\Delta\zeta=0.1$ showed that the heuristic model given by equation~(\ref{eq:b}) can only serve as a first-order approximation to the forcing dependence of $b$ (cf.~Fig.~\ref{fig:zeta}). We showed that $b$ scales with the normalised power of compressible modes in the velocity field, $\left<\Psi\right>$. A good approximation for $b$ is given by $b\approx\!\sqrt{D}\left<\Psi\right>$, where $D=3$ in 3D turbulence.

\item We compared the density PDFs in our models with observations in the \object{Perseus MC} by \citet{GoodmanPinedaSchnee2009}. \citet{GoodmanPinedaSchnee2009} obtained the largest density dispersion in all of the Perseus MC within a region that they call the Shell region. This Shell surrounds the massive star \object{HD 278942} suggesting that the Shell is an expanding HII region. Swept-up shells represent geometries that can be associated with compressive turbulence forcing, because an expanding spherically symmetric shell is driven by a fully divergent velocity field. This may explain why the Shell region in the Perseus MC exhibits the largest density dispersion among all of the subregions in the Perseus MC investigated by \citet{GoodmanPinedaSchnee2009}. We emphasise that the Shell region does not exhibit the highest RMS Mach number, but has an intermediate value among the examined subregions in the Perseus MC \citep{PinedaCaselliGoodman2008}. Furthermore, as pointed out by \citet{GoodmanPinedaSchnee2009} the density dispersion--Mach number relation of the form given by equation~(\ref{eq:sigsofmach}) for a fixed parameter $b$ is \emph{not} observed for the Perseus MC. This apparent contradiction with equation~(\ref{eq:sigsofmach}) for a fixed parameter $b$ is resolved, if different turbulence forcing mechanisms operate in different subregions of the Perseus MC, such that $b$ is a function of the mixture of solenoidal and compressive modes $\zeta$ as shown in Figure~\ref{fig:zeta}.

\item The turbulent density PDF is a key ingredient for the analytical models of the core mass function (CMF) and the stellar initial mass function (IMF) by \citet{PadoanNordlund2002} and \citet{HennebelleChabrier2008,HennebelleChabrier2009}, as well as for the star formation rate models by \citet{KrumholzMcKee2005}, \citet{KrumholzMcKeeTumlinson2009} and \citet{PadoanNordlund2009}, and the star formation efficiency model by \citet{Elmegreen2008}. We showed that the dispersion of the density probability distribution is not only a function of the RMS Mach number, but also depends on the nature of the turbulence forcing. All the analytical models above rely on integrals over the density PDF. Since the dispersion of the density PDF is highly sensitive to the turbulence forcing, we conclude that star formation properties derived in those analytical models are strongly affected by the assumed turbulence forcing mechanism.

\item The PDFs $p_s(s)$ of the logarithm of the density $s=\ln(\rho/\left<\rho\right>)$ are roughly consistent with log-normal distributions for both solenoidal and compressive forcings. However, the distributions clearly exhibit non-Gaussian higher-order moments, which are associated with intermittency. Including higher-order corrections represented by skewness and kurtosis is absolutely necessary to obtain a good analytic approximation for the PDF data, because the constraints of mass conservation (eq.~\ref{eq:mass-conservation}) and normalisation (eq.~\ref{eq:normalization}) of the PDF must always be fulfilled. Even stronger deviations from perfect log-normal distributions are expected if the gas is non-isothermal \citep[e.g.,][]{PassotVazquez1998,ScaloEtAl1998,LiKlessenMacLow2003}, magnetised \citep[e.g.,][]{LiEtAl2008} or self-gravitating \citep[e.g.,][]{Klessen2000,LiEtAl2004,FederrathGloverKlessenSchmidt2008,KainulainenEtAl2009}, which often leads to exponential wings or to power-law tails in the PDFs.

\item Non-Gaussian wings of the density PDFs are a signature of intermittent fluctuations, which we further investigated using centroid velocity increments (CVIs). We find strong non-Gaussian signatures for small spatial lags $\ell$ in the PDFs of the CVIs (Figure~\ref{fig:cvipdfs}). These PDFs exhibit values of the kurtosis significantly in excess of that expected for a Gaussian (see Figure~\ref{fig:cvipdfskurt}). Figure~\ref{fig:cvipdfskurt} can be compared with \citet[][Fig.~7]{HilyBlantFalgaronePety2008}, who analysed CVIs in the \object{Taurus MC} and in the \object{Polaris Flare}. The values of the kurtosis $\mathcal{K}$ measured in the Polaris Flare are consistent with exponential values ($\mathcal{K}=6$) for short spatial lags, which is also compatible with the results of solenoidal forcing. In contrast, compressive forcing yields values of the kurtosis twice as large at small lags, which indicates that compressive forcing exhibits stronger intermittency. The scaling of the CVI structure functions supports the conclusion that compressive forcing exhibits stronger intermittency compared to solenoidal forcing (see Figure~\ref{fig:cvisess} and Table~\ref{tab:cvis}). The scaling exponents of the CVI structure functions obtained for solenoidal forcing are in good agreement with the results by \citet{HilyBlantFalgaronePety2008} obtained in the Polaris Flare for the CVI structure functions up to the $6$th order using the extended self-similarity hypothesis.

\item We applied principal component analysis (PCA) to our models. A comparison of the PCA scaling exponents $\alpha_\mathrm{PCA}$ with the PCA study in the \object{Rosette MC} and in \object{G216-2.5} by \citet{HeyerWilliamsBrunt2006} showed that solenoidal forcing is consistent with the PCA scaling measured in G216-2.5 and with the PCA scaling measured in the \emph{outside} of the HII region (Zone II) surrounding the OB star cluster \object{NGC 2244} in the Rosette MC. On the other hand, the PCA scaling \emph{inside} this HII region (Zone I) is in good agreement with the PCA scaling obtained for compressive forcing (Table~\ref{tab:pca}). Similar to the Shell region in the Perseus MC, the HII region in the Rosette MC (Zone I) displays signatures of mainly compressive forcing. Recent numerical simulations by \citet{GritschnederEtAl2009} also show that ionisation fronts driven by massive stars can efficiently excite compressible modes in the velocity field.

\item The Fourier spectra of the velocity fluctuations showed that they follow power laws in the inertial range with $E(k)\propto k^{-1.86\pm0.05}$ for solenoidal forcing and $E(k)\propto k^{-1.94\pm0.05}$ for compressive forcing. Both types of forcing are therefore compatible with the scaling of velocity fluctuations inferred from observations and independent numerical simulations. The Fourier spectra of the logarithmic density fluctuations scale as $S(k)\propto k^{-1.56\pm0.05}$ for solenoidal forcing and $S(k)\propto k^{-2.32\pm0.09}$ for compressive forcing in the inertial range.

\item The inertial range scaling of the velocity and logarithmic density fluctuations inferred from the Fourier spectra was confirmed using the $\Delta$-variance technique.

\item We computed the sonic scale by integrating the velocity Fourier spectra. The sonic scale separates supersonic turbulent fluctuations on large scales from subsonic turbulent fluctuations on scales smaller than the sonic scale. We found a break in the density fluctuation spectrum $S(k)$ for compressive forcing roughly located on the sonic scale. The typical density fluctuations computed by integration of $S(k)$ over scales larger than the sonic scale are consistent with the logarithmic density dispersions derived from the probability density functions for solenoidal and compressive forcings. On the other hand, the typical density fluctuations on scales smaller than the sonic scale are significantly smaller for both forcing types, which may reflect the transition to coherent cores \citep[e.g.,][]{GoodmanEtAl1998}. Indeed, observations show that cores typically have transonic to subsonic internal velocity dispersions \citep[e.g.,][]{BensonMyers1989,AndreEtAl2007,KirkJohnstoneTafalla2007,WardThompsonEtAl2007,LadaEtAl2008,FosterEtAl2009,FriesenEtAl2009,BeutherHenning2009}. This can be understood if cores form near the sonic scale at the stagnation points of shocks in a globally supersonic turbulent ISM (cf.~\S~\ref{sec:sonicscale}).

\item We found that the correlations between the local densities and the local Mach numbers are typically quite weak (Figures~\ref{fig:2dpdfs} and~\ref{fig:slices}). However, this weak correlation shows that the local Mach number $M$ decreases with increasing density as $M(\rho)\propto \rho^{-0.06}$ for solenoidal forcing and $M(\rho)\propto \rho^{-0.05}$ for compressive forcing for densities above the mean density. This means that dense gas tends to have smaller velocity dispersions on average, consistent with observations of dense protostellar cores.

\end{enumerate}

\acknowledgements
We thank Robi Banerjee, Paul Clark, Simon Glover, Alyssa Goodman, Patrick Hennebelle, Alexei Kritsuk, Alex Lazarian, Daniel Price, Stefan Schmeja, and Nicola Schneider for interesting discussions and valuable comments on the present work. We thank the referee, Chris Brunt for suggesting a parameter study with different forcing ratios $\zeta$, and for clarifying the effects of projection-smoothing and intensity-weighting in observations of centroid velocity maps. The $\Delta$-variance tool used in this study was provided by Volker Ossenkopf and parallelised by Philipp Grothaus. We are grateful to Alyssa Goodman, Jaime Pineda, and Nicola Schneider for sending us their Perseus MC, and Cygnus~X raw data. C.~F.~acknowledges financial support by the International Max Planck Research School for Astronomy and Cosmic Physics (IMPRS-A) and the Heidelberg Graduate School of Fundamental Physics (HGSFP). The HGSFP is funded by the Excellence Initiative of the German Research Foundation DFG GSC 129/1. This work was partly finished while C.~F. was visiting the American Museum of Natural History as a Kade fellow. R.S.K.~and C.F.~acknowledge financial support from the German {\em Bundesministerium f\"{u}r Bildung und Forschung} via the ASTRONET project STAR FORMAT (grant 05A09VHA). R.S.K.\ furthermore acknowledges financial support from the {\em Deutsche Forschungsgemeinschaft} (DFG) under grants no.\ KL 1358/1, KL 1358/4, KL 1359/5, KL 1359/10, and KL 1359/11. R.S.K.\ thanks for subsidies from a Frontier grant of Heidelberg University sponsored by the German Excellence Initiative and for support from the {\em Landesstiftung Baden-W{\"u}rttemberg} via their program International Collaboration II (grant P-LS-SPII/18 ). M.-M.~M.~L. acknowledges partial support for his work from NASA Origins of Solar Systems grant NNX07AI74G. The simulations used computational resources from the HLRBII project grant h0972 at Leibniz Rechenzentrum Garching. The software used in this work was in part developed by the DOE-supported ASC~/~Alliance Center for Astrophysical Thermonuclear Flashes at the University of Chicago.

\bibliographystyle{aa}
% \bibliography{../../ms.bib}

\begin{appendix}

\normalsize{

\begin{figure*}[t]
\centerline{
\includegraphics[width=0.5\linewidth]{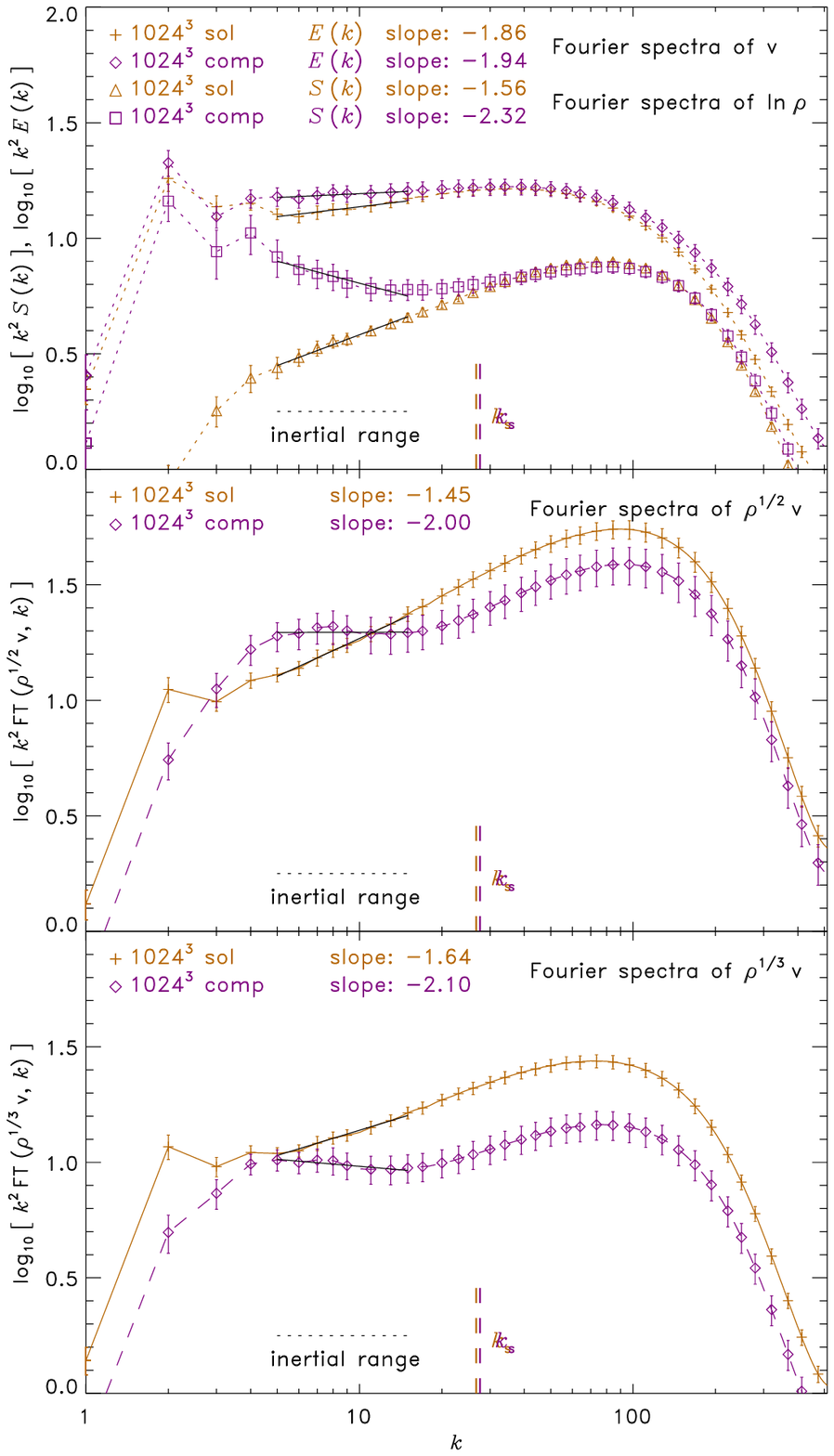}
\includegraphics[width=0.5\linewidth]{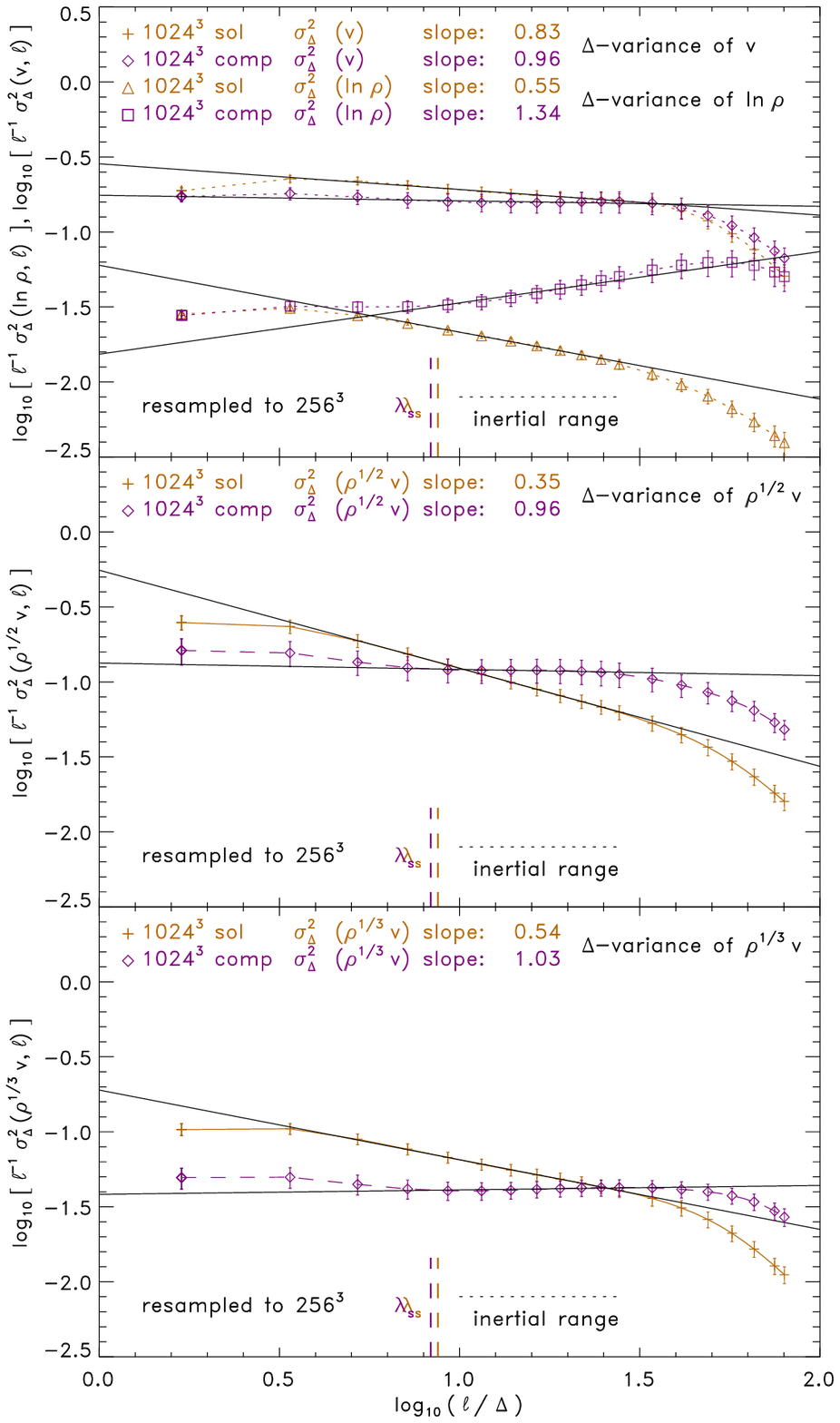}
}
\caption{\emph{Top panels:} Same as Figure~\ref{fig:spect}. \emph{Middle panels:} Same as top panels, but instead of the Fourier spectra and $\Delta$-variances of $v$, the Fourier spectra and $\Delta$-variances of the density-weighted velocity $\rho^{1/2}v$ are shown. The quantity $\rho^{1/2}v$ has physical reference to kinetic energy. \emph{Bottom panels:} Same as middle panels, but the Fourier spectra and $\Delta$-variances of the density-weighted velocity $\rho^{1/3}v$ are shown. The quantity $\rho^{1/3}v$ has physical reference to a constant kinetic energy dissipation within the inertial range \citep{KritsukEtAl2007,SchmidtFederrathKlessen2008}.}
\label{fig:spect_joint}
\end{figure*}

\section{Fourier spectra and $\Delta$-variance scaling of the combined quantities $\rho^{1/2 }v$ and $\rho^{1/3 }v$} \label{app:spectra_sqrtrho_rho3}
In this section we present the Fourier spectra and $\Delta$-variance results for the combined quantities $\rho^{1/2 }v$ and $\rho^{1/3 }v$. Usually, the pure velocity scaling is considered without density weighting. However, for highly supersonic turbulence it is interesting to investigate the scaling of combinations of density and velocity. Note that CVIs (\S~\ref{sec:cvis}) and PCA (\S~\ref{sec:pca}) also analyse convolutions of density and velocity statistics. Figure~\ref{fig:spect_joint} (top panel) shows a repetition of Figure~\ref{fig:spect} (scaling of $v$) together with the scaling of $\rho^{1/2 }v$ (middle panel) and $\rho^{1/3 }v$ (bottom panel) for direct comparison. Since Fourier spectra and $\Delta$-variance analyses always represent the mean squares of these quantities, $\rho^{1/2 }v$ corresponds to the scaling of the kinetic energy density $\rho\,v^2$. As shown by \citet{KritsukEtAl2007} \citep[see also][]{Henriksen1991,Fleck1996}, $\rho^{1/3 }v$ corresponds to a constant energy flux within the inertial range. This idea was first proposed by \citet{Lighthill1955}. Using the eddy turnover time $t_\ell$ as the typical evolution timescale of a turbulent fluctuation on scale $\ell$, the constancy of energy flux in the inertial range is defined as
\begin{equation}
\frac{\rho v^2}{t_\ell} \propto \frac{\rho v^2}{\ell/v} \propto \frac{\rho v^3}{\ell} \propto \mathrm{const}\;,
\end{equation}
which leads to the original \citet{Kolmogorov1941c} scaling (but now including density variations),
\begin{equation}
\rho^{1/3 }v \propto \ell^{1/3}
\end{equation}
for the quantity $\rho^{1/3 }v$. Using the extended self-similarity hypothesis \citep{BenziEtAl1993}, we showed in \citet{SchmidtFederrathKlessen2008} that the relative scaling exponents of $\rho^{1/3 }v$ provide a more universal scaling compared to the velocity scaling without density weighting. Figure~\ref{fig:spect_joint} (bottom panel) shows that the absolute scaling inferred from the Fourier spectra of $\rho^{1/3 }v$ is indeed close to the \citet{Kolmogorov1941c} scaling (scaling proportional to $k^{-5/3}$) for solenoidal forcing, which is in agreement with the results obtained in \citet{KritsukEtAl2007}. However, compressive forcing yields significantly steeper scaling (also for $\rho^{1/2 }v$), which is close to \citet{Burgers1948} turbulence (scaling proportional to $k^{-2}$). The corresponding results inferred from the $\Delta$-variance analyses are compatible with the Fourier spectra to within the uncertainties. Both quantities $\rho^{1/2 }v$ and $\rho^{1/3 }v$ show breaks in the scaling close to the sonic wavenumber $k_\mathrm{s}$ for compressive forcing.

\section{Convergence test for the structure functions of centroid velocity increments} \label{app:cviconvergence}
For an accurate and reliable determination of the structure function scaling, it must be verified that the number of data pairs used for sampling the structure functions was high enough to yield converged results. There is no general rule to determine a priori the number of data pairs necessary, because the required number of data pairs depends on the underlying statistics of the measured variable itself and on the desired structure function order. However, convergence can be tested by increasing the number of data pairs used for computing the structure functions. Showing that the computed structure functions do not change significantly by further increasing the number of data pairs demonstrates convergence. Furthermore, if convergence is verified for the highest order under consideration, then the structure functions of lower order are also converged. This is because the higher-order structure functions of a variable $q$ reflect the statistics of higher powers of $q$ than the lower order structure functions. This is reflected in the definition of the $p$th order structure function in equation~(\ref{eq:cvisf}).

\begin{figure}[!t]
\centerline{
\includegraphics[width=1.0\linewidth]{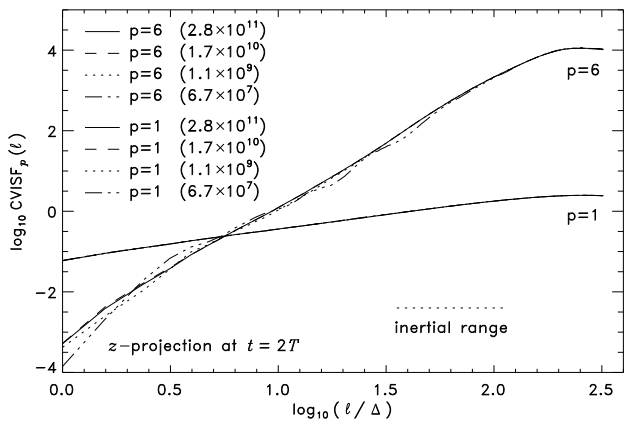}
}
\caption{The $1$st ($p=1$) and $6$th ($p=6$) order structure functions of the centroid velocity increments sampled with different numbers of data pairs is shown for a single snapshot at time $t=2\,T$ in $z$-projection for the case of compressive forcing. The number of data pairs used for sampling is given in brackets. The structure functions  of centroid velocity increments are statistically converged for $p\leq6$ for sample sizes of at least $1.7\times10^{10}$ data pairs per turbulent realisation and per projection as used throughout this study.}
\label{fig:cviconvergence}
\end{figure}

Figure~\ref{fig:cviconvergence} demonstrates convergence for the structure functions of CVIs with orders $p\leq6$ discussed in~\S~\ref{sec:cvisfs}. We only show the compressive forcing case for clarity, but we also verified convergence for the solenoidal forcing case with the same method. Figure~\ref{fig:cviconvergence} shows that sampling each structure function with roughly $1.7\times10^{10}$ data pairs is sufficient to yield converged results. The total number of data pairs used to construct the CVI structure functions shown in Figures~\ref{fig:cvis} and~\ref{fig:cvisess} was thus roughly $81\times3\times1.7\times10^{10}\approx4.1\times10^{12}$ from averaging over 81 realisations of the turbulence and three projections along the $x$, $y$ and $z$-axes for each of these realisations.

\section{Resolution study of the Fourier spectra and their dependence on the numerical scheme} \label{app:spectra}
The resolution and type of numerical method adopted to model supersonic turbulence are expected to critically affect the scaling of Fourier spectrum functions in the inertial range \citep[e.g.,][]{KleinEtAl2007,KritsukEtAl2007,PadoanEtAl2007}. In this section, we investigate the dependence of our Fourier spectra on the numerical resolution and on the numerical scheme used in the present study.

\subsection{Resolution study}

\begin{figure*}[!t]
\centerline{
\includegraphics[width=0.5\linewidth]{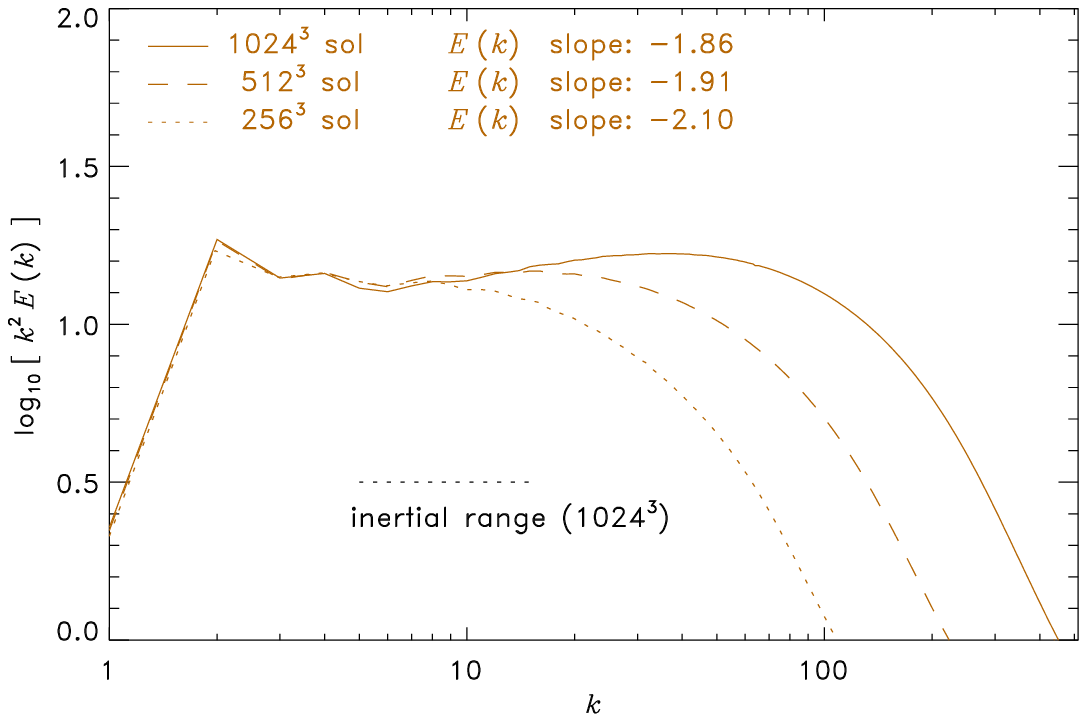}
\includegraphics[width=0.5\linewidth]{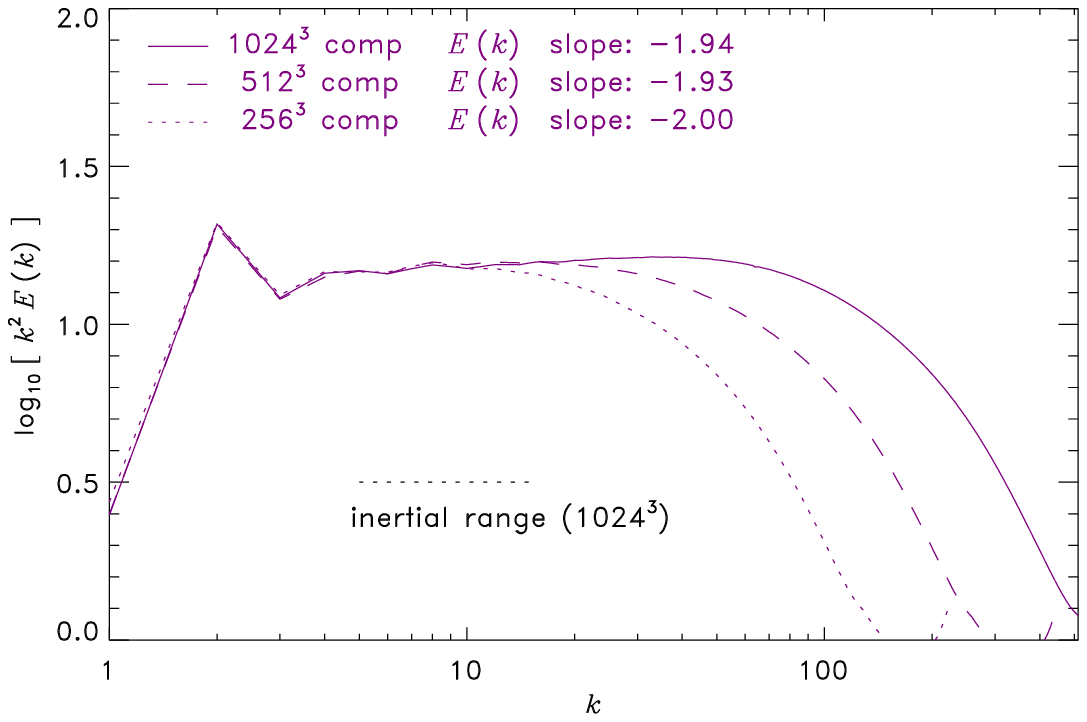}
}
\caption{Time-averaged velocity Fourier spectra $E(k)$ defined in equation~(\ref{eq:ft_e}) for numerical resolutions of $256^3$, $512^3$ and $1024^3$ grid points obtained with solenoidal forcing (\emph{left}) and compressive forcing (\emph{right}). The inferred inertial range scaling is converged to within less than 3\% at the typical resolution of $1024^3$ grid points used throughout this study for both types of forcing.}
\label{fig:spect_resol}
\end{figure*}

Figure~\ref{fig:spect_resol} shows velocity Fourier spectra $E(k)$ defined in equation~(\ref{eq:ft_e}) for numerical resolutions of $256^3$, $512^3$ and $1024^3$ grid points. The inertial range scaling is indeed affected by the numerical resolution. For solenoidal forcing, the inertial range scaling exponent $\beta$ at resolution of $256^3$ grid cells is roughly 13\% higher than the scaling exponent at a resolution of $1024^3$. However, the difference between the inertial range scaling at $512^3$ and $1024^3$ is less than 3\% for solenoidal forcing. For compressive forcing, the difference between the inertial range scaling exponents at resolutions of $512^3$ and $1024^3$ grid cells is less than 1\%. This result indicates that the systematic dependence of the inertial range scaling on the numerical resolution is less than 3\% for both solenoidal and compressive forcings. It should be emphasised that variance effects introduced by different realisations of the turbulence are typically on the order of 5--10\% (see error bars in Figure~\ref{fig:spect}), which is higher than the systematic errors introduced by resolution effects, as long as the numerical resolution is at least $512^3$ grid cells.

\subsection{Dependence on parameters of the piecewise parabolic method}
We used the piecewise parabolic method (PPM) \citep{ColellaWoodward1984} to integrate the equations of hydrodynamics (eqs.~\ref{eq:hydro1} and~\ref{eq:hydro2}). PPM improves on the finite-volume scheme originally developed by \citet{Godunov1959} by representing the flow variables with piecewise parabolic functions, which makes the PPM second-order accurate in smooth flows. However, PPM is also particularly suitable for the accurate modelling of turbulent flows involving sharp discontinuities, such as shocks and contact discontinuities. For that purpose, PPM uses a lower artificial viscosity controlled by the PPM diffusion parameter $K$. In three simulations with resolutions of $512^3$ grid cells, we varied the PPM diffusion parameter $K$ between $0.0$, $0.1$ and $0.2$. Note that $K=0.1$ is the value recommended by \citet{ColellaWoodward1984}, which was used for all production runs throughout this study. The PPM algorithm furthermore includes a steepening mechanism to keep contact discontinuities from spreading over too many cells. In one additional run at $512^3$, we switched off the PPM steepening algorithm to check its influence on our results.

\begin{figure*}[!t]
\centerline{
\includegraphics[width=0.5\linewidth]{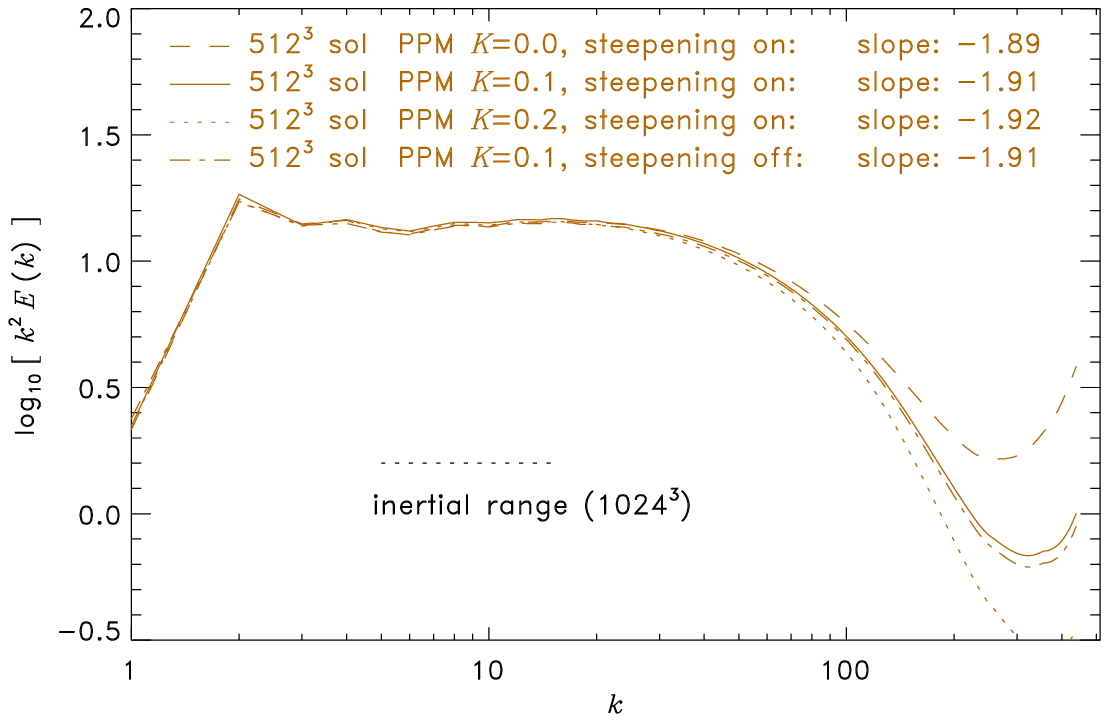}
\includegraphics[width=0.5\linewidth]{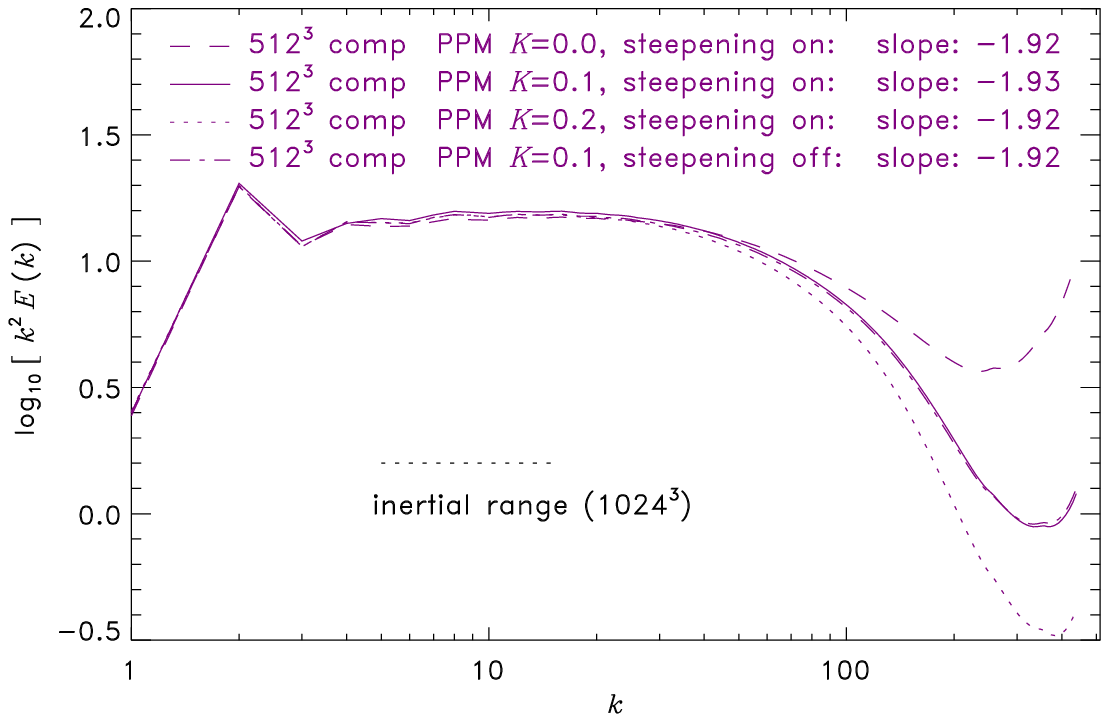}
}
\caption{Dependence of the time-averaged velocity Fourier spectra $E(k)$ on parameters of the piecewise parabolic method (PPM) \citep{ColellaWoodward1984} at fixed resolution of $512^3$ grid cells. Varying the PPM diffusion parameter $K$ between $0.0$, $0.1$ and $0.2$ affects the dissipation range at wavenumbers $k\gtrsim40$. However, the effect of varying the PPM diffusion parameter is negligible for $k\lesssim40$. Switching off the PPM steepening algorithm for contact discontinuities has also virtually no effect on the Fourier spectra at $k\lesssim40$.}
\label{fig:spect_ppm}
\end{figure*}

Figure~\ref{fig:spect_ppm} shows that the velocity spectra $E(k)$ decrease faster with increasing diffusion parameter $K$ for wavenumbers $k\gtrsim40$. It is expected that the scheme dissipates more kinetic energy on small scales with increasing $K$, because the PPM diffusion algorithm is designed to act on shocks only \citep[][eq.~4.5]{ColellaWoodward1984}. In contrast, Figure~\ref{fig:spect_ppm} demonstrates that the Fourier spectra at wavenumbers $k\lesssim40$ are hardly affected by the PPM diffusion algorithm for both solenoidal and compressive forcings. Note that \citet{KritsukEtAl2007} reported that their results for the inertial range scaling are highly sensitive to the choice of PPM diffusion parameter in the ENZO code. However, our results demonstrate that the choice of PPM diffusion parameter only affects the inertial range scaling within less than 1\%, which is clearly less than the influence of the numerical resolution and less than the typical snapshot-to-snapshot variations. Figure~\ref{fig:spect_ppm} furthermore demonstrates that the PPM contact discontinuity steepening has negligible effects for simulations of supersonic turbulence.

The results obtained here support our conclusion in~\S~\ref{sec:spectra} that the Fourier spectra at resolutions of $1024^3$ grid cells are robust for wavenumbers $k\lesssim40$.

}

\end{appendix}

\end{document}